\documentclass[prd,showpacs,superscriptaddress,nofootinbib,preprint,11pt]{revtex4}
\usepackage{amssymb,amsmath,amsfonts,epsfig,subfigure}

\newcommand{\beqs}{\begin{equation*}}
\newcommand{\beq}{\begin{equation}}

\newcommand{\eeqs}{\end{equation*}}
\newcommand{\eeq}{\end{equation}}

\newcommand{\beqas}{\begin{eqnarray*}}
\newcommand{\beqa}{\begin{eqnarray}}

\newcommand{\eeqas}{\end{eqnarray*}}
\newcommand{\eeqa}{\end{eqnarray}}

\newcommand{\eq}[2]{\begin{equation} #1 \label{#2} \end{equation}}

\newcommand{\eps}{\varepsilon}
\newcommand{\al}{\alpha}
\newcommand{\be}{\beta}
\newcommand{\ga}{\gamma}
\newcommand{\de}{\delta}
\newcommand{\om}{\omega}
\newcommand{\ka}{\kappa}
\newcommand{\la}{\lambda}
\newcommand{\si}{\sigma}

\newcommand{\Ga}{\Gamma}
\newcommand{\De}{\Delta}
\newcommand{\Om}{\Omega}

\newcommand{\La}{\Lambda}
\newcommand{\Si}{\Sigma}

\newcommand{\blist}{\begin{itemize}}

\newcommand{\elist}{\end{itemize}}

\providecommand{\href}[2]{#2}
\newcommand{\URL}[1]{{\href{#1}{{\tt{\small{#1}}}}}}

\DeclareFontFamily{OT1}{rsfs}{}
\DeclareFontShape{OT1}{rsfs}{m}{n}{ <-7> rsfs5 <7-10> rsfs7 <10->rsfs10}{} 
\DeclareMathAlphabet{\mycal}{OT1}{rsfs}{m}{n}

\DeclareMathOperator{\extdm}{d}
\newcommand{\extd}{\extdm \!}

\usepackage{slashed}
\usepackage{bbold}
\usepackage{cancel}
\usepackage{amsthm}
\usepackage{bm}
\usepackage{dcolumn}
\usepackage{epsfig}
\usepackage{graphicx}
\usepackage{graphics}
\usepackage[latin1]{inputenc}
\usepackage{latexsym}
\usepackage{rotating}
\usepackage{hyperref}

\usepackage{picture,epic}

\newcommand{\Tr}{\textrm{Tr}}

\newcommand{\cO}{\mathcal{O}}
\newcommand{\cT}{{\cal T}}

\newcommand{\cH}{\mathcal{H}}

\def\ric{R}

\newcommand{\weyl}{\Om} 
\newcommand{\lagA}{\iota} 
\newcommand{\lagB}{o} 
\newcommand{\lagC}{\zeta} 
\newcommand{\covD}{{\cal D}} 
\newcommand{\ternary}{\psi} 
\newcommand{\exponent}{b} 
\newcommand{\PMR}{J} 
\newcommand{\ECP}{\zeta}
\newcommand{\aff}{{\cal J}}
\newcommand{\vir}{{\cal L}}
\newcommand{\CS}{\textrm{CS\,}}
\newcommand{\Ric}{\slashed{R}}

\begin{document}

\title{Conformal Chern--Simons holography --- lock, stock and barrel}

\author{Hamid Afshar}
\email{afshar@ipm.ir}

\affiliation{Department of Physics, Sharif University of Technology, P.~O. Box 11365-9161, Tehran, Iran}
\affiliation{School of Physics, Institute for Research in Fundamental Sciences (IPM), P.~O. Box 19395-5531, Tehran, Iran}
\affiliation{Institute for Theoretical Physics, Vienna University of Technology, Wiedner Hauptstrasse 8-10/136, A-1040 Vienna, Austria, Europe}

\author{Branislav Cvetkovi{\'c}}
\email{cbranislav@ipb.ac.rs}
\affiliation{University of Belgrade, Institute of Physics, P.~O. Box 57, 11001 Belgrade, Serbia}

\author{Sabine~Ertl}
\email{sertl@hep.itp.tuwien.ac.at}
\affiliation{Institute for Theoretical Physics, Vienna University of Technology, Wiedner Hauptstrasse 8-10/136, A-1040 Vienna, Austria, Europe}

\author{Daniel Grumiller}
\email{grumil@hep.itp.tuwien.ac.at}
\affiliation{Institute for Theoretical Physics, Vienna University of Technology, Wiedner Hauptstrasse 8-10/136, A-1040 Vienna, Austria, Europe}

\author{Niklas Johansson}
\email{niklasj@hep.itp.tuwien.ac.at}
\affiliation{Institute for Theoretical Physics, Vienna University of Technology, Wiedner Hauptstrasse 8-10/136, A-1040 Vienna, Austria, Europe}

\date{\today}

\preprint{TUW-11-22}


\begin{abstract}
We discuss a fine-tuning of rather generic three dimensional higher-curvature gravity actions that leads to gauge symmetry enhancement at the linearized level via partial masslessness. 
Requiring this gauge symmetry to be present also non-linearly reduces such actions to conformal Chern--Simons gravity.
We perform a canonical analysis of this theory and construct the gauge generators and associated charges.
We provide and classify admissible boundary conditions.  
The boundary conditions on the conformal equivalence class of the metric render one chirality of the partially massless Weyl gravitons normalizable and the remaining one non-normalizable.
There are three choices --- trivial, fixed or free --- for the Weyl factors of the bulk metric and of the boundary metric.
This proliferation of boundary conditions leads to various physically distinct scenarios of holography that we study in detail, extending considerably the discussion initiated in Ref.~\cite{Afshar:2011yh}.
In particular, the dual CFT may contain an additional scalar field with or without background charge, depending on the choices above.
\end{abstract}

\pacs{04.60.Rt,04.20.Ha,11.25.Tq,11.15.Wx,11.15.Yc}

\maketitle
\tableofcontents

\section{Introduction}

Gravity in three dimensions belongs to the intriguing intersection between problems that are tractable and problems that seem relevant.
It has been used both as a toy model for classical and quantum gravity \cite{Staruszkiewicz:1963zz,Deser:1982vy,Deser:1982wh,Deser:1982a,Witten:1988hc,Banados:1992wn,Banados:1992gq,Strominger:1997eq,Carlip:1998uc} and as an adequate description of certain physical situations, such as the gravitational field close to cosmic strings \cite{Deser:1984tn}.
In recent years focus has been on 3-dimensional quantum gravity in anti-de Sitter space (AdS), as this allows novel insights into and applications of the AdS/CFT correspondence \cite{Witten:2007kt,Li:2008dq,Grumiller:2008qz,Ertl:2009ch} (see \cite{Grumiller:2010tj} for a more extensive list of recent Refs.).

An interesting class of pure gravity models depending solely on the metric $g$ is described by actions without derivatives of curvature:
\eq{
S[g]=\frac{1}{\ka^2}\,\int\!\extd^3x\sqrt{-g}\,L(g_{ab},\, R,\, R_{ab},\, R_{abcd})  + \frac{1}{2\ka^2\mu}\,\int\!\extd^3x\, \CS(\Ga) \,.
}{eq:gCS1}
Here $L$ is some scalar function of curvature invariants and $\CS(\Ga)$ is the gravitational Chern--Simons term, whose existence is a unique feature of gravity in $4n-1$ dimensions.
In three dimensions it reads
\eq{
\CS(\Ga) = \epsilon^{\la\mu\nu}\,\Ga^\si{}_{\la\rho}\,\Big(\partial_\mu\Ga^\rho{}_{\nu\si}+\tfrac23\,\Ga^\rho{}_{\mu\tau}\Ga^\tau{}_{\nu\si}\Big)\,.
}{eq:gCS2}
Besides the gravitational coupling constant $\ka^2=16\pi G_N$ and the Chern--Simons coupling constant $\mu$ there may be further coupling constants contained in $L$.
Since in three dimensions the Riemann tensor is determined uniquely from the Ricci tensor and the metric, the function $L$ can be simplified correspondingly.
Schouten identities further reduce the number of independent terms in the action \eqref{eq:gCS1}.
The most general higher curvature theory without derivatives of curvatures then contains a function $L$ that can be written as a formal power series involving only three curvature invariants:
the Ricci scalar $R$, the invariant $R_{(2)}=\Ric_{\mu\nu}\Ric^{\mu\nu}$, which is quadratic in the tracefree Ricci-tensor $\Ric_{\mu\nu}=R_{\mu\nu}-\frac13\,Rg_{\mu\nu}$, and the cubic curvature invariant $R_{(3)}=\Ric_{\mu\nu}\Ric^\mu_\tau\Ric^{\nu\tau}$ \cite{Paulos:2010ke}:
\eq{
L = \si R - 2 \La + \sum_{nmk}\,\la_{nmk} R^n R_{(2)}^m R_{(3)}^k\,.
}{eq:gCS5}
The coefficients $\La$ and $\la_{nmk}$ are coupling constants.
The sign $\si=\pm$ determines the sign of massive graviton energies, black hole masses and central charges (if applicable).
We are going to come back to the sign issue below.

We require the existence of an AdS solution
\eq{
g^{\textrm{\tiny AdS}}_{\mu\nu}\extd x^\mu\extd x^\nu = \ell^2\,\big(\extd\rho^2-\cosh^2{\!\!\rho}\, \extd t^2 +\sinh^2{\!\!\rho}\,\extd\varphi^2\big)
}{eq:gCS6}
of the classical equations of motion. The AdS radius $\ell$ in the line-element \eqref{eq:gCS6} is determined by the cosmological constant $\La$ and the remaining coupling constants $\la_{nmk}$.
Consistency with a holographic $c$-theorem is guaranteed if the coupling constants $\la_{nmk}$ are restricted by the following linear relations among them \cite{Paulos:2010ke}:\footnote{%
For each order $N$ there are $N-1$ such relations.
The number of independent coupling constants remaining at any order $N>1$ after imposing the conditions \eqref{eq:gCS6a} is given by the $x^N$ coefficient in the Taylor--MacLaurin expansion of $(1-x-x^2+x^4+x^5-x^7)/[(1-x)^2(1-x^2)(1-x^3)]$, which can be written as $(N-1)(N-5)/12+89/72+(-1)^N/8+2/9\,\cos{(2\pi N/3)}$ (see A001399 at The On-Line Encyclopedia of Integer Sequences \URL{http://oeis.org/}).
This implies that there is only one coupling constant per order $N<6$ and $N(N-6)/12 + {\cal O}(1)$ independent coupling constants per order in the limit of large $N$.}
\eq{
\sum_{N=n+2m+3k>1} \!\!\!\!\!\!\!\!\!\la_{nmk}\,(-4)^n(2/3)^m(-2/9)^k\,\binom{n}{r}= 0 \qquad  0\leq r< N-1\,.
}{eq:gCS6a}

The most prominent example of a theory of this type is (cosmological) topologically massive gravity (TMG) \cite{Deser:1982vy,Deser:1982wh,Deser:1982a,Deser:1982sv} with
\eq{
L_{\textrm{\tiny TMG}} = \si R - 2 \La\,.
}{eq:gCS3}
The coupling constant $\La$ contained in $L_{\textrm{\tiny TMG}}$ is the (negative) cosmological constant.
A more recent example is provided by Generalized Massive Gravity (GMG; without Chern--Simons term also known as ``New Massive Gravity'' or NMG) \cite{Bergshoeff:2009hq,Bergshoeff:2009aq}, with $L$ given by
\eq{
L_{\textrm{\tiny GMG}} = \si R - 2 \La + \frac{1}{m^2}\,\big(R_{(2)}-\frac{1}{24}\,R^2\big)\,.
}{eq:gCS4}
An additional coupling constant contained in $L_{\textrm{\tiny GMG}}$ is $\la_{010}=1/m^2$, which is allowed to be negative.
The relative factor $-1/24$ between the two terms in the bracket in \eqref{eq:gCS4} is fixed through \eqref{eq:gCS6a} for $N=2$.
Other examples are various extensions of GMG to cubic, quartic \cite{Sinha:2010ai} or quintic \cite{Paulos:2010ke} order, and Born--Infeld gravity \cite{Gullu:2010pc}
consistent with the physical requirement of ghost freedom spelled out in \cite{Deser:1998rj}.

There is something universal about the theories described by actions \eqref{eq:gCS1}--\eqref{eq:gCS6a}.
Namely, for generic values of the coupling constants the linearized fluctuations $h_{\mu\nu}$ around AdS \eqref{eq:gCS6} obey a fourth order partial differential equation \cite{Paulos:2010ke}, which in transversal gauge $\nabla_\mu h^{\mu\nu}=0$ can be written as
\eq{
({\cal D}^L{\cal D^R} {\cal D}^{M_1} {\cal D}^{M_2} h)_{\mu\nu} = 0\,,
}{eq:gCS7}
with mutually commuting first order differential operators introduced in Ref.~\cite{Li:2008dq}
\begin{align}\label{eq:gCS8}
\big({\cal D}^{L/R}\big)_\mu^\nu &= \de_\mu^\nu \pm \ell \,\varepsilon_\mu{}^{\tau\nu}\nabla_\tau \, ,\\  
\big({\cal D}^{M_{1,2}}\big)_\mu^\nu &= \de_\mu^\nu + \frac{1}{M_{1,2}}\,\varepsilon_\mu{}^{\tau\nu}\nabla_\tau\,.\label{eq:gCS8.5}
\end{align}

\begin{table}
\begin{center}
\begin{tabular}{l|ll|l}
 masses & acronym & dof & theory \\ \hline
$\infty,\, \infty$ & EHG & 0 & Einstein--Hilbert gravity (1915) \cite{Einstein:1915by,Hilbert:1915tx} \\
$\infty,\, \pm 1/\ell$ & $\chi$G & 0 & chiral gravity (2008) \cite{Li:2008dq} \\
$\infty,\, \pm 1/\ell$ & LOG & 1 & log gravity (2008) \cite{Grumiller:2008qz} \\
$\boldsymbol{\infty,\, 0}$ & {\bf CSG} & {\bf 0} & {\bf conformal Chern--Simons gravity (1982) \cite{Deser:1982wh,Horne:1988jf}} \\
$\infty,\, M_1$ & TMG & 1 & topologically massive gravity (1982) \cite{Deser:1982vy,Deser:1982wh} \\ \hline
$\pm 1/\ell,\, \pm 1/\ell$ & G$\chi$G & 0 & generalized chiral gravity (2009) \cite{Liu:2009pha} \\
$\pm 1/\ell,\, \pm 1/\ell$ & L$^2$G & 2 & log squared gravity (2009) \cite{Liu:2009pha,Grumiller:2010tj} \\
$\pm 1/\ell,\, \mp 1/\ell$ & LNG & 2 & log new massive gravity (2009) \cite{Liu:2009kc,Grumiller:2009sn} \\
$\pm 1/\ell,\, M_1$ & LGG & 2 & log generalized massive gravity (2009)\cite{Liu:2009pha,Grumiller:2010tj} \\
$\pm 1/\ell,\, 0$ & LPG & 2 (1) & log partially massless gravity \\
$0,\, 0$ & PMG & 2 (1) & partially massless gravity (2010) \cite{Grumiller:2010tj} \\
$0,\, M_1$ & GPG & 2 (1) & generalized partially massless gravity  (2009) \cite{Oliva:2009ip} \\
$M_1,\, M_1$ & MLG & 2 & massive log gravity (2010) \cite{Grumiller:2010tj} \\
$M_1,\, -M_1$ & NMG & 2 & new massive gravity (2009) \cite{Bergshoeff:2009hq,Bergshoeff:2009aq} \\
$M_1,\, M_2$ & GMG & 2 & generalized massive gravity (2009) \cite{Bergshoeff:2009hq,Bergshoeff:2009aq}
\end{tabular}
\end{center}
\caption{3-dimensional massive gravity menagerie. The figures in brackets in the second column indicate the number of degrees of freedom
in the linearized theory with AdS as background metric.}
\label{tab:3}
\end{table}

The mass scales $M_{1,2}$ are determined by the coupling constants in the action, see \cite{Paulos:2010ke,Grumiller:2010tj}.
At the linearized level the only differences between various models \eqref{eq:gCS1}-\eqref{eq:gCS6a} are the values of these masses and the AdS radius $\ell$.
A mode annihilated by ${\cal D}^{M_{1,2}}$ (${\cal D}^L$) [${\cal D}^R$] is called massive (left-moving) [right-moving] and is denoted by $h^{M_{1,2}}$ ($h^L$) [$h^R$].
The left- and right-moving modes are pure gauge in the bulk, whereas the massive modes constitute physical degrees of freedom in general.
Thus, for generic values of the coupling constants all these higher curvature theories contain two gauge modes and two massive spin-2 excitations, just like GMG.
At the linearized level the information contained in all the coupling constants $\La$ and $\la_{nmk}$ is reduced to only three numbers: the value of the AdS radius $\ell$ and the values of the two masses $M_{1,2}$ in \eqref{eq:gCS8.5}.
This tremendous reduction of parameters is what we referred to as ``universal'' above (see table \ref{tab:3} for a list of 3-dimensional massive gravity theories that belong to this class; we note that there exist also gravity models that do not have an Einstein- or cosmological constant term, $\sigma=\Lambda=0$, such as the ghost-free, finite, fourth order gravity introduced in \cite{Deser:2009hb}).

Another property that appears to be shared by generic models above is an instability.
At least one of the four modes has negative energy, and at least one has positive energy \cite{Paulos:2010ke}.
The only exception arises for certain fine-tunings, where all energies and central charges of the dual CFT can be non-negative.
However, in that case at least one massive mode degenerates with another mode and a logarithmic excitation (with negative energy) emerges \cite{Grumiller:2008qz,Grumiller:2010tj}.
It is possible to eliminate these logarithmic excitations, e.g.~by imposing certain boundary conditions \cite{Maloney:2009ck}.
This may lead to a consistent quantum theory of gravity along the lines of the chiral gravity conjecture \cite{Li:2008dq,Maloney:2009ck}.
Alternatively, if the logarithmic modes are not truncated one has a (non-unitary) gravity dual to a logarithmic CFT \cite{Grumiller:2008qz,Sachs:2008yi,Grumiller:2008es,Henneaux:2009pw,Maloney:2009ck,Skenderis:2009nt,Skenderis:2009kd,Grumiller:2009mw,Grumiller:2009sn,Grumiller:2010rm,Giribet:2010ed,Gaberdiel:2010xv,Grumiller:2010tj,Hosseiny:2011ct}.
In either case an interesting question arises, namely whether or not the corresponding fine-tuning of the coupling constants is stable under a renormalization group (RG) flow.

A straightforward way to check the stability of the fine-tuning of the coupling constants perturbatively is to perform a 1-loop analysis, calculate the $\beta$-functions and look for fixed points.
For the special case of TMG this analysis was performed by Percacci and Sezgin \cite{Percacci:2010yk} who provided evidence for a non-trivial fixed point.
(See \cite{Deser:1990bj} for the first analysis of the
renormalizability of TMG.) They found that the tuning required for chiral gravity, $\mu/\sqrt{|\La|}=\pm 1$, is not stable under RG flow, which is an indication that similar tunings may be unstable in generic models of type \eqref{eq:gCS1}.
Interestingly, and perhaps not unexpectedly, they also discovered that the dimensionless Chern--Simons coupling $\mu\ka^2$ is stable under RG flow.
One could generalize such an analysis to generic theories of type \eqref{eq:gCS1} --- or at least to GMG --- and to derive how the two mass scales $M_{1,2}$ in \eqref{eq:gCS7} behave under RG flow.
However, this involves a somewhat lengthy analysis.
Therefore, we pursue a different route: we look for symmetries in addition to diffeomorphisms or other features of a given model that may stabilize the fine-tuning.
A simple example is given by the choice $\mu\to\infty$ and $\la_{nmk}=0$, i.e., pure Einstein gravity.
This theory has no massive excitations, and thus is not continuously connected to ``nearby'' models in theory space with $\mu>>1$ and $|\la_{nmk}|<<1$, which generically have two massive modes as explained above.
However, the status of Einstein gravity as a toy model for quantum gravity is unclear, even in three dimensions --- see \cite{Witten:2007kt,Maloney:2007ud} and Refs.~therein.
This motivates us to seek 3-dimensional gravity models of type \eqref{eq:gCS1} different from Einstein gravity, with gauge symmetries in addition to diffeomorphisms.

At the linearized level partial masslessness \cite{Deser:1983mm,Deser:2001pe,Deser:2001us} provides an additional gauge symmetry, first encountered in NMG \cite{Bergshoeff:2009aq}.
Partial masslessness in this context means that at least one of the mass parameters $M_{1,2}$ in \eqref{eq:gCS7} vanishes.
The linearized equations of motion then exhibit an additional gauge symmetry acting on $h^0$, where $h^0$ is the partial massless mode annihilated by ${\cal D}^0:=\lim_{M_1\to 0} M_1 {\cal D}^{M_1}$. More specifically,
\begin{equation}\label{eq:PM}
\varepsilon_\mu{}^{\rho\la}\nabla_\rho \, h^0_{\la\nu} = 0\qquad\textrm{is\;invariant\;under} \qquad h^0_{\mu\nu} \to h^0_{\mu\nu} + 2 \Omega\,g^{\textrm{\tiny AdS}}_{\mu\nu} - 2\ell^2\nabla_\mu\nabla_\nu\Omega\,.
\end{equation}
The gauge symmetry \eqref{eq:PM} can be interpreted as a linearized Weyl rescaling of the metric $g^{\textrm{\tiny AdS}}$, together with an infinitesimal diffeomorphism.
The linearized Weyl factor is given by the function $\Omega$, which generates this symmetry.
This gauge enhancement reduces the number of linearized physical degrees of freedom and thus may stabilize a corresponding tuning of the coupling constants.
It is possible that partial masslessness is maintained under RG flow.
This can be checked by generalizing the analysis of \cite{Percacci:2010yk}.
However, it is not clear whether this gauge symmetry persists beyond the linearized approximation.
If it does persist for a certain tuning of the coupling constants this tuning is likely to be stable.
If it does not persist then non-linear effects are likely to destabilize the tuning.
To make a long story short, a canonical analysis along the lines of \cite{Blagojevic:2010ir} suggests that generically the latter case applies.
We have performed this analysis for GMG, see appendix \ref{app:B}, and conjecture that it generalizes to generic theories of type \eqref{eq:gCS1}, with possible exceptions for further fine-tunings of coupling constants.
Therefore, partial masslessness alone is unlikely to be sufficient for a stabilization of a tuning of coupling constants, unless one manages to lift the enhanced gauge symmetry at the linearized level to an enhanced gauge symmetry of the full theory.

Since none of the building blocks in \eqref{eq:gCS5} is Weyl invariant, the Chern--Simons term \eqref{eq:gCS2} is the only Weyl invariant Lagrange density available to us.
Thus, the only way to obtain a model where the enhanced gauge symmetry is manifest non-linearly is by taking the scaling limit 
\eq{
\mu\to 0, \qquad \frac{1}{\ka^2\mu}\to \frac{k}{2\pi} = \rm finite\,
}{eq:gCS9}
of \eqref{eq:gCS1}. This scaling limit reduces the action \eqref{eq:gCS1} to conformal Chern--Simons gravity (CSG) \cite{Deser:1982vy,Deser:1982wh,Deser:1982a,Horne:1988jf,Afshar:2011yh}
\eq{
S_{\textrm{gCS}}[g]=\frac{k}{4\pi}\,\int\!\extd^3x\, \CS(\Ga)\,.
}{eq:gCS10}
The action \eqref{eq:gCS10} differs from gauge-theoretic Chern--Simons actions insofar, as it should not be varied with respect to the connection $\Ga$, but rather with respect to the metric $g$, which enters the connection with first derivatives.
Thus, the equations of motion do {\em not} lead to (locally) flat connections, $R_{\mu\nu}=0$, but only to conformally flat connections, $C_{\mu\nu}=0$, where $C_{\mu\nu}$ is the Cotton tensor.
This property leads to an additional degree of freedom as compared to Einstein gravity, the partially massless modes.
In the bulk the theory \eqref{eq:gCS10} is not only diffeomorphism invariant but also invariant under Weyl rescalings
\eq{
g_{\mu\nu}\to e^{2\Om} g_{\mu\nu}\,.
}{eq:Weyl}
Thus, the invariance of linearized partial massless theories under linearized Weyl rescalings \eqref{eq:PM} is lifted to full gauge invariance \eqref{eq:Weyl}.
This property implies that the partially massless modes actually are pure gauge in the bulk.
Consequently, the theory defined by the action \eqref{eq:gCS10} is topological in the sense that it has zero physical bulk degrees of freedom.
Interesting physical properties emerge only if a boundary is introduced \cite{Witten:1988hc} --- for instance an asymptotic boundary, like in the AdS/CFT correspondence \cite{Maldacena:1997re,Aharony:1999ti}.
In that case the Einstein modes generate physical states at the boundary, called boundary gravitons, exactly as in 3-dimensional Einstein gravity.
Similarly, the partially massless modes may generate physical states at the boundary, which we denote as Weyl gravitons.

The model described by the bulk action \eqref{eq:gCS10}, CSG, has received considerably less attention than TMG in the past three decades. The purpose of the present work is to fill this gap and to study CSG comprehensively and in great detail, with particular focus on holography. We continue and generalize the discussion initiated in Ref.~\cite{Afshar:2011yh}, and substantiate the statements made in that paper.
These are our main new results as compared to \cite{Afshar:2011yh}:
\begin{itemize}
 \item We derive the gauge generators and boundary charges \eqref{eq:diffcharge}, \eqref{eq:weylcharge}.
 \item We generalize the boundary conditions set up in Ref.~\cite{Afshar:2011yh} to allow for curved and varying boundary metrics, \eqref{eq:bc1}-\eqref{eq:bc3}.
 \item We calculate the 1-, 2- and 3-point functions for asymptotic AdS holography, exploiting peculiar features of the Weyl graviton spectrum in Fig.~\ref{fig:1} on p.~\pageref{fig:1}.
 \item We considerably extend the discussion of generalized holography in several ways: 
1.~we allow for asymptotically non-AdS boundary conditions that effectively lead to a scalar field with background charge on the CFT side \eqref{eq:algebrabneqzero}; 
2.~we cover the case of non-chiral Weyl rescalings \eqref{eq:gCS222};
3.~we elaborate on semi-classical null vectors, derive a condition for the weight of the corresponding primary \eqref{eq:gCS233}, and show that the only TMG-like gravity model capable of obeying this condition is precisely the theory we are studying, CSG.
\end{itemize}

This paper is organized as follows. 
In section \ref{se:2} we study classical solutions, provide a canonical analysis, discuss the gauge symmetries and derive the canonical generators and associated charges.
In section \ref{se:leah} we address the boundary conditions and the transformations that preserve them.
In section \ref{se:3} we consider the simplest case of AdS holography and calculate 1-, 2- and 3-point functions.
In section \ref{se:genhol} we treat generalized holography where the Weyl factor is not asymptotically trivial and spacetime does not necessarily asymptote to AdS, with an extensive CFT discussion.
In section \ref{se:6} we provide an outlook and summarize open issues.
In appendix \ref{app:B} we analyze GMG canonically.
In appendix \ref{app:A} we display the Dirac brackets in the reduced phase space of CSG.
In appendix \ref{se:3.6} we discuss the uniqueness of our boundary conditions.
In appendix \ref{app:gauge} we show that certain simplifying transformations have vanishing boundary charges.
In appendix \ref{app:EOM} we mention known classical solutions and provide an asymptotic analysis of the equations of motion.
In appendix \ref{app:Weyl} we collect Weyl rescaling formulas.

Before starting we mention some conventions.
Latin indices refer to the (anholonomic) local Lorentz frame, Greek indices refer to the (holonomic) coordinate frame;
the middle alphabet letters $(i,j,k,...;\mu,\nu,\la,...)$ run over 0,1,2;
the first alphabet letters $(a,b,c,...;\al,\be,\ga,...)$ run over 1,2;
the metric in the local Lorentz frame is $\eta=\textrm{diag}\,(-,+,+)$;
hatted indices like $\hat y$ are local Lorentz indices;
indices are raised and lowered with the corresponding metrics and converted between anholonomic and holonomic frames using the dreibein (triad) or its inverse, e.g.~$T^\mu=e_i{}^\mu T^i$;
the totally antisymmetric tensor $\eps^{ijk}$ and the related tensor density $\epsilon^{\mu\nu\rho}$ are normalized as %
$\epsilon^{\rho t\varphi}=\epsilon^{ty\varphi}=\epsilon^{012}=1=-\epsilon_{012}$.
Note that we shall, according to convenience, use two different holographic coordinates: the Gaussian normal coordinate $\rho$ and $y = 2e^{-\rho}$.
Since the coordinate change between the two flips parity, the whole action changes sign and the theory under study is not the same in the two 
formulations. To compensate for this we let $\eps^{\mu\nu\la}$ transform as a \emph{tensor} as opposed to a pseudotensor. This ensures that 
we compute quantities valid for the same theory in both coordinate systems, and is the reason for having the unusual relation $\epsilon^{\rho t\varphi}=\epsilon^{ty\varphi}$.
The 2-dimensional $\epsilon$-symbol in light-cone gauge ($x^+ = \varphi + t =u $, $x^-= \varphi - t=-v$) is fixed as $\epsilon^\pm{}_{\pm}=\pm \,1$;
we omit the wedge products between the forms. The boundary theory lives either on the cylinder or the torus, so we always have periodicity in the angular coordinate $(\varphi\sim\varphi + 2\pi)$. Relatedly, our background metric is always global AdS$_3$ as given in \eqref{eq:gCS6}.

\section{Canonical analysis}\label{se:2}

A detailed canonical analysis for TMG --- which includes as a limiting case CSG \eqref{eq:gCS10} --- was performed by Carlip \cite{Carlip:2008qh}.
(See also \cite{Deser:1991qk}, where the 2+1 decomposition was performed for the first time.) However, the limit $\mu\to 0$ is not smooth in TMG, since the number of physical degrees of freedom changes.
In this section we provide a canonical analysis for CSG \eqref{eq:gCS10}.
Starting from a covariant first order formulation, we discuss the field equations and gauge symmetries of the theory defined by the action \eqref{eq:gCS10} in section \ref{se:2.1}.
In section \ref{se:2.2} we switch to the Hamiltonian formulation, derive and classify all constraints, and show that there are no local physical degrees of freedom, as expected.
Finally, in section \ref{se:2.3} we construct the canonical generators of gauge transformations and the associated boundary charges, which are the key results of this section.

\subsection{Field equations and gauge symmetries}\label{se:2.1}

It is useful to employ the vielbein formalism, since this will convert the action \eqref{eq:gCS10} (up to boundary terms) directly into first order form.
\begin{equation}\label{2.1}
S_{\textrm{gCS}}^{(1)}=\frac{k}{2\pi}\,\int\Big[\om^i\extd\om_i +\tfrac{1}{3}\eps_{ijk}\,\om^i\om^j\om^k + \la^i T_i\Big] = \frac{k}{2\pi}\,\int\extd^3x\,{\cal L}\, .
\end{equation}
Here $\la^i$ is the Lagrange multiplier 1-form that ensures vanishing torsion, $T^i=\extd e^i+\eps^i{}_{jk}\,\om^j e^k=0$.
In component notation torsion is given by $T^i{}_{\mu\nu}=\partial_\mu e^i{}_\nu-\partial_\nu e^i{}_\mu + \eps^i{}_{jk}\,(\om^j{}_\mu e^k{}_\nu-\om^j{}_\nu e^k{}_\mu)$.
The metric $g$ is constructed from the dreibein 1-forms $e^i$ as usual, $g=\eta_{ij}\,e^i\otimes e^j$.
The dualized spin-connection 1-form $\om^i$ defines the (dualized) curvature 2-form, $R^i=\extd\om^i+\eps^i{}_{jk}\,\om^j\om^k$.
The standard curvature 2-form is then obtained by dualizing with the $\eps$-tensor, $R^{ij}=-\eps^{ij}{_k}R^k$.

Before discussing the field equations we recall some relations between various actions.
The first order action \eqref{2.1} differs from our starting point \eqref{eq:gCS10} by a total derivative bulk term, the dreibein-winding, and a boundary term \cite{Kraus:2005zm}.
\eq{
S_{\textrm{gCS}}^{(1)} = S_{\textrm{gCS}} + \frac{k}{12\pi}\,\int \Tr\big(e^{-1}\extd e\big)^3 - \frac{k}{4\pi}\,\int_{\partial M}\!\!\!\Tr\big(\om\extd e e^{-1}\big)\, .
}{eq:angelinajolie}
Here $e$ is the dreibein interpreted as a matrix-valued 0-form.
Starting instead from Chern--Simons gauge theory with an $SO(3,2)$ gauge connection 1-form $A$,
\eq{
S_{\textrm{CS}} = \frac{k}{4\pi}\,\int\Tr\big(A\extd A + \tfrac23\,A^3\big)\,,
}{eq:lalapetz}
one recovers the first order action \eqref{2.1} --- as well as the requirement that the dreibein must be invertible --- for a specific partial gauge fixing \cite{Horne:1988jf}, thus breaking $SO(3,2)\to SL(2,\mathbb{R})_L\times SL(2,\mathbb{R})_R\times U(1)_{\rm Weyl}$.
The second order action \eqref{eq:gCS10} is manifestly Lorentz invariant, but not diffeomorphism invariant at a boundary.
By contrast, the first order action \eqref{eq:angelinajolie} is manifestly diffeomorphism invariant, but not Lorentz invariant at a boundary.
It is possible to add further boundary terms to the bulk action, provided they are Lorentz-, diffeomorphism- and Weyl-invariant.
We shall add such a term in section \ref{se:3.2} in order to obtain a well-defined Dirichlet boundary value problem.

The variation of the action \eqref{2.1} with respect to $e^i$ and $\om^i$ yields the field equations
\begin{subequations}
 \label{2.2}
\begin{align}
& \covD\la_i:=\extd\la_i+\eps_{ijk}\,\om^j\la^k=0 \,,      \label{2.2a}\\
& R_i+\frac 12\,\eps_{imn}\,\la^m e^n=0 \, .                \label{2.2b}
\end{align}
\end{subequations}
The torsion constraint $T_i=0$ gives the dualized connection $\om^i$ in terms of the dreibein $e^i$.
It is useful to define the Schouten 1-form $L_m$,
\eq{
L_m:=\Big(\ric_{mn}-\frac{1}{4}\eta_{mn}R\Big)e^n = L_{mn}\,e^n\,,
}{eq:schouten}
where $\ric_{mn}=-\eps^{kl}{_m}R_{kln}$, $R=-\eps^{ijk}R_{ijk}$, and $R_{ijk}$ is obtained from the (dual) curvature 2-form $R_i$ using the dreibein, $R_{ijk}=R_{i\,\mu\nu} \, e_j{}^\mu e_k{}^\nu$.
Solving \eqref{2.2b} allows to express the Lagrange multiplier 1-form $\la_m$ in terms of the Schouten 1-form: $\la_m= - 2\,L_m$.
After that, equation \eqref{2.2a} takes essentially the same form as the field equations in the metric formulation,
\eq{
C_{ij}=0\,.
}{2.3a}
Here  $C_{ij}=\eps_i{}^{mn}\covD_m L_{nj}$ is the (anholonomic) Cotton tensor.
Since the 3-dimensional Cotton tensor vanishes if and only if the metric is conformally flat, see e.g.~\cite{Garcia:2003bw}, any conformally flat metric solves the field equations \eqref{2.3a}, as mentioned already in the introduction.
Another consequence of the field equations is symmetry of the Lagrange multiplier components, $\la_{mn}=\la_{nm}=-2\,L_{mn}$.

By construction, local translations (diffeomorphisms) and local Lorentz rotations are gauge symmetries of the theory \eqref{2.1}.
They are parametrized by $\xi^\mu$ and $\theta^i$, respectively.
In local coordinates $x^\mu$ we have $e^i=e^i{_\mu}\extd x^\mu$, $\om^i=\om^i{_\mu}\extd x^\mu$, $\la^i=\la^i{_\mu}\extd x^\mu$, and local Poincar\'e transformations take the standard form
\begin{subequations}
\label{2.5}
 \begin{align}
\de_P e^i{_\mu}&=\eps^i{}_{jk}e^j{}_{\mu}\theta^k
   +(\partial_\mu\xi^\nu)e^i{_\nu}+\xi^\nu\partial_\nu e^i{}_\mu   \, ,  \\
\de_P\om^i{_\mu}&=\covD_\mu\theta^i+(\partial_\mu\xi^\nu)\om^i{_\nu}
   +\xi^\nu\partial_\nu\om^i{}_\mu  \, ,\\
\de_P\la^i{_\mu}&=\eps^i{}_{jk}\la^j{}_{\mu}\theta^k
   +(\partial_\mu\xi^\nu)\la^i{_\nu}+\xi^\nu\partial_\nu \la^i{}_\mu\,.
\end{align}
\end{subequations}
In addition the action \eqref{2.1} has an extra gauge symmetry, which in the second order formalism corresponds to Weyl rescaling  of the metric \eqref{eq:Weyl}.
Its action on the variables in first order formulation is given by
\begin{subequations}
 \label{2.6}
\begin{align}
\de_W e^i{_\mu}&=\weyl\, e^i{_\mu}  \, ,\\
\de_W\om^i{_\mu}&=\eps^{ijk}e_{j\mu}e_k{^\nu}\partial_\nu\weyl  \, ,\\
\de_W\la^i{_\mu}&=2\,\covD_\mu(e^{i\nu}\partial_\nu\weyl)-\weyl\, \la^i{_\mu}\,.
\end{align}
\end{subequations}
The transformation parameter of infinitesimal Weyl rescalings is denoted by $\weyl$.

\subsection{Hamiltonian and constraints}\label{se:2.2}

In local coordinates $x^\mu$ the Lagrange density ${\cal L}$ related to the action \eqref{2.1} reads
\eq{
{\cal L}=\epsilon^{\mu\nu\rho}\big[\om^i{_\mu}\partial_\nu\om_{i\rho}+\frac{1}{3}\eps_{ijk}\om^i{_\mu}\om^j{_\nu}\om^k{_\rho}+\frac{1}{2}\la^i{_\mu}T_{i\nu\rho}\big]\,.
}{eq:gCS12}
Introducing the canonical momenta $p_I=\partial{\cal L}/\partial\partial_0q^I=(\pi_i{^\mu},\Pi_i{^\mu},
p_i{^\mu})$ corresponding to the Lagrangian variables
$q^I=(e^i{_\mu},\om^i{_\mu},\la^i{_\mu})$, we find the primary constraints:
\begin{subequations}
\label{3.1}
\begin{align}
\phi_i{^0} &:=\pi_i{^0}\approx 0\,,\qquad\,\,
  \phi_i{^\al}:=\pi_i{^\al}-\epsilon^{0\al\be}\la_{i\be}\approx 0\,, \\
\Phi_i{^0}&:=\Pi_i{^0}\approx 0\, ,\qquad
  \Phi_i{^\al}:=\Pi_i{^\al} -\epsilon^{0\al\be}\om_{i\be}\approx 0 \,,\\
p_i{^\mu} &\approx 0\,.
\end{align}
\end{subequations}
The canonical Hamiltonian density ${\cal H}_c=p_I\partial_0 q^I - {\cal L}$ is given by
\begin{subequations}
\begin{align}
{\cal H}_c &= e^i{}_0{\cal H}_i+\om^i{}_0{\cal K}_i+\la^i{_0}{\cal T}_i+\partial_\al B^\al \,,  \\
{\cal H}_i &= -\epsilon^{0\al\be}\covD_\al\la_{i\be} \,, \\
{\cal K}_i &= -\epsilon^{0\al\be}\big(R_{i\al\be} + \eps_{ijk}e^j{}_\al \la^k{}_\be\big)\,,            \\
{\cal T}_i &= -\frac{1}{2}\epsilon^{0\al\be}T_{i\al\be}        \,,       \\
 B^\al &= \epsilon^{0\al\be}\big(\om^i{_0}\om_{i\be}+e^i{}_0 \la_{i\be}\big)\,.
\end{align}
\end{subequations}
We recall that the covariant derivative $\covD$ acts as defined in \eqref{2.2a}.
Going over to the total Hamiltonian
\eq{
H_T = \int \extd^2x \,{\cal H}_T\,,
}{eq:gCS17}
with the total Hamiltonian density
\begin{equation}\label{eq:gCS13}
{\cal H}_T=e^i{}_0{\cal H}_i+\om^i{}_0{\cal K}_i+\la^i{}_0{\cal T}_i +\lagA^i{}_\mu\phi_i{}^\mu+\lagB^i{}_\mu\Phi_i{}^\mu +\lagC^i{}_\mu p_i{^\mu}+\partial_\al B^\al \,,
\end{equation}
we find that the consistency conditions of the primary constraints $\pi_i{}^0$, $\Pi_i{}^0$ and $p_i{}^0$ yield secondary constraints:
\eq{
{\cal H}_i\approx0\,, \qquad {\cal K}_i\approx0\,, \qquad {\cal T}_i\approx 0\,.
}{eq:gCS14}
Thus, as expected the canonical and total Hamiltonians are sums over constraints, up to a boundary term $\partial_\al B^\al$.
The consistency of the remaining primary constraints $\chi_I:=(\phi_i{}^\al, \Phi_i{}^\al, p_i{}^\al)$ leads to the determination of the multipliers $(\lagA^i{}_\al,\lagB^i{}_\al,\lagC^i{}_\al)$.
However, we find it more convenient to continue our analysis in the reduced phase space formalism.
Using the second class constraints $\chi_I$, we can eliminate the momenta $(\pi_i{^\al},\Pi_i{^\al},p_i{^\al})$ and construct the reduced phase space $R_1$, in which the basic nontrivial Dirac brackets take the form
\begin{equation}\label{3.3}
\{e^i{_\al}(x),\la^j{_\be}(x')\}^*_1 \,=-\epsilon_{0\al\be}\eta^{ij}\de^{(2)}(x-x') \,= 2\, \{\om^i{_\al}(x),\om^j{_\be}(x')\}^*_1\,.
\end{equation}
Thus, $e^i{}_\al$ and $\la^i{}_\al$ effectively become canonical pairs, while half of the spin-connection components $\om^i{}_\al$ becomes the canonical partner of the other half.
The remaining Dirac brackets are the same as the corresponding Poisson brackets, for instance $\{e^i{}_\mu(x),\pi_j{}^\nu(x')\}^*_1=\de_i^j\de_\mu^\nu\de^2(x-x')$.
The Dirac brackets between the constraints are summarized in appendix \ref{app:A}.
In $R_1$, the total Hamiltonian takes the simpler form
\eq{
{\cal H}_T = {\cal H}_c + \lagA^i{}_0\phi_i{}^0 + \lagB^i{}_0\Phi_i{}^0 + \lagC^i{}_0 p_i{^0}\,.
}{eq:gCS22}
This result can also be obtained more directly from the Faddeev--Jackiw method \cite{Faddeev:1988qp}, in full analogy to TMG at the critical point \cite{Grumiller:2008pr}.
The consistency conditions of the secondary constraints read:
\begin{subequations}
 \label{3.5}
\begin{align}
&\{{\cal H}_i,H_T\}^*_1\approx -\frac12\,\epsilon^{\mu\nu\rho}\la_{i\mu}\la_{\nu\rho} \,,\\
&\{{\cal K}_i,H_T\}^*_1\approx 0\,,\\
&\{{\cal T}_i,H_T\}^*_1\approx \frac12\,(\det{e})\,\eps_{ijk}\la^{jk}\,.
\end{align}
\end{subequations}
Assuming a non-degenerate dreibein, $\det{(e)}\neq 0$, the relations \eqref{3.5} yield the following ternary constraints:
\eq{
\ternary^\mu = \epsilon^{\mu\nu\rho}\,\la_{\nu\rho} \approx 0\,.
}{eq:gCS24}
They are obviously compatible with the symmetry of the Lagrange multiplier components required by the Lagrangian equations of motion.
The consistency  condition of the ternary constraint $\ternary^0$ is identically satisfied:
\eq{
\{\ternary^0,H_T\}\approx 0\,.
}{eq:gCS15}
To interpret the consistency condition for $\ternary^\al$, we introduce the notation
\eq{
\pi_i{^0}{}':=\pi_i{^0}+\la_i{^k}p_k{^0}\, , \qquad
\lagC^i{_\mu}{}':=\lagC^i{_\mu}-\lagA^k{_\mu}\la_k{^i}\,.
}{eq:gCS16}
The $(\pi_i{^0},p_i{^0})$ piece of the Hamiltonian can be written in the form
$\lagA^i{_0}\pi_i{^0} + \lagC^i{_0}p_i{^0} = \lagA^i{_0}\pi_i{^0}{}' + \lagC^i{_0}{}'p_i{^0}$.
The consistency of $\ternary^\al$ imposes a condition on the two components of the Lagrange multiplier $\lagC'_{\be 0}=\lagC_{m0}{}'e^m{_\be}$ of $\lagC_i{^0}{}'$:
\eq{
\{\ternary^\al,\,H_T\}^*_1=\epsilon^{\al\mu\nu}\,\lagC'_{\mu\nu}\approx 0\,.
}{eq:gCS18}
This restricts the Lagrange multipliers $\lagC'_{0\be}$ to be symmetric and completes the consistency procedure.
We have found all constraints of the theory \eqref{2.1}.

The dimension of the phase space $R_1$ is 36.
It is spanned by $(e^i{}_\al$, $\la^i{}_\al$, $\om^i{}_\al$, $e^i{}_0$, $\la^i{}_0$, $\om^i{}_0$, $\pi^i{}_0$, $p^i{}_0$, $\Pi^i{}_0)$.
We could now determine the dimension of the physical phase space by classifying all constraints into first and second class.
However, for simplicity we perform first a further reduction of the phase space $R_1$ to a smaller phase space $R_2$, by imposing the second class constraints $\zeta_I:=(\ternary^\al,p^{\al 0})$.
The constraints $\ternary^\al$ allow to express $\la_{\al 0}$ in terms of the other canonical variables.
The constraints $p^{\al 0}$ fix the corresponding momenta. 
This effectively reduces the dimension of the phase space to 32.
It is spanned by $(e^i{}_\al$, $\la^i{}_\al$, $\om^i{}_\al$, $e^i{}_0$, $\la_{00}$, $\om^i{}_0$, $\pi^i{}_0$, $p^{00}$, $\Pi^i{}_0)$. 
The Dirac brackets in $R_2$ retain the same form as in $R_1$ (see again appendix \ref{app:A}), while the final form of the total Hamiltonian density in $R_2$ is
\begin{subequations}
\label{eq:gCS19}
\begin{align}
({\cal H}_T)_{R_2}&=({\cal H}_c)_{R_2}+\lagA^i{_0}\pi_i{^0}+\lagB^i{_0}\Pi_i{^0}+\lagC_{00}p^{00} \,,\\
({\cal H}_c)_{R_2}&=e^i{_0} \big({\cal H}_i+\la_{i\al}{\cal T}^\al\big) +\om^i{_0}{\cal K}_i+\la_{00} {\cal T}^0\,.
\end{align}
\end{subequations}

We classify now the remaining constraints in the reduced phase space $R_2$ into first and second class.
Among the primary constraints those that appear in ${\cal H}_T$ with arbitrary multipliers are first class,
$\pi_i{^0}{}$, $\Pi_i{^0}$ and $p^{00}$, while the remaining ones are second class.
Going to the secondary constraints, we use the following simple theorem:
If $\phi$ is a first class constraint, then $\{\phi,H_T\}$ is also a first class constraint.
The proof relies on using the Jacobi identity.
The theorem implies that the secondary constraints ${\cal H}_i+\la_{i\al}{\cal T}^\al$, ${\cal K}_i$, ${\cal T}^0=e_i{^0}{\cal T}^i$ and the ternary constraint $\ternary^0$ are first class.
This can also be verified by direct computation, using the results of appendix \ref{app:A}.
The complete classification of constraints in $R_2$ is summarized in table \ref{tab:1}.
\begin{table}
\begin{center}
\doublerulesep 1.8pt
\begin{tabular}{||l|l|l||}
                                                       \hline\hline
\rule{0pt}{12pt}
&~First class \phantom{x}&~Second class \phantom{x}     \\
                                                      \hline
\rule[-1pt]{0pt}{15pt}
\phantom{x}Primary &$\pi_i{^0}{},\,\Pi_i{^0},\,p^{00}$
            &   \\
                                                      \hline
\rule[-1pt]{0pt}{15pt}
\phantom{x}Secondary\phantom{x} &${\cal H}_i+\la_{i\al}{\cal T}^{\al},\,{\cal K}_i,\,{\cal T}^0$
            &${\cal T}^\al$                                \\
                                                      \hline
\rule[-1pt]{0pt}{15pt}
\phantom{x}Ternary\phantom{x}
                   &$\ternary^0$ &
                        \\
                                                      \hline\hline
\end{tabular}
\end{center}
\caption{Classification of constraints in partially reduced phase space $R_2$}
\label{tab:1}
\end{table}
According to our results, we have a 32-dimensional phase space with 15 first class and two second class constraints.
In conclusion, the dimension of the physical phase space is zero degrees of freedom per space-point, and thus the theory \eqref{2.1} has no local physical degrees of freedom, as expected on general grounds.

\subsection{Gauge generators and boundary charges}\label{se:2.3}

The canonical gauge generators are constructed using the procedure of Castellani \cite{Castellani:1981us}.
The generator of Poincar\'e gauge transformations 
\begin{subequations}
 \label{6.2}
\eq{
G_P=\frac{k}{2\pi}\,(G_1+G_2)
}{eq:GP}
has the following standard form:
\begin{align}
G_1&=\dot\xi^\mu\left(e^i{}_\mu\pi_i{}^0
       +\la^i{_\mu}p_i{}^0+\om^i{}_\mu\Pi_i{}^0\right)    \nonumber \\
&\quad +\xi^\mu\big[e^i{}_\mu{\cal H}_i+\la^i{_\mu}{\cal T}_i
  +\om^i{}_\mu{\cal K}_i+(\partial_\mu e^i{}_0)\pi_i{}^0 +(\partial_\mu\la^i{_0})p_i{^0}+(\partial_\mu\om^i{}_0)\Pi_i{}^0\big] \,, \\
G_2&=\dot{\theta^i}\Pi_i{}^0+\theta^i\big[{\cal K}_i
  -\eps_{ijk}\,(e^j{}_0\pi^{k0}+\la^j{}_0p^{k0}
  +\om^j{}_0\Pi^{k0})\big] \,.
\end{align}
\end{subequations}
The gauge transformations generated by $G_P$ correspond \emph{on shell} to the Poincar\'e gauge transformations \eqref{2.5}.

The generator of the extra symmetry in $R_2$ is given by
\begin{align}\label{eq:Wgen}
G_W&=\frac{k}{2\pi}\,\Big(2\ddot{\weyl}\,p^{00}+\dot\weyl\,\big[\eps_{ijk}e^{i0}e^j{_0}\Pi^{k0}+2{\cal T}^0-2(e_i{^0}\covD_0 e^i{_0})p^{00}\big]\nonumber\\
&\quad+\weyl\,\big[\epsilon^{0\al\be}\la_{\al\be}+\pi_0{^0}-\eps_{ijk}\covD_\al(e^{i\al}e^j{_0}\Pi^{k0})-2\covD_\al{\cal T}^\al +2\covD_\al(e_i{^\al}\covD_0 e^i{_0})p^{00}\big]\Big)\,.
\end{align}
The action of $G_W$ on some function $\phi$ of the canonical variables in the reduced phase space $R_2$ is given by the Dirac bracket operation $\de_W\phi=\frac{2\pi}{k}\,\{\phi,G_W\}_2^*$:
\begin{subequations}
\label{eq:gCS80}
\begin{align}
\de_W e^i{_\mu} &= \weyl\, e^i{_\mu}\,,\\
\de_W\om^i{_\mu} &= \eps^{ijk}e_{j\mu}e_k{^\nu}\partial_\nu\weyl\,,\\
\de_W\la^i{_\al} &= 2\covD_\al(e^{i\nu}\partial_\nu\weyl) - \weyl\, \la^i{_\al}\,,\\
\de_W\la_{00} &= 2e_{i0}\covD_0\left(e^{i\nu}\partial_\nu\weyl\right)\,. 
\end{align}
\end{subequations}
This behavior is in accordance with the infinitesimal Weyl rescalings \eqref{2.6}.

Smearing the generator $G_P$ with a vector field $\xi$, varying it with respect to the fields and integrating over a spacelike hypersurface $\Si$ with boundary $\partial\Si$ yields 
\begin{equation}\label{eq:diffvar}
\int_{\Si} \extd^2 x \, \de G_P[\xi^\mu] =\frac{k}{2\pi}\,\int_{\partial \Si}\!\!\! \extd \varphi \, \Big[ \xi^\mu \big(e^i{}_\mu\, \de \la_{i\varphi}  + \la^i{}_\mu\, \de e_{i\varphi}  + 2 \om^i{}_\mu\, \de \om_{i\varphi}\big) + 2\theta^i\, \de\om_{i\varphi} \Big]  
 + {\rm regular}\, . 
\end{equation}
The variation $\de$ denotes the difference between two states in the theory, both satisfying a given set of boundary conditions (see section \ref{se:leah} below).
The `regular' terms do not require the introduction of boundary terms and thus do not contribute to the boundary charges.
Note that the term proportional to the parameter of Lorentz transformations $\theta^i$  
in Eq.~\eqref{eq:diffvar} usually vanishes at an asymptotic boundary; 
we shall demonstrate later, however, that this ceases to be the case when considering specific sets of asymptotic boundary conditions.
To get differentiable charges $Q_P$ we must add a boundary piece to the generator, $\tilde{G}_P = G_P + \Ga_P$, where
\begin{equation}\label{eq:diffcharge}
\de Q_P[\xi^\mu] = \int\limits_0^{2\pi} \extd \varphi \, \de \Ga_P = -\frac{k}{2\pi}\,\int\limits_0^{2\pi} \extd \varphi \, \Big[ \xi^\mu \big(
e^i{}_\mu\, \de \la_{i\varphi}  + \la^i{}_\mu\, \de e_{i\varphi}  + 2 \om^i{}_\mu\, \de \om_{i\varphi} \big) +  2\theta^i\, \de\om_{i\varphi} \Big]\, .
\end{equation}

Similarly, smearing, varying and integrating the expression for the Weyl generator \eqref{eq:Wgen} yields
\eq{
\int_{\Si} \extd^2 x \, \de G_W [\Om] 
= - \frac{k}{\pi}\, \int\limits_0^{2\pi} \extd \varphi\,
(e^{i \mu} \partial_\mu \Om)\,\de e_{i\varphi} + {\rm regular} \, .
}{eq:deGW}
Again, we must add a boundary piece to the generator, $\tilde{G}_W = G_W + \Ga_W$.
The asymptotic Weyl charges $Q_W$ are then given by
\eq{
\de Q_W[\Om] = \int\limits_0^{2\pi} \extd \varphi \, \de \Ga_W = \frac{k}{\pi}\, \int\limits_0^{2\pi} \extd \varphi\, (e^{i \mu} \partial_\mu \Om)\,\de e_{i\varphi} \, .
}{eq:weylcharge}

At a later stage we shall require the conservation of the diffeomorphism and Weyl charges.
In order to do this we first need to consider boundary conditions.

\section{Boundary conditions}\label{se:leah}

In this section we state and classify the boundary conditions in section \ref{se:state}, which fall into conditions on the conformal class of the metric and on the Weyl factor. 
We then consider the asymptotic symmetry group in section \ref{se:ASG}, which consists of all gauge transformations preserving the boundary conditions modulo trivial gauge transformations. 

The motivation for choosing the specific boundary conditions below comes from consistency conditions on the first variation of the action, invariance of the full action under symmetry/gauge transformations and a consistency requirement on the Brown--York stress tensor. 
It will turn out that after these requirements are fulfilled, one can additionally simplify the boundary conditions by suitable gauge fixings. 
The detailed discussion of these issues is deferred to the two appendices \ref{se:3.6} and \ref{app:gauge}.

\subsection{Statement and classification of boundary conditions}\label{se:state}

We assume that the manifold $M$ has a (connected) boundary $\partial M$, which may or may not be an asymptotic one.
It is convenient to parametrize the boundary such that one of the coordinates, $y$, is constant on it.
With no loss of generality we assume that $y=0$ at the boundary. In its vicinity we write the metric $g_{\mu\nu}$ as
\eq{
g_{\mu\nu} = e^{2\phi(x^+,\,x^-,\,y)}\,\bar{g}_{\mu\nu} = e^{2\phi(x^+,\,x^-,\,y)} \big( g_{\mu\nu}^{\rm AAdS} + h_{\mu\nu}\big)\,,
}{eq:bc1}
and impose the condition that the metric $\bar{g}_{\mu\nu}$ be asymptotically AdS. More specifically,
with the leading metric
\eq{
g_{\mu\nu}^{\rm AAdS}\,\extd x^\mu\extd x^\nu = \frac{e^{2\zeta(x^+,x^-)}\extd x^+\extd x^- + \extd y^2}{y^2}\,,
}{eq:bc2}
we require that the subleading state-dependent part $h_{\mu\nu}$ take the form
\eq{
\left(\begin{array}{lll}
h_{++}={\cal O}(1/y) & h_{+-} = {\cal O}(1) & h_{+y} = {\cal O}(1) \\
              & h_{--} = {\cal O}(1)        & h_{-y} = {\cal O}(1) \\
              &                             & h_{yy} = {\cal O}(1)
\end{array}\right)\,.
}{eq:bc3}
The expression ${\cal O}(1/y)$ implies that the asymptotic behavior of the corresponding quantity diverges at most with $1/y$ close to $y=0$, and ${\cal O}(1)$ means that the corresponding quantity is finite or zero close to $y=0$.
The boundary conditions \eqref{eq:bc2} with \eqref{eq:bc3} restrict the conformal equivalence class of the metric.
The Weyl factor $\phi$ is required to obey the boundary condition
\eq{
\phi = b \ln y + f(x^+,x^-) + {\cal O}(y^2)\, . 
}{eq:gCS30}
Some remarks are in order. The reason for our split into boundary conditions on the conformal class and on the Weyl factor is the enhanced gauge invariance of CSG: if $g$ is a solution to the equations of motion \eqref{2.3a} then also $e^{2\phi}\,g$ is a solution. The boundary conditions on the conformal class of the metric, \eqref{eq:bc2}--\eqref{eq:bc3}, are chosen such that AdS is allowed as a background and that most of the linearized excitations around AdS, solutions of \eqref{eq:gCS7} with $M_1=0$, $M_2=\infty$, are also admissible. 

However, as explained in appendix \ref{se:3.6}, we cannot consistently allow all such excitations, and neither can we permit all classical solutions of CSG.
If the Weyl factor did diverge stronger than logarithmically near the boundary or if the logarithmic term in \eqref{eq:gCS30} was not a constant $b$ but rather some function of $x^\pm$, then the Weyl charges \eqref{eq:weylcharge} no longer would be finite.
Thus, we cannot relax the boundary conditions \eqref{eq:gCS30} any further.
In this sense the boundary conditions \eqref{eq:bc1}--\eqref{eq:gCS30} are as loose as possible.
However, as we shall demonstrate in later sections, further consistency requirements like the conservation of the asymptotic boundary charges can restrict the boundary conditions even more.
As a technical simplification we shall often use Gaussian coordinates where $h_{\mu y}=0$ for all $\mu$, with no loss of essential features. This choice fixes some of the trivial gauge freedom.

It also turns out that there are additional gauge transformations that transform the functions $\zeta(x^+, x^-)$ and $f(x^+, x^-)$. 
In fact, only the combination $f + \zeta$ is gauge invariant and either $\zeta(x^+, x^-)$ or $f(x^+, x^-)$ can be set to zero by these transformations.
To see that the corresponding transformations actually are pure gauge requires the computation of the asymptotic charges, and is a little involved. 
In order to focus in the main text on the relevant physics, we work out these details in appendix \ref{app:gauge}.
There it is also shown that although some asymptotic charges depend parametrically on the parameter $b$, allowing it to vary does not change any charges.
Therefore, it is an arbitrary but fixed constant. %

To save space appendix \ref{app:gauge} presupposes some material and notation explained in detail in section \ref{se:genhol}, and is therefore better read after that section.

From now on in the main text we shall use $\zeta = 0$ as a gauge fixing condition, and let $f$ carry the physical information. 
This results in many technical simplifications since the equations of motions are insensitive to $f$.
Similarly, we also set $b$ to an arbitrary but fixed constant.

The boundary conditions on the Weyl factor $\phi$ lead to three cases:
\begin{enumerate}
 \item[I.] Trivial Weyl factor $\phi=\rm const.$ ($=0$~with no loss of generality)
 \item[II.] Fixed Weyl factor $\phi\neq \rm const.$
 \item[III.] Free Weyl factor $\phi$ not fixed completely by boundary conditions
\end{enumerate}
The calculations in the first order formulation require the translation of the boundary conditions \eqref{eq:bc1}--\eqref{eq:bc3} into boundary conditions on vielbein $e^i$, dualized spin-connection $\om^i$ and Lagrange multiplier $\la^i$.
It is sufficient to provide the boundary conditions on the vielbein, since the other quantities follow straightforwardly from the torsion constraint and the equation of motion \eqref{2.2b}. The appropriate Fefferman--Graham expansions for the vielbein in the most general case is
\eq{
e^i{}_\mu = e^{\phi} \,\frac{1}{y}\, \begin{pmatrix} 1 &0&0\\0 & 1 &0 \\ 0&0&1 \end{pmatrix}
 + e^{\phi}\, \begin{pmatrix} 0&0&0\\\bar e_{(1) +}^{\, -} &0 &0 \\ 0&0&0 \end{pmatrix} 
 + e^{\phi}\, y \,\begin{pmatrix} \bar e_{(2)+}^+ &\bar e_{(2)-}^+&\bar e_{(2)y}^+\\ \bar e_{(2)+}^- & \bar e_{(2)-}^- & \bar e_{(2)y}^- \\ \bar e_{(2)+}^y&\bar e_{(2)-}^y&\bar e_{(2)y}^y \end{pmatrix} + \ldots
}{eq:Feffe}
The variation of the vielbein $\de e$ follows directly from varying \eqref{eq:Feffe}.
It depends on $\de\bar e$ and $\de f$.
Case II and case I can be obtained from case III by setting to zero appropriate quantities in \eqref{eq:Feffe} and its variation.
As a technical simplification one can use Gaussian coordinates $\bar e_{(2)\mu}^{\hat y} = \bar e_{(2)\hat y}^\mu = \de \bar e_{(2)\mu}^{\hat y} = \de \bar e_{(2)\hat y}^\mu = 0$.

An important detail of the asymptotic analysis is that for some cases the parameter of Lorentz transformations $\theta^i$ may contribute at finite order in the small $y$ expansion to the last term in the diffeomorphism charges \eqref{eq:diffcharge}.
Notably, without this term --- which by itself is not Lorentz-invariant --- the total result for the diffeomorphism charges may fail to be Lorentz-invariant.
We shall encounter and highlight these issues in section \ref{se:genhol}.

\subsection{Asymptotic symmetry group}\label{se:ASG}

The asymptotic symmetry group is generated by a combination of diffeomorphisms generated by a vector field $\xi$ 
and Weyl rescalings generated by a scalar field $\Om$:
\eq{
\xi^\la\partial_\la g_{\mu\nu} + g_{\mu\la}\partial_\nu\xi^\la + g_{\la\nu}\partial_\mu\xi^\la + 2\Om g_{\mu\nu} = \delta g_{\mu\nu}\, .
}{eq:asg1}
Here $\de g$ refers to the allowed variations of \eqref{eq:bc1}.

In the simplest case~I the diffeomorphisms that preserve the boundary conditions are exactly as in 3-dimensional Einstein gravity \cite{Brown:1986nw}, i.e., 
\begin{subequations}
 \label{eq:asg2}\label{eq:gCS28}
\begin{align}
\xi^\pm &= \eps^\pm(x^\pm)-\frac{y^2}{2}\,\partial_\mp^2\eps^\mp(x^\mp)+{\cal O}(y^3)\, ,\\
\xi^y &= \frac{y}{2}\, \partial\cdot\eps + {\cal O}(y^3) \, .
\end{align}
In addition the boundary conditions are preserved by Weyl rescalings that vanish asymptotically quadratically in $y$:
\eq{
\Om = {\cal O}(y^2) \, .
}{eq:gCS31}
\end{subequations}
We introduced the notation $\partial\cdot\eps=\partial_+\eps^+(x^+)+\partial_-\eps^-(x^-)$.
The higher order terms in $y$, including the Weyl rescalings, comprise the trivial gauge transformations, which are modded out in the asymptotic symmetry algebra. The asymptotic symmetry algebra then is generated by vector fields $\eps^\pm\partial_\pm$, whose Fourier-components produce two copies of the Virasoro algebra.

In case~II the boundary conditions are preserved by diffeomorphisms generated by the same vector fields as in \eqref{eq:asg2}, but they have to be accompanied by Weyl rescalings of the form
\begin{equation}
\Om = -\frac b2\,\partial\cdot\eps - \eps\cdot\partial f + {\cal O}(y^2)\,, 
\label{eq:gCS29}
\end{equation}
where $\eps\cdot\partial = \eps^+(x^+)\partial_+ + \eps^-(x^-)\partial_-$. The reason for this is that the diffeomorphisms \eqref{eq:asg2} preserve the form 
\eqref{eq:bc1}, but transform the Weyl factor $\phi$. Since this is a fixed function in case II, a Weyl rescaling \eqref{eq:gCS29} must be applied to cancel 
this transformation. In the end, we have the same asymptotic symmetry algebra as in case~I, but instead of arising from pure diffeomorphisms, the two Virasoro
generators correspond to Fourier coefficients of the above combination of diffeomorphisms and Weyl rescalings.
The compensating Weyl rescaling depends on the quantity $b$. However, as shown in appendix \ref{app:gauge}
we can set it to zero with no loss of generality in this case. 

In case~III the transformations that preserve the boundary conditions are given by
\begin{subequations}
 \label{eq:asg5}
\begin{align}
\xi^\pm &= \eps^\pm(x^\pm)-\frac{y^2}{2}\, \partial_\mp^2 \eps^\mp(x^\mp)  + {\cal O}(y^3)\,,\\
\xi^y &= \frac{y}{2}\, \partial\cdot\eps\, + {\cal O}(y^3)\,, \\
\Om &= f_\Om(x^+,x^-) + {\cal O}(y^2)\,. \label{eq:3Weyl}
\end{align}
\end{subequations}
The asymptotic symmetry algebra could then be extended as compared to previous cases, since it may contain the scalar field $f_\Om(x^+,x^-)$.
That this indeed happens will be shown in section \ref{se:genhol}. The function $f_\Om(x^+,x^-)$ could be further restricted (e.g.~to be chiral), depending on which consistency conditions one would like to impose. We shall address these issues in detail in the next two sections. In case III there are charges that depend on 
$b$ so we keep it as a fixed but arbitrary number. 

\section{Asymptotic AdS holography}\label{se:3}

In this section we provide the first steps towards a detailed holographic description of CSG \eqref{eq:gCS10} and fill the gaps in \cite{Afshar:2011yh}.
We focus on the case~I where the metric is asymptotically AdS and the boundary metric is flat. 
Cases II and III are treated in the next section. 
In subsection \ref{se:3.2} we calculate the 1-point functions, Brown--York stress tensor and partially massless response function.
Equipped with these results we calculate the charges in subsection \ref{se:3.3} and compare with corresponding results from the canonical analysis.
In subsection \ref{se:3.modes} we discuss salient features of normalizable and non-normalizable graviton modes, with particular focus on Weyl gravitons.
In subsection \ref{se:3.3a} we calculate all 2- and 3-point functions and determine the central charges.

\subsection{1-point functions}\label{se:3.2}

We calculate the response functions using the standard AdS/CFT dictionary for calculating 1-point functions \cite{Aharony:1999ti}.
To this end we need the first variation of the on-shell action.
In this subsection we use the results from \cite{Grumiller:2008qz,Ertl:2009ch}, e.g.,
\begin{equation}\label{eq:gCS42}
\de S_{\rm gCS}\big|_{\rm EOM} = \frac{k}{2\pi}\,\int_{\partial M}\!\!\!\extd^2x\,\epsilon^{\al\be}\,\Big(-R^{\ga n}{}_{\be n}\,\de\ga_{\al\ga}+K_\al{}^\ga\,\de K_{\ga\be}-\frac12 \Ga^\ga{}_{\de\al}\,\de\Ga^\de{}_{\ga\be}\Big)\,.
\end{equation}
Here $\ga_{\al\be}$ is the induced metric at the boundary $\partial M$, $K_{\al\be}$ is extrinsic curvature and $R^{\ga n}{}_{\be n}$ is the Gauss--Codazzi expression for the Riemann-tensor contracted with two unit normal vectors.
We are going to be explicit about all these quantities in a moment.
We add a boundary term to the bulk action \eqref{eq:gCS10} that leads to a well-defined Dirichlet boundary value problem \cite{Guica:2010sw}.
This defines the total action of CSG.
\eq{
S_{\rm CSG} =  S_{\rm gCS} + \frac{k}{2\pi}\,\int_{\partial M}\!\!\!\extd^2x\sqrt{-\ga}\,k^{\al\be}_L k_{\al\be}^R \, ,
}{eq:gCS67}
The quantities $k_{\al\be}^{L/R}$ are chiral projections of extrinsic curvature
\eq{
k^{L/R}_{\al\be}=\frac12\,\big(\de_\al^\ka\pm \eps_\al{}^\ka\big)\big(K_{\ka\be}-\frac12\,K\ga_{\ka\be}\big) \, ,
}{eq:gCS68}
with the properties
\begin{equation}\label{eq:gCS69}
k^{L/R}_{\al\be} =k^{L/R}_{\be\al} \, , \qquad
k^{L/R}_{\al\be}\ga^{\al\be} = k^{L/R}_{\al\be}k^{\al\be}_{L/R} = 0 \, ,\qquad
K_{\al\be} = k_{\al\be}^L + k_{\al\be}^R + \frac12\,\ga_{\al\be}K \,.
\end{equation}
Besides the boundary metric $\ga_{\al\be}$ we are keeping fixed $k^L_{\al\be}$.
If we wanted to keep fixed $k^R_{\al\be}$ instead we would have to subtract the same boundary term in \eqref{eq:gCS67} rather than adding it;
this feature explains how it is possible to have a boundary term that does not contain an $\eps$-tensor in a parity-odd theory.
The choice \eqref{eq:gCS67} turns out to be compatible with the boundary conditions \eqref{eq:bc3}.
Moreover, the boundary term in \eqref{eq:gCS67} is invariant under Weyl rescalings \eqref{eq:Weyl} and manifestly invariant under Lorentz transformations and diffeomorphisms along the boundary.
Thus, it passes all consistency tests.
From now on we consider exclusively the total action \eqref{eq:gCS67}.

It is useful to employ Gaussian normal coordinates for the calculation ($e^{\rho}\propto 1/y$)
\eq{
\extd s^2 = \extd \rho^2 + \ga_{\al\be}\,\extd x^\al\extd x^\be\,.
}{eq:gCS43}
As we focus here on case~I, the metric is asymptotically AdS according to our boundary conditions \eqref{eq:bc1}--\eqref{eq:bc2}.
The boundary metric $\ga$ then must allow for the following Fefferman--Graham expansion in the limit of large $\rho$:
\eq{
\ga_{\al\be} = \ga_{\al\be}^{(0)}\,e^{2\rho} + \ga_{\al\be}^{(1)}\,e^\rho + \ga_{\al\be}^{(2)} + \dots
}{eq:gCS44}
The ellipsis denotes terms that vanish in the limit $\rho\to\infty$.
At the moment we are not going to specify the expansion matrices $\ga^{(0,1,2)}$, but let us mention their respective roles:
$\ga^{(0)}$ is the boundary metric, $\ga^{(1)}$ describes Weyl gravitons and their sources, and $\ga^{(2)}$ contains information about the left- and right-moving massless boundary gravitons.
The appearance of $\ga^{(1)}$ is the only difference to the situation studied by Brown and Henneaux in their seminal paper \cite{Brown:1986nw}.
In terms of the expansion \eqref{eq:gCS44} the first variation of the on-shell action \eqref{eq:gCS42} reads
\eq{
\de S_{\rm CSG}\big|_{\rm EOM} = \frac{1}{2}\,\int_{\partial M}\!\!\!\extd^2x\sqrt{-\ga^{(0)}}\,\Big(T^{\al\be}\,\de\ga^{(0)}_{\al\be} + \PMR^{\al\be}\,\de\ga^{(1)}_{\al\be}\Big)\,.
}{eq:gCS45}
Note that no terms of the form $\de\ga^{(2)}$ (or higher order) remain after taking the limit $\rho\to\infty$.
The response functions $T^{\al\be}$ and $\PMR^{\al\be}$ are Brown--York stress tensor and partially massless response, respectively.
We are going to calculate them now.

In Gaussian normal coordinates the expressions for extrinsic and Gauss--Codazzi curvature are rather simple, $K_{\al\be} = \frac12\,\partial_\rho \ga_{\al\be}$ and $R^{\al n}{}_{\be n}=-\partial_\rho K^\al{}_\be-K^\al{}_\ga K^\ga{}_\be$, respectively.
Indices are lowered and raised with the boundary metric $\ga^{(0)}$, and also the trace is defined with respect to the boundary metric, $\Tr\,\ga_{(1)}=\ga^{(1)}_{\al\be}\ga^{\al\be}_{(0)}$.
The asymptotic expansions of various geometric quantities are collected at the end of appendix \ref{app:EOM}.
Since the boundary metric is flat, the non-covariant term in \eqref{eq:gCS42} containing the Christoffel symbols vanishes asymptotically, and we obtain
\begin{align}
\label{eq:T}
 T^{\al\be} &= \frac{k}{2\pi}\,\eps^{\al\ga}\,\big(\ga^{\be\,(2)}_{\;\,\ga} -  \frac{1}{4}\,\ga_{(1)}^{\be\de}\ga_{\de\ga}^{(1)}\big) - \frac{k}{8\pi}\,\big(\ga^{\al\ga}_{(1)}\ga_\ga^{(1)\,\be}-\frac12\,\ga^{\al\be}_{(1)}\,\Tr\,\ga_{(1)}\big) \nonumber \\
&\quad + \frac{k}{32\pi}\,\ga^{\al\be}_{(0)}\,\big(\Tr\,(\ga_{(1)})^2-\frac12\,(\Tr\,\ga_{(1)})^2\big) + (\al\leftrightarrow\be)\,, \\ 
 \PMR^{\al\be} &= \frac{k}{8\pi}\,\big(\de^\al_{\ga}-\eps^{\al}{}_{\ga}\big)\,\big(\ga^{\ga\be}_{(1)}-\frac12\,\ga^{\ga\be}_{(0)}\,\Tr\,\ga_{(1)}\big) + (\al\leftrightarrow\be)\,. 
\label{eq:J}
\end{align}
Examining the index structure of the terms bilinear in $\ga^{(1)}$ reveals that they disappear for the boundary conditions \eqref{eq:bc3}.
As we comment on this situation in appendix \ref{se:3.6} we display these terms in \eqref{eq:T}, but in fact we have
\eq{
T^{\al\be} = \frac{k}{2\pi}\,\eps^{\al\ga}\,\ga^{\be\,(2)}_{\;\,\ga} + (\al\leftrightarrow\be)\,.
}{eq:goodT}
We remark in passing that the Brown--York stress tensor $T^{\al\be}$ and the partially massless response function $\PMR^{\al\be}$ are finite already 
before adding the boundary term in \eqref{eq:gCS67}, i.e., without holographic renormalization.
Moreover, both tensors are traceless, $T^\al{}_\al=\PMR^\al{}_\al=0$. The partially massless response function $\PMR^{\al\be}$ additionally is null, $\PMR^{\al\be}\PMR_{\al\be}=0$. 
Interestingly, $\ga^{(1)}$ appears both as source and as vacuum expectation value.
This feature is a consequence of the fact that normalizable partially massless modes (Weyl gravitons) and non-normalizable partially massless modes (sources) have the same asymptotic behavior, as we discuss in detail in appendix \ref{se:3.6}.
For now, let us just note that our case~I boundary conditions result in a well defined variational principle.

In the light-cone coordinates used in \eqref{eq:bc2} the boundary metric is anti-diagonal, $\ga^{(0)}_{+-}=\frac12$, $\ga^{(0)}_{\pm\pm}=0$.
In these coordinates only the following components of the 1-point functions are non-vanishing:
\begin{align}
 T_{\pm\pm} &= \mp\frac{k}{\pi}\,\ga^{(2)}_{\pm\pm} \label{eq:gCS48}\,,\\
 \PMR_{++} &= \frac{k}{2\pi}\,\ga_{++}^{(1)} \,.\label{eq:gCS49}
\end{align}
Solving the equations of motion \eqref{2.3a} asymptotically with case~I boundary conditions (see appendix \ref{app:EOM}) establishes the conservation equations
\eq{
\partial_{\mp} T_{\pm\pm} = 0\, .
}{eq:gCS46}
For a traceless tensor the covariant version of \eqref{eq:gCS46} is the usual covariant conservation equation, $\nabla_\al T^{\al \be}=0$ (see e.g.~Eq.~(6.8) in \cite{Grumiller:2002nm}).
Another consequence of the asymptotic solution to the equations of motion is a differential relation between the partially massless response function $\PMR_{++}$ and the anti-holomorphic flux component of the stress energy tensor $T_{--}$:
\eq{
\partial_-^2 \PMR_{++} = \frac{\pi}{k}\,\PMR_{++} T_{--}\, .
}{eq:response}

We shall see in section \ref{se:genhol} that cases~II and~III can lead to 1-point functions with considerably different properties.
In particular, the Brown--York stress tensor no longer is conserved.
In appendix \ref{se:3.6} we demonstrate that loosening the boundary conditions \eqref{eq:bc3} can also lead to non-conservation, even for case~I.

\subsection{Charges}\label{se:3.3}

In this section we calculate the asymptotic charges $Q$.
Since the allowed Weyl rescalings have to go to zero asymptotically \eqref{eq:gCS31} we expect no Weyl charges, but only the standard diffeomorphism charges
\eq{
Q[\xi] = \int_{\partial \Sigma}\!\!\!\extd x\sqrt{\si}\,t^\al T_{\al\be} \xi^\be \, .
}{eq:gCS50}
Here $\Sigma$ is some constant time hypersurface in $M$ and $\partial \Sigma$ its intersection with the boundary $\partial M$, an $S^1$.
The induced metric is $\si$ and $t^\al$ is the (timelike) unit normal.
The vector field $\xi$ is required to behave asymptotically as in \eqref{eq:gCS28}.
In the light-cone gauge used in the previous section we obtain explicitly
\eq{
Q = \int\extd x^+ T_{++}\eps^+ - \int\extd x^- T_{--}\eps^- \,.
}{eq:gCS51}
We show now that the charges are finite and conserved.
Using $x^\pm=\varphi\pm t$ we find
\eq{
\partial_t Q = -2 \int\limits_0^{2\pi}\extd\varphi \,\big(\eps^+\partial_- T_{++} + \eps^-\partial_+ T_{--}\big) = 0\,,
}{eq:gCS52}
owing to the conservation \eqref{eq:gCS46} of the Brown--York stress tensor.
Mass and angular momentum for stationary solutions then become
\begin{align}
M &= \int\extd x^+ T_{++} + \int\extd x^- T_{--} = -\frac{k}{\pi}\,\int\extd x^+\ga_{++}^{(2)} +  \frac{k}{\pi}\,\int\extd x^-\ga_{--}^{(2)}\label{eq:gCS57} \,,\\
J &=  \int\extd x^+ T_{++} - \int\extd x^- T_{--} = \frac{k}{\pi}\,\int\extd x^+\ga_{++}^{(2)} +  \frac{k}{\pi}\,\int\extd x^-\ga_{--}^{(2)} \,. \label{eq:gCS58}
\end{align}
For the BTZ black hole \cite{Banados:1992wn} [see \eqref{eq:BTZ}] we obtain
\eq{
M_{\rm BTZ}= 2k r_+ r_-\,, \qquad J_{\rm BTZ}= k(r_+^2+r_-^2)\,,
}{eq:gCS59}
where $|r_+| \geq |r_-|$ are the inner and outer horizon radii, respectively. Note however that $r_\pm$, and therefore $M_{\rm BTZ}$, 
can have either sign. Effectively, the role of angular momentum and mass are interchanged as compared to Einstein gravity.
However, for positive $k$ the angular momentum $J_{\rm BTZ}$ is non-negative for all BTZ black holes.
This suggests to interchange the roles of ``time'' and ``angular coordinate'' in the dual CFT.

As a consistency check we compute the corresponding charges using the canonical formalism.
Inserting the asymptotic expansions \eqref{eq:Feffe} and corresponding expansions for $\om$, $\la$ and their variations into the result for the diffeomorphism charges \eqref{eq:diffcharge} [with $\xi^\rho$ as in \eqref{eq:gCS28}] yields
\eq{\begin{split}
\de Q_P[\xi^\rho] = -\frac{k}{\pi}\,\int\limits_0^{2\pi} \extd \varphi \,
\left[ \xi^+ \, \de e_{(2)-}^+  + \xi^-\, \de e_{(2)+}^-  + \xi^y \cO(1) \right] 
=-\frac{k}{\pi}\,\int\limits_0^{2\pi} \extd \varphi \, \left[ \eps^+ \, \de \ga^{(2)}_{++}  + \eps^-\, \de \ga^{(2)}_{--} \right] \, .
\end{split}
}{eq:varexp}
The boundary charge  from the canonical analysis \eqref{eq:varexp} agrees with the Brown--York result \eqref{eq:gCS51}, using the expression for the boundary stress-tensor \eqref{eq:gCS48}.

Let us now turn to the asymptotic charge corresponding to the Weyl rescalings \eqref{eq:weylcharge}.
For our boundary conditions $\de e_{i\varphi} = \cO(y^{-1})$ and $e^{i\mu} = \cO(y)$.
Thus the fall-off behavior of $\Omega$ renders the generator integrable and the allowed Weyl rescalings \eqref{eq:gCS31} are trivial gauge transformations.
As expected, the Weyl charges vanish for case~I.

\subsection{Graviton modes}\label{se:3.modes}

For many purposes --- e.g.~the calculation of 2- and 3-point correlators in the section \ref{se:3.3a} --- it is convenient to have explicit expressions for the normalizable and non-normalizable complex graviton modes in momentum space:\footnote{%
The modes below are displayed in transversal gauge and refer to fluctuations on the global AdS background as given in \eqref{eq:gCS6}, but in light-cone coordinates $u= x^+=t+\varphi$, $v=-x^-=t-\varphi$ and $e^{-\rho}\propto y$.}
\eq{
h_{\mu\nu}^{L/R/0} = e^{-ihu-i\bar h v}\,F_{\mu\nu}^{L/R/0}(\rho)\, .
}{eq:gCS63}
We call the $L_0$, $\bar L_0$ eigenvalues $h$, $\bar h$ ``weights'' and denote them by $(h,\,\bar h)$.
Expressions for the $SL(2,\mathbb{R})_{L/R}$ generators $L_{0\, \pm}$, $\bar L_{0\, \pm}$, the six Killing vectors of the AdS background \eqref{eq:gCS6}, can be found e.g.~in \cite{Li:2008dq}.
An explicit construction of the boundary gravitons (and their sources) $h^{L/R}$ for arbitrary weights was provided in Ref.~\cite{Grumiller:2009mw}.
The partially massless Weyl gravitons $h^0$ were discussed (though not exhaustively) in Ref.~\cite{Grumiller:2010tj}.
The modes can be classified into those that are regular at $\rho=0$ and those that are singular at $\rho=0$.
The latter are not a small perturbation of the background metric, which is regular at $\rho=0$, and thus we disregard these modes here.\footnote{%
These modes are important, however, to describe scalar quasi-normal modes in a BTZ black hole background \cite{Sachs:2008gt,Afshar:2010ii}.}
The regular modes can be divided into normalizable modes --- these are modes compatible with the boundary conditions \eqref{eq:bc3} --- and non-normalizable ones.
For instance, the partially massless primary with weights $(3/2,\,-1/2)$,
\eq{
h^0_{\mu\nu}(3/2,\,-1/2) = \big(g^{\textrm{\tiny AdS}}_{\mu\nu} - \nabla_\mu\nabla_\nu\big) \frac{\sinh^2\!\rho}{2\cosh\rho}\,e^{-3iu/2+iv/2}\,,
}{eq:gCS36}
is normalizable, and so are all its descendants. 
However, the partially massless primary with weights $(-1/2,\,3/2)$,
\eq{
j^0_{\mu\nu}(-1/2,\,3/2) = \big(g^{\textrm{\tiny AdS}}_{\mu\nu} - \nabla_\mu\nabla_\nu\big) \frac{\sinh^2\!\rho}{2\cosh\rho}\,e^{iu/2-3iv/2}\,,
}{eq:gCS37}
is non-normalizable, and so are all its descendants that are not also descendants of the other primary \eqref{eq:gCS36}.
As usual in the AdS/CFT correspondence the non-normalizable modes act as sources for the corresponding operators, while the normalizable modes appear as vacuum expectation value.

Note that both primaries \eqref{eq:gCS36} and \eqref{eq:gCS37} (and hence all descendants) are pure gauge; they are explicitly written
as an infinitesimal Weyl rescaling plus a diffeomorphism. These gauge transformations are not in the asymptotic symmetry group however, 
since the Weyl factor diverges too fast. So while the action of these gauge transformations might yield normalizable states when acting on the vacuum, 
they do not when acting on general states satisfying the boundary conditions. 
The Weyl gravitons are thus produced by large gauge transformations, but the corresponding generators are not part of the asymptotic symmetry algebra.
This is quite different from the boundary gravitons:
They arise as descendants of the vacuum by action of Virasoro generators, which do form part of the asymptotic symmetry algebra.

Using the results of \cite{Grumiller:2009mw} in the partially massless limit we find a simple relation between the flux components of the partially massless graviton modes that is valid for all weights $(h,\,\bar h)$:
\eq{
\big(\bar h^2-\frac14\big) h_{++}^0(h,\,\bar h) = \big(h^2-\frac14\big) h_{--}^0(h,\,\bar h)\, .
}{eq:gCS64}
We discuss now implications of the relation \eqref{eq:gCS64}.
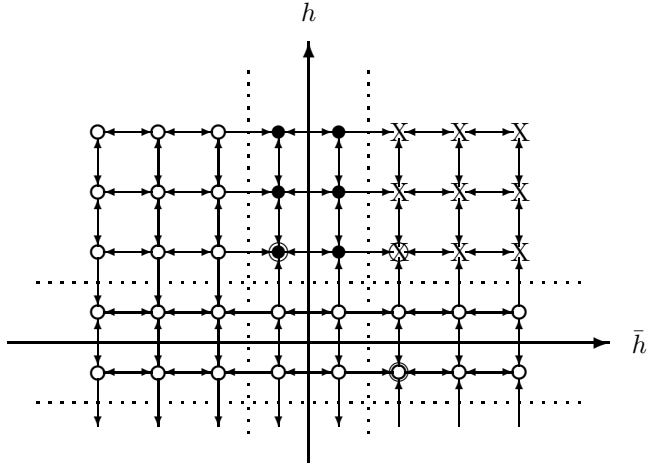
\begin{figure}
\begin{center}
\setlength{\unitlength}{0.8mm}
\begin{picture}(100,70)(-10,20)
\thicklines
%
\matrixput(5,55)(10,0){3}(0,10){3}{\circle{2}} 
\matrixput(5,35)(10,0){3}(0,10){2}{\circle{2}} 
\matrixput(35,55)(10,0){2}(0,10){3}{\circle*{2}} 
\matrixput(53.5,53.6)(10,0){3}(0,10){3}{X} 
\matrixput(35,35)(10,0){2}(0,10){2}{\circle{2}} 
\matrixput(55,35)(10,0){3}(0,10){2}{\circle{2}} 
%
%
\dottedline{2}(-5,50)(85,50) 
\dottedline{2}(-5,30)(85,30) 
\dottedline{2}(30,25)(30,85) 
\dottedline{2}(50,85)(50,25) 
\put(40,20){\vector(0,1){70}} 
\put(-10,40){\vector(1,0){100}} 
\put(40,95){\makebox(0,0){$h$}}
\put(95,40){\makebox(0,0){$\bar h$}}
\thinlines
%
\matrixput(6,35)(10,0){2}(0,10){2}{\vector(1,0){8}}
\matrixput(36,35)(10,0){4}(0,10){2}{\vector(1,0){8}}
\matrixput(6,55)(10,0){7}(0,10){3}{\vector(1,0){8}}
%
\matrixput(44,55)(-10,0){1}(0,10){3}{\vector(-1,0){8}}
\matrixput(74,55)(-10,0){2}(0,10){3}{\vector(-1,0){8}}
\matrixput(24,55)(-10,0){2}(0,10){3}{\vector(-1,0){8}}
\matrixput(74,35)(-10,0){2}(0,10){2}{\vector(-1,0){8}}
\matrixput(44,35)(-10,0){4}(0,10){2}{\vector(-1,0){8}}
%
\matrixput(5,36)(10,0){5}(0,10){1}{\vector(0,1){8}}
\matrixput(5,56)(10,0){8}(0,10){2}{\vector(0,1){8}}
\matrixput(55,26)(10,0){3}(0,10){3}{\vector(0,1){8}}
\matrixput(35,46)(10,0){2}(0,10){1}{\vector(0,1){8}}
%
\matrixput(5,74)(10,0){3}(0,-10){5}{\vector(0,-1){8}}
\matrixput(35,74)(10,0){5}(0,-10){2}{\vector(0,-1){8}}
\matrixput(35,44)(10,0){5}(0,-10){1}{\vector(0,-1){8}}
\matrixput(35,34)(10,0){2}(0,10){1}{\vector(0,-1){8}}
%
\put(35,55){\circle{3}}
\put(55,35){\circle{3}}
\put(55,55){\circle{3}}
\end{picture}
\caption{Spectrum of Weyl gravitons. Physical modes are denoted by full circles, pure gauge modes by X, and non-normalizable modes by empty circles. Primaries are additionally encircled.
Arrows denote the action of ladder operators $L_+$ (up), $L_-$ (down), $\bar L_+$ (right) and $\bar L_-$ (left). Dotted lines are semi-permeable barriers.
}
\label{fig:1}
\end{center}
\end{figure}
\begin{itemize}
 \item If $\bar h=\pm 1/2$ then the mode has only a component $h_{++}^{0}$.
Closer inspection reveals that it grows like $e^\rho$ and is thus compatible with the boundary conditions \eqref{eq:bc3}.
All these modes are generated from the normalizable primary \eqref{eq:gCS36} and its $\bar L_+$-descendent by acting on them (repeatedly) with $L_+$.
 \item Similarly, if $h=\pm 1/2$ we recover the non-normalizable primary \eqref{eq:gCS37} and its non-normalizable descendants.
See Fig.~\ref{fig:1} where all these modes are displayed.
 \item If $|h|, |\bar h|\neq 1/2$ then the partially massless mode necessarily is either non-normalizable ($h_{\pm\pm}^0\propto e^\rho$) or trivial ($h_{\pm\pm}^0\propto {\cal O}(e^{-\rho})$).
 \end{itemize}
Most importantly, the relation \eqref{eq:gCS64} is actually all we need to know about the partially massless modes to extract the 2- and 3-point correlators in section \ref{se:3.3a}.

\subsection{2- and 3-point functions}\label{se:3.3a}

We start with the 2-point functions involving solely the stress tensor, $\langle T T \rangle$.
The only interesting information contained in these 2-point functions are the values of the central charges, since everything else is fixed uniquely by conformal invariance.
We adopt the techniques of \cite{Grumiller:2009mw} to calculate the 2-point functions by reducing them to 2-point functions known from Einstein gravity (see section 4.1 in that work).
The only difference to the Einstein gravity result is the appearance of the operator ${\cal D}^0$ acting on one of the modes in the boundary integral that determines the 2-point functions.
Exploiting the properties
\eq{
\big({\cal D}^0 h^{R/L}\big)_{\mu\nu}= \eps_\mu{}^{\rho\la}\nabla_\rho h_{\la\nu}^{R/L} = \pm h_{\mu\nu}^{R/L} \,,
}{eq:gCS55}
we obtain immediately
\eq{
c_{R/L} = \pm c_{\rm BH} = \pm \frac{3}{2\hat G_N} \,,
}{eq:gCS53}
where $c_{\rm BH}$ is the Brown--Henneaux central charge evaluated for AdS radius $\ell=1$.
In order to match the corresponding overall coupling constants correctly we have to identify $16\pi\hat G_N$ in Einstein's theory with $2\pi/k$.
Thus we obtain
\eq{
c_{R/L} = \pm 12 k \,.
}{eq:gCS54}
The result \eqref{eq:gCS54} coincides with the scaling limit \eqref{eq:gCS9} of the result for the central charges in TMG \cite{Kraus:2005zm,Solodukhin:2005ah}.

Mixed 2-point functions vanish, $\langle\PMR T\rangle= 0$, since stress tensor $T$ and partially massless response $\PMR$ have different conformal weights.
The 2-point correlator between two partially massless excitations, $\langle\PMR\PMR\rangle$, is determined exclusively by the holographic counterterms calculated in \cite{Skenderis:2009nt}.
This is so, because the boundary expression without holographic counterterms that determines this correlator contains the operator $\eps^{\mu\rho\la}\nabla_\rho$ acting on one of the partially massless modes, which is annihilated by it [see \eqref{eq:PM}].
Thus, only the counterterms constructed in \cite{Skenderis:2009nt} contribute.
They are (at most) finite in the present case.
Actually, the contribution from these counterterms is on-shell equivalent to the contribution coming from the boundary term in \eqref{eq:gCS67}, so no further boundary terms are required.
We then obtain
\begin{equation}\label{eq:gCS65}
\langle\PMR\PMR\rangle \sim  \frac{k}{2\pi}\,\int_{\partial M}\!\!\!\extd^2 x\sqrt{|\ga|}\,k_L^{\al\be}k_{\al\be}^R \sim \frac{k}{4\pi}\,\int_{\partial M}\!\!\!\extd^2 x\sqrt{|\ga|}\,K_{\al\be}(\psi)K^{\al\be}(\psi) \,.
\end{equation}
Here $\psi$ is a (non-normalizable) Weyl graviton and $K$ is a corresponding deviation of extrinsic curvature.
We can now proceed similarly to \cite{Grumiller:2009mw} and use the large weight expansion as well as the asymptotic expansion \eqref{eq:gCS44} to evaluate the correlator \eqref{eq:gCS65} in momentum space, yielding
\eq{
\langle\PMR\PMR\rangle \propto \int_{\partial M}\!\!\!\extd^2 x \,\ga_{++}^{(1)}(h,\bar h) \,\ga_{--}^{(1)}\, .
}{eq:gCS66}
The source contribution $\ga_{--}^{(1)}$ is independent from the weights.
By virtue of the universal relation \eqref{eq:gCS64} the vacuum expectation value acquires a second order pole in one of the weights, $\ga_{++}^{(1)}=\ga_{--}^{(1)}\, h^2/\bar h^2$.
Fourier-transformation then establishes
\eq{
\langle\PMR(z,\bar z)\PMR(0,0)\rangle = \frac{c_0\,\bar z}{2z^3}\, ,
}{eq:gCS60}
with $z=\varphi+it$ and some normalization constant $c_0$ that can be deduced by tracing all the proportionality constants in the calculations above.
In fact, even though Skenderis, Taylor and van Rees excluded the partially massless case from their considerations, our result \eqref{eq:gCS60} is consistent with the scaling limit \eqref{eq:gCS9} of their results.
This allows us to read off the normalization constant from (7.13) of their work \cite{Skenderis:2009nt}:
\eq{
c_0 = 4 k\, .
}{eq:gCS61}

3-point functions require the evaluation of the third variation of the action.
Up to boundary terms it is given by \cite{Maloney:2009ck,Grumiller:2009mw}
\eq{
\de^{(3)} S_{\textrm{CSG}} = \frac{k}{2\pi}\,\int\extd^3x\sqrt{-g}\,\Big[\big({\cal D}^0 h\big)^{\mu\nu}\,\de^{(2)} R_{\mu\nu} + h^{\mu\nu}\,\De_{\mu\nu}\Big]\,.
}{eq:f73}
Here $\de^{(2)} R$ is the second variation of the Ricci tensor, ${\cal D}^0$ is defined in \eqref{eq:gCS55}, and we have used the definition
\eq{
\De_{\mu\nu}=\varepsilon_\mu{}^{\si\kappa}\,\de\Ga^\rho{}_{\kappa\nu}(h)\,\big({\cal D}^L{\cal D}^R h\big)_{\si\rho} \,.
}{eq:f76}
The properties \eqref{eq:gCS55} together with the fact that $\De_{\mu\nu}$ vanishes if $h$ contains only left- and right-moving boundary gravitons again allow to reduce correlators of the stress tensor with itself to correlators known from Einstein gravity.
Namely, the Einstein gravity result for the third variation contains only the first term in \eqref{eq:f73}, but without the operator ${\cal D}^0$.
The latter has $h^{R/L}$ as eigenmodes with eigenvalues $\pm 1$, which is compatible with the identification performed in the calculation of 2-point correlators \eqref{eq:gCS53}.
Thus, the results for the 3-point correlators $\langle T T T \rangle$ are consistent with the conformal Ward identities, as may have been anticipated on general grounds.
3-point correlators that involve an odd number of partially massless insertions $\PMR$ vanish, $\langle\PMR T T\rangle=\langle\PMR\PMR\PMR\rangle=0$, since Weyl gravitons have half-integer weights.
Thus, all 3-point correlators can be obtained using shortcuts, except for the correlator $\langle\PMR\PMR T \rangle$.
Exploiting again the results of \cite{Skenderis:2009nt} we show now that this correlator is non-vanishing.
Defining $|\PMR\rangle:=\lim_{z,\bar z\to 0}\PMR(z,\bar z)|0\rangle$ and $\langle\PMR|:=\lim_{z,\bar z\to\infty}\langle 0|\PMR(z,\bar z)z^3\bar z^{-1}$ yields
\eq{
\langle\PMR| T^R(z) |\PMR\rangle = \frac{3c_0}{4z^2}\qquad  \langle\PMR| T^L(\bar z)|\PMR\rangle = -\frac{c_0}{4\bar z^2}\,.
}{eq:gCS56}
The result \eqref{eq:gCS56} does not provide a check of the conformal Ward identities, but rather uses them as an input.

\section{Generalized holography}\label{se:genhol}

In this section we consider cases~II and III of the boundary conditions defined in section \ref{se:leah}.
For all these cases it is practical to note that under Weyl rescalings \eqref{eq:Weyl} the CSG action \eqref{eq:gCS67} transforms with a boundary term:
\eq{
\De S_{\textrm{CSG}}[\Om] = \frac{k}{4\pi}\,\int_{\partial M}\!\!\!\extd^2x\sqrt{-\gamma}\, n_{\mu}\varepsilon^{\mu\nu\lambda}\,g^{\rho\si}\,(\partial_\si\Omega)(\partial_{\nu}g_{\rho\lambda})\,,
}{eq:gCS38}
where $n^\mu$ is the unit normal vector. Note that $\Omega$ can be either finite or infinitesimal in \eqref{eq:gCS38}.
This relation allows us to write the gravitational Chern--Simons action of a metric $g_{\mu\nu} = e^{2\phi}\bar{g}_{\mu\nu}$ as
\eq{
S_{\textrm{CSG}}[g] = S_{\textrm{CSG}}[\bar{g}] + S_{\textrm{B}}[\bar{g}, \phi] \, ,
}{eq:an1}
where the boundary term $S_{\textrm{B}}[\bar{g}, \phi]$ is given by Eq.~\eqref{eq:gCS38} with $g\to \bar{g}$ and $\Omega \to \phi$.
We record this term for the convenient case when $\bar{g}_{\mu\nu}$ is expressed in Gaussian normal coordinates
\begin{equation}\label{eq:an2}
g_{\mu\nu} \extd x^\mu\extd x^\nu \,= \,
 e^{2\phi} \, \bar{g}_{\mu\nu} \extd x^\mu\extd x^\nu \,=\, e^{2\phi}\left( \extd \rho^2 + \bar{\ga}_{\al\be}\,\extd x^\al\extd x^\be\right)\,.
\end{equation}
We then have
\eq{
S_{\textrm{B}}[\bar{g}, \phi] = \frac{k}{4\pi}\,\int_{\partial M}\!\!\!\extd^2x  \, \epsilon^{\al\be}\,\bar{\ga}^{\ga\de}\,(\partial_\de \phi)(\partial_{\al} \bar{\ga}_{\ga\be})\,.
}{eq:an3}

In section \ref{se:3.4} we study case~II and fill the gaps of the corresponding discussion in Ref.~\cite{Afshar:2011yh}.
The properties of the boundary CFT do not differ appreciably from case~I.
In section \ref{se:3.5} we study case~III.
The boundary CFT then contains an additional scalar field, whose properties depend on certain choices that we explain in section \ref{se:5}.

\subsection{Case~II: Weyl factor fixed}\label{se:3.4}

A $y$-dependent Weyl rescaling can turn an asymptotic boundary to one at finite distance, or even to a curvature singularity.
Interestingly, if the Weyl factor $\phi$ is a function of $y$ only, $f=0$ in \eqref{eq:gCS30}, then we recover all the results from the previous sections that dealt with case~I.
This can be seen directly from the boundary term \eqref{eq:an3} in the action. 
If the Weyl factor $\phi$ depends only on $y$ the boundary contribution \eqref{eq:an1} vanishes. Consequently,
both the action and the boundary conditions are identical
to case~I if we send $g_{\mu\nu} \to \bar{g}_{\mu\nu}$. 

Now let us turn to boundary response functions. Note first that there is no trivial gauge transformation that takes the metric into Gaussian normal coordinates.
In fact, the best we can do is to use ``conformal'' Gaussian normal coordinates \eqref{eq:an2}.
This is, however, enough to make extensive use of our results for case~I.
Indeed, for the present case the most general deformation of $g_{\mu\nu}$ that preserves the boundary conditions is a deformation of $\bar{g}_{\mu\nu}$ that satisfies the same conditions as in case~I.
Therefore, the first term of the action \eqref{eq:an1} gives an identical contribution as for case~I.

When performing the computations it is convenient to work with a Fefferman--Graham expansion of $\bar{\ga}_{\al\be}$\ .
\eq{
\bar{\ga}_{\al\be} = e^{2\rho}\bar\ga_{\al\be}^{(0)} + e^\rho\bar\ga_{\al\be}^{(1)} + \bar\ga_{\al\be}^{(2)} + \dots
}{eq:FGbar}
All boundary indices should be raised and lowered with $\bar\ga_{\al\be}^{(0)}$, and $\varepsilon^{\al\be}$ is normalized by its determinant.
The variation of the boundary term is then
\eq{
\de S_{\rm B} = \frac{1}{2}\int_{\partial M}\!\!\! \extd^2 x \sqrt{-\bar\ga^{(0)}} \, T^{\al\be}_{\rm an} \, \de \bar\ga^{(0)}_{\al\be} \, ,
}{eq:an4}
with
\eq{
T^{\al\be}_{\rm an} = \frac{k}{4\pi} \varepsilon^{\al\ga}\partial^\be \partial_\ga f + (\al \leftrightarrow \be)\,.
}{eq:an5}
In the total Brown--York stress tensor this piece is added to the analog of the case~I result \eqref{eq:goodT}:
\eq{
T^{\al\be} = \frac{k}{2\pi}\,\eps^{\al\ga}\,\left(\bar\ga^{\be\,(2)}_{\;\,\ga}  + \frac{1}{2}\partial^\be \partial_\ga f \right) + (\al\leftrightarrow\be) \,,
}{eq:an6}
or, in light-cone components,
\eq{
T_{\pm\pm} = \mp\frac{k}{\pi}\,(\bar{\ga}^{(2)}_{\pm\pm} + \frac{1}{2}\partial^2_{\pm} f ) \, .
}{eq:an8}
The contribution $T^{\al\be}_{\rm an}$ is state independent and responsible for the anomalous conservation law
of the stress energy tensor expected for a CFT on a curved background\footnote{As noted in section \ref{se:leah} and appendix \ref{app:gauge}, 
the presence of a non-trivial $f$ is gauge equivalent to a non-trivial boundary metric.}. This effect was explained in Ref.~\cite{Kraus:2005zm}, and
we recover the expected anomalous result
\eq{
\nabla_{\al} T^{\al\be} = \ga_{(0)}^{\al\be}\varepsilon^{\de\ga}\partial_{\de}\partial_{\la}\Ga^\la{}_{\al\ga}[\ga^{(0)}]\, .
}{eq:an7}
On the gravity side, the reason for the anomalous conservation \eqref{eq:an7} is that the action is not diffeomorphism invariant: it transforms
by a boundary term. This anomaly vanishes only for flat boundary metrics. Indeed, requiring the stress tensor to be conserved implies
\eq{
f(x^+, x^-) = f_+(x^+) + f_-(x^-) + ({\rm const.}) \, x^+ x^-\, .
}{eq:an7.5}
The last term is not periodic in $\varphi = (x^+ + x^-)/2$ and is thus disallowed,
leaving only flat boundary metrics, albeit in different coordinate systems. Restricting to
this class of Weyl factors $\phi$ produces a class of anomaly free CFTs.
Since the variation \eqref{eq:an4} does not contain any response to $\ga^{(1)}$, the result for $\PMR^{\al\be}$ can be read off directly
from its case~I counterpart \eqref{eq:J}.

The additional term $T^{\al\be}_{\rm an}$ gives in general a contribution to the Virasoro charges.
Equation \eqref{eq:gCS51} still holds, and we can just plug in the result \eqref{eq:an6}.
As an example we record here the contributions for the mass $M$ and angular momentum $J$ coming from the anomalous piece:
\begin{align}\label{eq:gCS81}
M_{\rm an} &= -k\int\limits_0^{2\pi}\extd\varphi\,(\partial^2_+ + \partial^2_-)\,f \, ,\\
J_{\rm an} &= -k\int\limits_0^{2\pi}\extd\varphi\,(\partial^2_+ - \partial^2_-)\,f = 0 \, .
\end{align}
The whole tower of charges is conserved if and only if $f$ is a solution to the massless 2-dimensional Klein--Gordon equation
\eq{
\partial_+\partial_- f = 0\, .
}{eq:gCS82}
From the canonical analysis, however, the potential non-conservation of the charges is invisible.
The appropriate Fefferman--Graham expansions for the dreibein and its variation are given by the expressions \eqref{eq:Feffe}. 
Since the conformal factor itself does not vary between the metrics satisfying the boundary conditions of case~II, we just send $\de e^i{}_\mu \to e^{\phi} \de \bar e^{\,i}{}_\mu$. Computing the expression \eqref{eq:diffcharge} produces now exactly the same result as before, i.e.,
\eq{
Q_P[\xi^\rho] = \int\limits_0^{2\pi} \extd \varphi \, \de \Ga_P =
-2\int\limits_0^{2\pi} \extd \varphi \, \left[ \eps^+ \, \de \bar{\ga}^{(2)}_{++}  + \eps^-\, \de \bar{\ga}^{(2)}_{--} \right] \,,
}{eq:varexp2}
without the anomalous piece.
The difference between the Brown--York and the canonical result comes about because the former is based upon the variation of the action \eqref{eq:gCS67},
which is not diffeomorphism invariant at the boundary, while the latter is based upon the canonical analysis of the action \eqref{2.1},
which is manifestly diffeomorphism invariant. The fact that the charges are conserved does not mean that the anomaly is not present in this formulation.
As explained in Ref.~\cite{Kraus:2005zm}, in the first order formulation the anomaly manifests itself as a Lorentz anomaly, leading to an antisymmetric piece in the stress tensor.

In precise parallel to case~I the asymptotic charge corresponding to the Weyl rescalings vanishes.
In the expression \eqref{eq:deGW} we now have $\de e_{i\varphi} = \cO(y^{\exponent-1})$ and $e^{i \mu} = \cO(y^{2-\exponent})$, with the constant $\exponent$ from the Weyl factor \eqref{eq:gCS30}.
The fall-off behavior of $\Omega$ again shows that the allowed Weyl rescalings \eqref{eq:gCS31} are trivial gauge transformations.

\subsection{Case III: Weyl factor set free}\label{se:3.5}

Case III has several similarities to case~II, but also some essential differences.
For the same reason as before we have to restrict the Weyl factor $\phi$ as in \eqref{eq:gCS30}, but with $f$ now being free rather than fixed. 
As explained in appendix \ref{app:gauge} we keep the constant $\exponent$ arbitrary but fixed.

The gauge transformations preserving the boundary conditions now include the asymptotic
diffeomorphisms \eqref{eq:gCS28}, and the allowed Weyl rescalings are enhanced to include all $\Om$ of the form
\eq{
\Om = 
f_\Om(x^+,\,x^-) + {\cal O}(y^2)\,.
}{eq:gCS27}
with an arbitrary function $f_\Om(x^+,\,x^-)$.

When computing boundary response functions it is important to parametrize the variations suitably: to identify the stress tensor one must know which deformation corresponds to a variation of the boundary metric. Since now also variations of the Weyl factor
$\phi$ are allowed, the variation of the induced metric at the boundary has two pieces.
In conformal Gaussian normal coordinates \eqref{eq:an2} we have
\eq{\begin{split}
\de \left( g_{\mu\nu}\extd x^\mu \extd x^\nu \right)&= 2 \de\phi \,  g_{\mu\nu}\extd x^\mu \extd x^\nu
+ e^{2\phi}\left(\de \bar{\ga}_{\al\be}\,\extd x^\al\extd x^\be \right) =\\
&= e^{2\phi}\left( 2 \de\phi\, \extd\rho^2  + [2\bar{\ga}_{\al\be}\, \de\phi + \de \bar{\ga}_{\al\be} ]\,\extd x^\al\extd x^\be \right)\, .
\end{split}
}{eq:free1}
Together with the prefactor $e^{2\phi}$ the two terms in the square brackets make up the variation of the induced metric at the boundary.
However, we define the stress-tensor as the response to the second of these variations, i.e., to a variation where $\de \phi = 0$.
We expand the conformal class of the boundary metric $\bar{\ga}_{\al\be}$ as in \eqref{eq:FGbar}, and the variation of the Weyl factor $\de \phi$ as
\eq{
\de \phi =
\de \exponent \log y + \de f + \ldots \, .
}{eq:free3}
Our boundary conditions are 
$\de \exponent = 0$. To compute the first variation of the on-shell action let us note that the only difference from case~II will be the contribution from $\de \phi$.
It is clear from the forms \eqref{eq:an1} and \eqref{eq:an3} of the action that this term takes the form
\begin{align}\label{eq:free4}
\frac{k}{4\pi}\,\int_{\partial M}\!\!\!\extd^2x  \, \epsilon^{\al\be}\,\bar{\ga}^{\ga\de}\,(\partial_\de \de \phi)(\partial_{\al} \bar{\ga}_{\ga\be}) = 0\, .
\end{align}
To get something non-vanishing one would need an $x^-$-dependent term of order $y^{-1}$ in \eqref{eq:free3}. 
Therefore, the variational principle is well-defined. The full on-shell variation of the action then becomes
\begin{equation}\label{eq:free5}
\de S_{\rm CSG}\big|_{\rm EOM} = \frac{1}{2}\,\int_{\partial M}\!\!\!\extd^2x\sqrt{-\bar\ga^{(0)}}\,\Big(T^{\al\be}\,\de\bar\ga^{(0)}_{\al\be} + \PMR^{\al\be}\,\de\bar\ga^{(1)}_{\al\be} 
\Big)\,,
\end{equation}
with the boundary stress-tensor $T^{\al\be}$ given by \eqref{eq:an6} and the partially massless response function $\PMR^{\al\be}$ by \eqref{eq:J}.

We compare below these response functions to the asymptotic charges corresponding to diffeomorphisms and Weyl rescalings obtained from the canonical analysis.
We start with the interpretation of the Weyl charge, which for case~III finally becomes non-trivial.
Two metrics that differ in the prefactor $\phi(x^+, x^-)$ have $\de e_{i\phi} = \cO(y^{-2})$, which together with $\Om = \cO(1)$ gives a finite contribution in Eq.~\eqref{eq:deGW}.
In fact, for two metrics differing in $\phi$ by a deformation
\eq{
\de \phi = 
\de f(x^+,\,x^-) + \ldots
}{eq:free2}
we get
\eq{
\int \extd^3 x \, \de G_{W}  =\frac{k}{\pi}\, \int\limits_0^{2\pi} \extd \varphi
\, \de f \, \partial_\varphi \Om + {\rm regular} \, .
}{eq:free7}
To obtain differentiable charges we thus need to add a boundary piece
\eq{
\de Q_W[\Om] = -\frac{k}{\pi} \int\limits_0^{2\pi} \extd \varphi \, \de \left( f \, \partial_\varphi \Om \right) 
}{eq:free8}
to the generator. Note that it is clear that the charge \eqref{eq:free8} is not conserved for arbitrary functions $f$ and $\Om$.
It is conserved if and only if $\partial_t  (f \partial_\varphi \Om )$
is a total $\varphi$-derivative.
Let us remark how this can be interpreted from a Brown--York perspective. 
The boundary response \eqref{eq:free5} contains no term that corresponds to this charge. Instead, it is related to a ``boundary-of-the-boundary'' term that we threw away.
Indeed, using the variation \eqref{eq:gCS38} of the action under a Weyl rescaling $\Om$, we see that for
$g_{\mu\nu} = e^{2\phi} \bar{g}_{\mu\nu}$ the corresponding ``response'' is
\eq{
\begin{split}
\de S_{\rm CSG} &= \frac{k}{2\pi}\,\int_{\partial M}\!\!\!\extd\; (f \extd \Omega) =\frac{k}{2\pi}\,\int \!\extd\varphi \extd t
\left[ \partial_t (f \partial_\varphi \Om ) -  \partial_\varphi (f \partial_t \Om ) \right]  \\
&= \frac{k}{2\pi}\,\int \!\extd\varphi \extd t \,\, \partial_t (f \partial_\phi \Om ) \,.
\end{split}
}{eq:free9}
In the last step we used that the $\varphi$-direction is periodic. The conservation of the Weyl charge
\eqref{eq:free8} is therefore equivalent to demanding that the action is also invariant under Weyl rescaling
without neglecting boundary terms in the $t$-direction.

Let us consider explicitly the case when the stress tensor is conserved, i.e., when $f = f_+(x^+) + f_-(x^-)$.
(As noted earlier, in the first order formalism this corresponds to having no Lorentz anomaly.)
To preserve this form, we must also have $\Omega = \Om_+(x^+) + \Om_-(x^-)$.
Fourier expanding these functions as
\eq{
f_\pm = \frac{f_0}{2} + \frac{p_f}{2}(t \pm \varphi) + \sum_{n\neq 0} f_\pm^{(n)} e^{-in(t \pm \varphi)}\, ,
}{eq:free10}
\eq{
\Om_\pm = \frac{\Om_0}{2} + \frac{p_\Om}{2}(t \pm \varphi) + \sum_{n\neq 0} \Om_\pm^{(n)} e^{-in(t \pm \varphi)}\, ,
}{eq:free11}
one finds straightforwardly that the conservation of the Weyl charge is equivalent to requiring
\eq{
f_+^{(n)} \Om_-^{(n)} = f_-^{(n)} \Om_+^{(n)}\qquad \forall n \neq 0  \,.
}{eq:free12}
This requirement means that the Weyl rescaling $\Omega$ preserves a functional relation
between $f_+$ and $f_-$ of the form
\eq{
f_+^{(n)} = C_n f_-^{(n)}
}{eq:free13}
for some constants $C_n$ (the same applies for $\Om$). Particularly simple choices are
\eq{
 f_- = 0\, ,  \qquad f_+ = 0\, ,  \qquad\mbox{or} \qquad f_+(x) = C\, f_-(x)\,.
}{eq:free14}
Any such choice leads to an infinite tower of conserved charges.
For instance, the first choice in Eq.~\eqref{eq:free14} leads to the tower
\eq{
Q_W[\Om = -e^{in(t+\varphi)}] = 2i k\, n f_+^{(n)} \, .
}{eq:free15}
The generators $\aff_n = \tilde{G}_W[\Om = -e^{in(t+\varphi)}]$ have Poisson brackets
\begin{align}
& i\,\{\aff_n, \aff_m\}^\ast_1 = 2k n\,\de_{n+m,0} \label{eq:gCS89} 
\end{align}
which can be shown by noting that $\phi = \frac{1}{3}\log(\det e) + \dots$, where the ellipsis denotes field-independent terms.

Let us now turn to the Virasoro charges.
At this stage it seems somewhat artificial to accompany the diffeomorphisms in \eqref{eq:gCS28} by the
compensating Weyl rescaling \eqref{eq:gCS29}, since the transformation preserves the boundary condition also without it.
With it, however, the metric stays unchanged at leading order.
We therefore still define the transformation $\de_\xi$ as a combination of the diffeomorphism \eqref{eq:gCS28} and the
Weyl rescaling \eqref{eq:gCS29}.
Let us anyhow start by analyzing the contribution to the charge coming from the diffeomorphisms.
Putting the expression \eqref{eq:Feffe} into the variation \eqref{eq:diffcharge} produces
\eq{
\de Q_P[\xi^\rho] = -\frac{k}{\pi}\int\limits_0^{2\pi} \extd \varphi \, \Big[ \xi^+ \, \de C_+  + \xi^-\, \de C_-  + \xi^y\, \de C_y + \theta^{\hat y}[\xi^\rho]\,\de\om_{\hat y\varphi}\Big]\, ,
}{eq:diffch3}
with
\begin{align}
&\de C_{\pm} = \left[ \de \bar e_{(2)\pm}^{\, \mp} + \partial_{\pm} \partial_\varphi \de f - (\partial_{\pm} f) \partial_\varphi \de f \right]+ {\cal O}(y)\, ,\\
&\de C_y = \frac{1-\exponent}{y}\partial_\varphi \de f + {\cal O}(1)\, .
\end{align}
The Lorentz parameter and the variation of the connection read
\begin{align}
&\theta^{\,\hat y}[\xi^\rho] = \frac12\,\big(\partial_+\eps^+-\partial_-\eps^-\big) + {\cal O}(y^2)\,,   \\
&\theta^{\,\pm}[\xi^\rho] = \pm\, y \, \partial^2_{\mp}\,\eps^{\mp} + {\cal O}(y^2) \qquad \text{and} \qquad\om^{\hat y}{}_\varphi = \partial_t\phi + {\cal O}(y)\,.
\end{align}
Using these expressions it is straightforward to obtain the diffeomorphism and Weyl contributions to the charges:
\begin{align}
&\de Q_P[\xi^\rho] = -\frac{k}{\pi}\int\limits_0^{2\pi} \extd \varphi \, \Big( \eps^+ \, \de \bar e_{(2)+}^{\, -}  + \eps^- \, \de \bar e_{(2)-}^{\, +}
- \big[ (\eps \cdot \partial f) + \frac{\exponent}{2} \, (\partial \cdot \eps) \big] \, \partial_\varphi \de f \Big) \, , \label{eq:gCS93.5}\\
&\de Q_W[\xi^\rho] = -\frac{k}{\pi}\int\limits_0^{2\pi} \extd \varphi \, \big[ (\eps \cdot \partial f) + \frac{\exponent}{2} \, (\partial \cdot \eps) \big] \, \partial_\varphi \de f  \, .  \label{eq:gCS94}
\end{align}
The second expression results from combining \eqref{eq:free8} and \eqref{eq:gCS29}, and we have eliminated $\varphi$-derivatives in both expressions. 
Note that both contributions on their own depend on $\exponent$ in a nontrivial way.

Adding the contributions, $\de Q[\xi^\rho] = \de Q_P[\xi^\rho] + \de Q_W[\xi^\rho]$, cancels the terms proportional to $\exponent$  and the 
terms bilinear in $\phi$ and $\de \phi$. For the diffeomorphisms compensated by the Weyl rescaling \eqref{eq:gCS29} we thus obtain the charges.
\eq{
\de Q[\xi^\rho] = -\frac{k}{\pi} \int\limits_0^{2\pi} \extd \varphi \, \Big[\eps^+ \, \de \bar e_{(2)+}^{\, -}  + \eps^- \, \de \bar e_{(2)-}^{\, +}\Big]\, .
}{eq:gCS88}
The result \eqref{eq:gCS88} coincides with the one for cases~I and II.
Thus, the charges \eqref{eq:gCS88} are conserved, as expected from the discussion of case~II below Eq.~\eqref{eq:varexp2}.

As remarked, the diffeomorphism charges \eqref{eq:gCS93.5} contain a term bilinear in $\phi$ and $\de \phi$. It is not obvious that this term
is integrable, nor that it is conserved. Assuming certain restrictions on the function $f$ it is however possible to obtain integrable
and conserved charges also for the pure diffeomorphism. We shall investigate such possibilities in the next subsection.

\subsection{CFT interpretation}\label{se:5}

In Ref.~\cite{Afshar:2011yh} we conjectured that CSG with boundary conditions \eqref{eq:bc1}-\eqref{eq:bc3}, \eqref{eq:gCS27}, \eqref{eq:free2}, $f=f(x^+)$ and $f_- =0 $ is dual to a CFT with enhanced symmetries.
Below we substantiate and generalize this conjecture, following the suggestions of Ref.~\cite{Afshar:2011yh}. 

The enhanced symmetries of the dual CFT are generated by the Virasoro operators $L_n$, $\bar L_n$ as well as the affine $\hat u(1)$-generators $\aff_n$.
The Virasoro generators are defined as combinations of diffeomorphisms \eqref{eq:gCS28} and
compensating Weyl rescalings \eqref{eq:gCS29}: $L_n=\tilde G_P[\eps^+=e^{inx^+}]+\tilde G_W[\Om=\Om(\eps^+)]$ and $\bar L_n=\tilde G_P[\eps^-=-e^{-inx^-}]+\tilde G_W[\Om=\Om(\eps^-)]$.
As usual, we convert the Poisson brackets into commutators by the prescription $i\{q,\,p\}=[\hat q,\,\hat p]$.
The non-zero commutators (including also the Weyl generators $\aff_n = \tilde{G}_W[\Om = -e^{in(t+\varphi)}]$) are
\begin{subequations}
\label{eq:algebra}
\begin{align}
 [L_n,\,L_m] &= (n-m)\,L_{n+m} + \frac{c_L}{12}\,(n^3-n)\,\de_{n+m,0} \, ,\\
 [\bar L_n,\,\bar L_m] &= (n-m)\,\bar L_{n+m} + \frac{c_R}{12}\,(n^3-n)\,\de_{n+m,0}\, ,\\
 [\aff_n,\,\aff_m] &= c_0\,n\,\de_{n+m,0} \, .
\end{align}
\end{subequations}
The commutator $[L_n,\,\aff_m]$ vanishes due to the peculiar way the Virasoro generators $L_n$, $\bar L_n$ arise.
By construction they act trivially on the 3-dimensional Weyl factor.
The values of the central charges are determined by the Chern--Simons level $k$ from the action \eqref{eq:gCS10} [or, more accurately, from the action \eqref{eq:gCS67}]:
\eq{
 c_R = - c_L = 6\, c_0 = 12\,k\, .
}{eq:gCS86}
Of course, the value of $c_0$ is defined only with respect to a given normalization of the generators $\aff_n$.

Let us now consider the pure diffeomorphism charges for this case. 
We treat first the simpler case $\exponent = 0$. 
Since $f(x^+)$ is now not completely general, but only a function of
$x^+$, and since we furthermore assume $\exponent = 0$, the compensating Weyl rescaling in \eqref{eq:gCS29} reads
\eq{
\Om = -\frac b2\,\partial\cdot\eps - \eps\cdot\partial f + {\cal O}(y^2) = - \eps^+ \, \partial_+ f\, . 
}{eq:gCS92.5}
Note in particular that the transformations parametrized by $\eps^-$ do not need a compensating rescaling. They leave $f$ invariant. 
Thus the problematic terms in \eqref{eq:gCS93.5} are zero for the $\eps^-$ transformations and the $\bar{L}$-algebra generates
pure diffeomorphisms. The $\eps^+$ transformations on the other hand need a compensating rescaling to leave $f$ invariant.
Dropping this compensation, however, only sends $f$ to a new function of $x^+$, preserving the boundary conditions.

We define now Sugawara-shifted generators $\vir_n$ that generate the pure diffeomorphisms parametrized by $\eps^+$.  
We shift the generators according to ($::$ denotes normal ordering):
\eq{
L_n\to\vir_n = L_n + \frac{1}{4k}\,\sum_{m\in\mathbb{Z}} :\aff_m \aff_{n-m}:\, .
}{eq:gCS93}
To show that the generators $\vir_n$ produce only diffeomorphisms it is sufficient to check that the compensating Weyl charge \eqref{eq:gCS94} by virtue of \eqref{eq:free15} can be written as
\begin{align}\label{eq:gCS95}
\de Q_W = \frac{k}{\pi}\,\int\limits_0^{2\pi}\extd\varphi\, \de f \partial_\varphi ( e^{inx^+}\partial_+ f) \, 
&= \, k\,\de \Big(\sum_{m\in\mathbb{Z}} m(n-m) f_+^{(m)}f_+^{(n-m)}\Big) \Rightarrow \\
\tilde G_W &= - \frac{1}{4k}\,\sum_{m\in\mathbb{Z}} :\aff_m \aff_{n-m}:\, .
\end{align}
In the last equality we have converted classical expressions into quantum operators, with corresponding ordering ambiguities.
We have fixed the latter by requiring normal ordering.
Since $L_n$ is a sum of a diffeomorphism and a holomorphic Weyl rescaling with Weyl charge \eqref{eq:gCS95}, the shifted Virasoro operators $\vir_n$ in \eqref{eq:gCS93} generate by construction solely diffeomorphisms.
The definition \eqref{eq:gCS93} together with the old algebra \eqref{eq:algebra} establish a new algebra that contains an affine $\hat u(1)_R$ generated by $\aff_n$.
The non-zero commutators are given by
\begin{subequations}
\label{eq:algebranew}
\begin{align}
 [\vir_n,\,\vir_m] &= (n-m)\,\vir_{n+m} - \big(k - \frac{1}{12}\big)\,(n^3-n)\,\de_{n+m,0} \label{eq:CFT1new}\, ,\\
 [\bar L_n,\,\bar L_m] &= (n-m)\,\bar L_{n+m} +  k\,(n^3-n)\,\de_{n+m,0}\, ,\\
 [\aff_n,\,\aff_m] &= 2k\,n\,\de_{n+m,0} \, ,\\
 [\vir_n,\,\aff_m] &= -m\,\aff_{n+m}\, .
\end{align}
\end{subequations}
Note in particular that the last commutator is now non-vanishing and shows that $\aff_n$ behaves as an operator with the appropriate conformal weights $(1,0)$.
As compared to cases I and II, in case III the holomorphic Weyl factor constitutes an additional free chiral boson in the theory.
The shift $c_L\to c_L+1$ (and no shift of $c_R$) is precisely what one would expect from a free chiral boson.
Note that there is no corresponding shift of the right central charge or the right Virasoro generators. 

At this point a comment regarding the normal ordering procedure is apposite. We determined the algebra \eqref{eq:CFT1new} of diffeomorphisms 
using the results in \eqref{eq:algebra} for the commutators involving $L_n$ and $\aff_n$. The resulting central charge depends crucially on the 
ordering prescription of the term bilinear in $\aff$ in the generators \eqref{eq:gCS93}. Thus the appearance of the shift in the central charge is a purely quantum mechanical effect.
Indeed, simply computing the Poisson brackets of the corresponding constraints misses the contribution. Similarly, by changing the normal ordering prescription, i.e., by choosing another vacuum state for the quantum theory, it is possible to shift the central charge by the same magnitude, but with the opposite sign. 

For the second choice in \eqref{eq:free14} essentially the same discussion applies, with suitable changes $L\leftrightarrow R$.
Therefore
\eq{
[\bar \vir_n,\,\bar \vir_m] = (n-m)\,\bar\vir_{n+m} + \big(k - \frac{1}{12} \big)\,(n^3-n)\,\de_{n+m,0}\, .
}{eq:algebranewLR}

It is worthwhile recalling the peculiar way in which the chiral boson arises on the gravity side.
In the bulk there is no scalar field, but only the Weyl factor $\phi$ in \eqref{eq:bc1}.
The bulk equations of motion \eqref{eq:gCS1} do not restrict $\phi$ at all!
The whole dynamics of $\phi$ emerges through consistency conditions imposed at the boundary.
If the stress-energy tensor is postulated to be conserved then the flatness condition \eqref{eq:an7.5} must hold, which is equivalent to demanding that $\phi$ obeys the massless Klein-Gordon equation at $\partial M$.
If additionally the Weyl charges are required to be conserved then $\phi$ is restricted further, as we have shown above.
Thus, the whole dynamics of the scalar field arises solely through boundary and consistency conditions, and not through an interplay between bulk and boundary dynamics.

Let us now turn to the case $\exponent \neq 0$.
Despite the fact that theories with different $\exponent$ can be mapped to each other by a $y$-dependent Weyl rescaling, it is worthwhile to treat this case explicitly.
The compensating Weyl rescaling now acquires a $\exponent$-dependent piece, and with it the algebra \eqref{eq:algebra} remains unchanged.
However, the presence of $\exponent\neq 0$ influences the algebra of pure diffeomorphisms.
We show now how this algebra changes and assume for simplicity again $f_-=0$. 
Note first that for the $\eps^-$ transformations, 
a compensating Weyl rescaling is required --- without it $f$ transforms by a function of $x^-$ violating the boundary conditions. 
For the $\eps^+$ transformations, using the same construction as with $b=0$ we obtain the following generators of diffeomorphisms:
\eq{
\vir_n = L_n + \frac{1}{4k}\,\sum_{m\in\mathbb{Z}}:\aff_m{} \aff_{n-m} : - \frac{ib}{2}\,n\,\aff_n
}{eq:gCS100}
which leads to the algebra
\begin{subequations}
\label{eq:algebrabneqzero}
\begin{align}
 [\vir_n,\,\vir_m] &= (n-m)\,\vir_{n+m} - \big(k-\frac{1+6kb^2}{12}\big)\,(n^3-n)\,\de_{n+m,0} \label{eq:CFTbneqzero}\, ,\\
 [\bar L_n,\,\bar L_m] &= (n-m)\,\bar L_{n+m} + k\,(n^3-n)\,\de_{n+m,0}\, ,\\
 [\aff_n,\,\aff_m] &= 2k\,n\,\de_{n+m,0} \, ,\\
 [\vir_n,\,\aff_m] &= -m\,\aff_{n+m} - ibk\,n^2\,\de_{n+m,0}\, . \label{eq:CFTbneqzeronew}
\end{align}
\end{subequations}
We note that a scalar field with background charge $Q$ with action
\eq{
S_Q = \frac{1}{4\pi\ell_s^2}\,\int\extd^2x\sqrt{g}\,g^{\al\be}\partial_\al X\partial_\be X + \frac{Q}{4\pi\ell_s}\,\int\extd^2x\sqrt{g}\,X R
}{eq:gCS102}
leads to the same algebraic structure (see e.g.~section 4.14 in \cite{Kiritsis:2007}): 
The holomorphic part of the stress tensor is
\eq{
T = -\frac{1}{\ell_s^2} : \partial X \partial X : + \frac{Q}{\ell_s}\,\partial^2 X
}{eq:gCS101}
whose Fourier-components coincide with the last two terms in \eqref{eq:gCS100} upon identifying $Q=\sqrt{k}b$ and $\ell_s^2= 4k$.
The central charge of the scalar field is the shifted by
\eq{
c = 1 + 6Q^2 = 1 + 6kb^2
}{eq:gCS103}
which coincides precisely with the shift of the left central charge in \eqref{eq:CFTbneqzero}.
Also the new central charge proportional to $ibn^2$ appearing in \eqref{eq:CFTbneqzeronew} arises in the corresponding commutator of the chiral scalar field current with the Virasoro generators, and is indeed proportional to $iQn^2$; 
(the proportionality constants depend on the normalizations for the currents; note also that we can introduce a term linear in $n$ in the central charge in \eqref{eq:CFTbneqzeronew} by a shift $\aff_0\to\aff_0+\rm const.$).
The appearance of this central charge implies an anomaly in the chiral conservation of the current.

Let us now turn to other choices of the constants $C_n$ in \eqref{eq:free13}.
For general $C_n$ the allowed Weyl rescalings have Fourier modes of the form $\Om = -e^{-inx^-} - C_{-n} e^{inx^+}$.
Computing the corresponding charge is straightforward and yields
\eq{
Q[\Om = -e^{-inx^-} - C_{-n} e^{inx^+}] = 2kin\,(|C_n|^2 - 1)f_{-}^{(n)}
}{eq:gCS200}
where we used $C_n = C_{-n}^*$ following from the reality of $f$. Note that the charge vanishes for $|C_n| = 1$.
One particularly interesting class of choices therefore is 
\eq{
C_n C_{-n} = |C_n|^2 = 1 \qquad \forall n
}{eq:restriction}
for which all Weyl charges vanish and the corresponding generators commute with each other. More generally it is possible to 
have a subset of the constants $C_n$ fulfilling $|C_n| = 1$. In this case only the corresponding charges are trivial. 

In the general case of nonzero charges we define the normalized generators
\eq{
\aff_n = \frac{1}{\sqrt{|C_n|^2 -1 }}\tilde{G}_W[\Om = -e^{-inx^-} - C_{-n} e^{inx^+}]\, .
}{eq:gCS222}
These fulfill the commutation relations \eqref{eq:gCS89}.
Note the curious fact that the charge corresponding to the normalized generators vanishes, but their commutator stays finite, if we take $|C_n| \to 1$.
It is evident from the definition \eqref{eq:gCS222} that this limit is singular.

After having understood the Weyl charges in the most general case, let us turn to the diffeomorphisms. Our boundary conditions now include
the relations \eqref{eq:free13}, so any diffeomorphism must respect them. The diffeomorphisms that are compensated by Weyl rescalings
are still allowed since these leave $f$ unaffected, but for the pure diffeomorphisms the relations \eqref{eq:free13} impose restrictions on the Fourier coefficients of the parameters as well as on the constants $C_n$. 
Expanding the latter as
\eq{
\eps^\pm = \sum_{m} \left(\eps_m^{(+)} e^{-im(t + \varphi)} + \eps_m^{(-)} e^{-im(t - \varphi)} \right)
}{eq:gCS224}
and assuming that the parameters $\exponent$ and $p_f$ vanish results in the condition
\eq{
C_n = d \, \al^n \quad \forall n \neq 0;  \qquad \eps^{(+)}_n = -\al^n \eps^{(-)}_n \quad  \forall n 
}{eq:gCS225}
for some constants $\al$ and $d$.
Thus only a specific combination of the generators of the $L$ and $\bar{L}$ algebras generate allowed diffeomorphism.
To derive this result one notes that the function $f$ transforms as
\eq{
\de f = \eps \cdot \partial f 
}{eq:gCS226}
by a pure diffeomorphism and requires that the relation \eqref{eq:free13} holds also with $f$ replaced by $\de f$.
The fact that $C_n = C_{-n}^*$ now implies that $d \in \mathbb{R}$ and $\al = e^{i\theta}$ for some $\theta \in \mathbb{R}$.
We see from \eqref{eq:free10} and \eqref{eq:free13} that a non-zero $\theta$ corresponds to sending $x^+ \to x^+ + \theta$. 

The linear combinations of diffeomorphism generators that are consistent with the boundary conditions are thus
\eq{
{\cal L}_n  \equiv \tilde{G}_P[\eps^- = -e^{-inx^-}; \,  \eps^+ = e^{in(x^+ - \theta)}]\, .
}{eq:gCS227}
The term bilinear in $f$ and $\de f$ in the corresponding charge becomes
\eq{
\de Q_P^{{\rm bilinear}} = k(1-d^2)\de\left( \sum_m m(n-m)f^{(m)}_- f^{(n-m)}_-   \right)\, .
}{eq:gCS228}
Expressing the Fourier components $f^{(m)}_- $ in terms of our normalized generators ${\cal J}_m$ we obtain
\eq{
\de Q_P^{{\rm bilinear}}  = \frac{1}{4k}\,\sum_{m\in\mathbb{Z}} :\aff_m \aff_{n-m}:
}{eq:gCS229}
where again normal ordering was introduced. Computing the full algebra gives again a Virasoro algebra with 
equal and opposite contributions to the central charge originating from the $L$ and $\bar{L}$ algebras. Thus the only remaining contribution
comes from the Sugawara sum.
\eq{
 [\vir_n,\,\vir_m] = (n-m)\,\vir_{n+m} + \frac{1}{12}\,(n^3-n)\,\de_{n+m,0} 
}{eq:gCS230}
Interestingly this result is independent of the quantity $d$. 
So even though the Weyl charges \eqref{eq:gCS200} vanish for the case $d=1$, and even though the bilinear term \eqref{eq:gCS228} vanishes for this case, the limit from $d\neq 1$ yields a contribution in the diffeomorphism algebra. 
We attribute this feature to the singularity of this limit.

This concludes our discussion of the extended symmetries of the CFTs dual to CSG in case~III. 
As a last piece of information regarding the CFT interpretation let us note that the 1-loop torus partition function of this theory has been 
calculated\footnote{An analytic continuation of CSG to Euclidean signature was used in that calculation.
Analytically continued Chern--Simons theories were discussed recently in Ref.~\cite{Witten:2010cx}.
Here we display only the 1-loop contribution to the partition function.
The classical partition function should be given by $Z_{\rm CSG}^{\rm 0-loop}= (q/\bar q)^{k/2}$, though it is not clear to us how to derive this result in the Euclidean path integral formulation for CSG.} with the result \cite{Bertin:2011jk}, 
\eq{
Z_{\rm CSG}^{\rm 1-loop}(q,\bar q)=\prod_{n=2}^\infty \frac{1}{|1-q^n|^2} \prod_{m = 0}^\infty \frac{1}{\big(1-q^{m+3/2}\bar q^{-1/2}\big)\big(1-q^{m+3/2}\bar q^{1/2}\big)}\,.
}{eq:gCS91}
Here $q=e^{i\tau}$ depends on the modular parameter $\tau=\vartheta+i\beta$ of the boundary torus, which in turn depends on angular potential $\vartheta$ and inverse temperature $\beta$. 
The first factor is the partition function of the Virasoro vacuum representation, which arises already in Einstein gravity.
The remaining factor describes the partially massless excitations/Weyl gravitons, whose spectrum is depicted in Fig.~\ref{fig:1} on p.~\pageref{fig:1}.

It is an interesting feature of the partition function \eqref{eq:gCS91} that it correctly takes into account null states in the graviton spectrum.
As pointed out in Ref.~\cite{Castro:2011ui}, for many gravity models the appearance of null states is a quantum effect recognizable
in general only when the Dirac brackets are promoted to quantum operators, and then through the Klein--Gordon norm expressed in terms
of boundary charges. The path integral techniques used for computing 1-loop partition functions might miss this feature. 

For CSG there is however a null state that is captured by the 1-loop partition function \eqref{eq:gCS91}.
This null state, which appears at level two as a descendant of a primary state of weight $h=-1/2$, is present semi-classically, i.e., for arbitrary large values of the central charge.
This explains why it can be spotted by semi-classical methods, like the heat kernel evaluation of 1-loop determinants extracted from the saddle-point approximation in the Euclidean path integral.

We address now the issue when semi-classical null states can exist in TMG with asymptotic AdS boundary conditions, assuming with no loss of generality $\mu\geq 0$ and $\ell=1$.
We focus on the right-moving sector.
The corresponding weights of normalizable massive primaries in TMG are given by \cite{Li:2008dq}
\eq{
\bar h = -\frac{1}{2} + \frac{\mu}{2} \geq -\frac{1}{2}\,.
}{eq:gCS237}
The inequality in \eqref{eq:gCS237} is saturated only for CSG.
On the CFT side there is a null state at level $n$ in the Verma module of a primary state of weight $\bar h$ if \cite{diFrancesco}
\eq{
\bar h = \bar h_{p,q} = \frac{\big((m+1)p - mq \big)^2 - 1}{4m(m+1)}\,, \qquad m = -\frac{1}{2} \pm  \frac{1}{2} \sqrt{ \frac{c-25}{c-1}}
}{eq:gCS231}
for some positive integers $p,q$ with $pq \leq n$. If the central charge $c$ is large, expanding in its inverse yields
\eq{
\bar h = \bar h_{p,q} = \frac{c\,(1-p^2)}{24} + \frac{(p^2-1)}{24} + \frac{p(p-q)}{2} +
\cO(1/c)\,.
}{eq:gCS232}
In order to have a null state in the Verma module as $c \to \infty$ with finite weight $|\bar h|<\infty$ we must have $p=1$. 
Then the possible weights are
\eq{
\bar h = \frac{1-q}{2} \leq 0\, .
}{eq:gCS233} 
The case $q=1$, $\bar h=0$ is trivial and just captures the fact that the $\bar L_{-1}$ descendant of the SL$(2, \mathbb{R})\times$SL$(2, \mathbb{R})$ invariant vacuum vanishes.
The only non-trivial case possible in TMG is $q=2$ and $\bar h=-1/2$, since otherwise the two inequalities \eqref{eq:gCS237}, \eqref{eq:gCS233} are incompatible with each other.
Indeed, as seen in Fig.~\ref{fig:1} on p.~\pageref{fig:1} there is a pure gauge state that is also a primary
appearing as a descendant of the $\bar{h} = -1/2$ primary. This is the null state bound to appear at this level. 
Since this state is pure gauge with respect to the gauge transformations that are
fixed by corresponding ghost determinants in the path integral method, this state and its descendants are divided out correctly in this approach \cite{Bertin:2011jk}.

In conclusion, the semi-classical methods used to derive the partition function \eqref{eq:gCS91} appear to work reliably and produce the correct null state expected from CFT considerations.
We have thus a reasonable matching between properties on the gravity and the CFT side.

\section{Discussion}\label{se:6}

In this work we performed a holographic analysis of conformal Chern--Simons gravity with the total action
\eq{
S_{\textrm{CSG}}[g]=\frac{k}{4\pi}\, \int\!\extd^3x\, \epsilon^{\la\mu\nu}\,\Ga^\si{}_{\la\rho}\,\Big(\partial_\mu\Ga^\rho{}_{\nu\si}+\tfrac23\,\Ga^\rho{}_{\mu\tau}\Ga^\tau{}_{\nu\si}\Big) 
+ \frac{k}{4\pi}\,\int_{\partial M}\!\!\!\extd^2x\sqrt{-\ga}\,\Big(K^{\al\be}K_{\al\be} - \frac{1}{2}K^2 \Big) \,.
}{eq:concl}

Our boundary conditions in section \ref{se:leah} can be made either weaker or stronger, depending on the desired applications.
For instance, to accommodate the solitonic background \eqref{eq:soliton} we have to include a suitable $1/y$-contribution corresponding to the non-normalizable Weyl graviton zero-mode. 
In NMG these solitons have the same energy as the AdS ``groundstate'' \cite{Perez:2011qp}.
It would be of interest to check a corresponding statement for CSG using the result for the canonical diffeomorphism charges \eqref{eq:diffcharge}. 
Imposing instead Brown--Henneaux boundary conditions eliminates the Weyl gravitons and thus leads to a theory where all linearized excitations have positive conformal weights.
This theory thus has a chance of being dual to a unitary CFT, at least for certain values of the level $k$.
In fact, the situation could be similar to Einstein gravity upon redefining the Virasoro generators $L_n\to-L_{-n}$, which makes both central charges positive and equal in magnitude. 
We also mention that other classes of boundary conditions are possible, where the asymptotic line-element is not conformally related to AdS.
Such boundary conditions can then accommodate e.g.~warped AdS solutions \eqref{eq:eom2}, where the asymptotically leading term in the metric is not invertible, but has a 1-dimensional kernel.

The holographic discussion in section \ref{se:3} showed consistency with standard AdS/CFT considerations and did not involve extra-ordinary features, except for the interpretation of the Weyl gravitons.
Namely, even though the Weyl gravitons are descendants of the vacuum in the sense that they are generated by large gauge transformations as in \eqref{eq:gCS36}, these transformations are not part of our asymptotic symmetry group. 
In fact, they generate non-normalizable modes when acting on general states that satisfy our boundary conditions. 
This is a remarkable property of CSG that deserves further study.

We encountered also several non-standard features in section \ref{se:genhol}.
Most of them can be traced back to the fact that CSG allows for geometries that are not diffeomorphic to each other, but nevertheless are gauge-equivalent.
This property is also encountered in higher-spin gravity theories (see e.g.~\cite{Ammon:2011nk,Castro:2011fm} and Refs.~therein), so in that sense CSG might serve as a toy model for higher-spin gravity.
In the present case, however, the enhanced gauge symmetry has nothing to do with higher spins but is just Weyl invariance.
It should be possible to study a higher-spin generalization of CSG along the lines of the present work, essentially by taking the scaling limit \eqref{eq:gCS9} of the higher-spin generalization of TMG \cite{Chen:2011vp,Bagchi:2011vr,Bagchi:2011td, Chen:2011xx}.

An interesting outcome is the emergence of a scalar field in the dual CFT, which is not restricted at all by the bulk equations of motion.
Thus, the dynamics of this scalar field in the CFT is determined solely by boundary- and consistency conditions.
While in the asymptotic AdS scenario studied in section IV the central charge arises through classical contributions (boundary terms) on the gravity side, the presence of the boundary scalar field leads to the expected quantum shift of the central charge by one unit.
The total central charge in the Virasoro algebra \eqref{eq:CFTbneqzero} then consists of classical {\em and} quantum contributions. Even though these contributions cancel if the Chern--Simons level takes a certain value, this does not eliminate the corresponding charges. Therefore, the
Virasoro generators remain non-trivial also for this tuning.
Another striking property of CSG is the possibility to have a CFT dual in a non-asymptotically-AdS setup.
We have shown that the scalar field then acquires a background charge and the generators of diffeomorphisms experience a twisted Sugawara-shift \eqref{eq:gCS100}.

Similar features to those described above are likely to occur in any
Weyl invariant theory of gravity. Therefore,
one line of future research would be to extend the present analysis to
Weyl invariant gravity theories including matter.
A particularly interesting class of such theories have recently been
constructed \cite{Chu:2009gi, Chu:2010fk}, and correspond
to the ABJM \cite{Aharony:2008ug} models coupled to conformal supergravity.

Comparing semi-classical partition functions calculated on the gravity side with CFT partition functions requires the applicability of the semi-classical approximation, since otherwise null states that exist on the CFT side may not be spotted on the gravity side \cite{Castro:2011ui}.
We have shown that CSG is unique insofar as null states exist even semi-classically.
If these null states persist on the CFT side regardless of the value of the Chern--Simons level $k$, then the weights of the primary corresponding to Weyl gravitons must receive quantum corrections that can be deduced from the relation \eqref{eq:gCS231}.
For small (fractional) levels $k$ it may be that additional null states exist that are not taken into account in the 1-loop partition function \eqref{eq:gCS91}.
It is possible that one may identify specific unitary CFT duals for some isolated values of $k$ where such additional null states arise.

Finally, it would be of interest to generalize our results, where applicable, to supersymmetric CSG \cite{Deser:1982sv,Deser:1983sw, vanNieuwenhuizen:1985cx} and to 4-dimensional conformal gravity, see \cite{Mannheim:1988dj,Lu:2011zk,'tHooft:2011we,Maldacena:2011mk} and references therein.

\acknowledgments

We thank Glenn Barnich, Alejandra Castro, Stanley Deser, Matthias Gaberdiel, Michael Gary, Per Kraus, Roman Jackiw, Alex Maloney, Miguel Paulos, Achilleas Porfyriadis, Rady Rashkov, Simon Ross, Andy Strominger, Stefan Theisen, Ricardo Troncoso, Peter van Nieuwenhuizen, Dima Vassilevich, Frank Wilczek, Edward Witten and Thomas Zojer for discussions. HA thanks Hessamaddin Arfaei for his support and encouragement as well as ITP members and secretaries in Vienna for their hospitality.

HA is supported by the Ministry of Science, Research and Technology in
Iran and project P22000-N16 of the Austrian Science Fund (FWF). BC is supported by the Serbian Science Foundation, Grant No.~171031.
SE, DG and NJ are supported by the START project Y435-N16 of the FWF
and by the FWF project P21927-N16. HA and BC acknowledge financial support from the START project
Y435-N16 of the FWF. NJ acknowledges financial support from the Erwin-Schr\"odinger
Institute (ESI) during the workshop ``Gravity in three dimensions''.

\begin{appendix}

\section{Canonical analysis in generalized massive gravity}\label{app:B}

\subsection{Action and field equations}

In this appendix we examine the canonical structure of GMG using the first order formulation analog to section \ref{se:2}.
Basic dynamical variables are triad fields $e^i{_\mu}$, Lorentz connection $\om^i{_\mu}$ and
Lagrange multipliers $\la^i{_\mu}$ and $f^i{_\mu}$ \cite{Blagojevic:2010ir}.
The action of GMG is given by $S_{\rm GMG}=1/\ka^2\,\int\extd^3 x\, {\cal L}_{\rm GMG}$, where $\ka^2=16\pi G_N$.
The second order Lagrange density can be read off from \eqref{eq:gCS1} with \eqref{eq:gCS4}.
In first order form it reads
\begin{subequations}
\begin{align}\label{eq:appb1}
{\cal L}_{\rm GMG}&=\epsilon^{\mu\nu\rho}\left(\si e^i{_\mu}R_{i\nu\rho}
    -\frac{1}{3}{\mit\La}_0\eps^{ijk}e_{i\mu}e_{j\nu}e_{k\rho}\right)\nonumber \\&+
    \frac 1\mu\,{\cal L}
    +\frac{1}{m^2}\,{\cal L}_K\, ,
\end{align}
where $\si=\pm 1$ and the first order gravitational Chern--Simons Lagrange density ${\cal L}$ is defined in \eqref{2.1}.
The term ${\cal L}_K$ is conveniently represented in the first order formalism as \cite{Blagojevic:2010ir}
\eq{
{\cal L}_K=\frac{1}{2}\epsilon^{\mu\nu\rho} f^i{_\mu}R_{i\nu\rho}-e\,{\cal V}_K\, ,\qquad {\cal V}_K:=\frac{1}{4}( f_{ik} f^{ik}-f^2)\,,
}{eq:appb2}
\end{subequations}
where $f:=f^i{_i}$.
Variation with respect to the fields yields constraints and equations of motion.
The former can be solved algebraically for the Lagrange multipliers $f^i{_\mu}$  and $\la^i{_\mu}$:
\begin{subequations}
\eq{
f_{ij}=2L_{ij}\,,\qquad \la_{ij}=\frac{2}{m^2}C_{ij}+\frac{2}{\mu}L_{ij}\,.
}{eq:appb3}
The equations of motion of GMG take essentially the same form as the field equations in the metric formulation \cite{Bergshoeff:2009hq,Bergshoeff:2009aq},
\eq{
G_{ij}-{\mit\La}_0\eta_{ij}+\frac 1{\mu}C_{ij}-\frac 1{2m^2}K_{ij}=0\,.
}{eq:appb4}
The anholonomic Cotton tensor $C_{ij}$ is defined below \eqref{2.3a}, $K_{ij}:={\cal T}_{ij}-2\eps^{mn}{}_j\covD_m C_{in}$ and ${\cal T}_i{}^\mu:=e_i{}^\mu{\cal V}_K-\frac12 (f_{ik}f^{k\mu}-ff_i{}^\mu)$.
\end{subequations}

\subsection{Hamiltonian and constraints}

\paragraph{Primary constraints.}
From the definition of the canonical momenta $(\pi_i{^\mu},$ $\Pi_i{^\mu},$ $p_i{^\mu},$ $P_i{^\mu})$, conjugate to the Lagrangian variables ($e^i{_\mu}, \om^i{_\mu},\la^i{_\mu},f^i{_\mu})$,
respectively, we obtain the primary constraints:
\begin{subequations}
\begin{align}
& \phi_i{^0} :=\pi_i{^0}\approx 0 \, ,  & 
& \phi_i{^\al} :=\pi_i{^\al}-\epsilon^{0\al\be}\la_{i\be}\approx 0\, , \\  
& \Phi_i{^0} :=\Pi_i{^0}\approx 0\, ,   & 
& \Phi_i{^\al} :=\Pi_i{^\al}-\epsilon^{0\al\be}\left(
  \si e_{i\be}+\frac{1}{m^2} f_{i\be}+\frac{1}{\mu}\om_{i\be}\right)\approx 0\,,  \\
& p_i{^\mu} \approx 0\, , && P_i{^\mu}\approx 0\, . \label{A.2}
\end{align}
\end{subequations}
From the Poisson bracket algebra of the primary constraints it follows that the constraints $(\phi_i{^\al},\Phi_i{^\al},p_i{^\al},P_i{^\al})$ are second class.
Thus, we eliminate the momenta $(\pi_i{^\al},\Pi_i{^\al},p_i{^\al},P_i{^\al})$ and construct the reduced phase space $R_1$, in which the basic nontrivial Dirac brackets take the following form ($\de:=\de^{(2)}(x-x^\prime)$):
\begin{subequations}
\begin{align}
&\{e^i{_\al},\la^j{_\be}\}^*_1=\eta^{ij}\epsilon_{0\al\be}\de\, ,  &  
&\{\om^i{_\al},f^j{_\be}\}^*_1=m^2\eta^{ij}\epsilon_{0\al\be}\de\,, & \\
&\{\la^i{_\al},f^j{_\be}\}^*_1=-2m^2\si\eta^{ij}\epsilon_{0\al\be}\de\,, & 
&\{f^i{_\al},f^j{_\be}\}^*_1=-\frac{2m^4}{\mu}\epsilon_{0\al\be}\de\,. &
\end{align}
\end{subequations}
The remaining Dirac brackets are the same as the corresponding Poisson brackets.

After noting that the term $e\,{\cal V}_K$ is \emph{bilinear} in the variables
$e^i{_0}$ and $f^i{_0}$, one can conveniently represent the canonical
Hamiltonian as
\begin{subequations}
\label{eq:appb5}
\eq{
\cH_c= e^i{}_0\cH_i+\om^i{}_0{\cal K}_i+f^i{_0}{\cal R}_i+\la^i{_0}\cT_i
         +\frac{1}{m^2}\,e\,{\cal V}_K+\partial_\al D^\al\, ,
}{eq:gCS83}
where
\begin{align}
\cH_i &=-\epsilon^{0\al\be}\left(\si R_{i\al\be}
        -{\La_0}\eps_{ijk}e^j{}_\al e^k{}_\be+{\cal D}_\al\la_{i\be}\right)\,,\\
{\cal K}_i &=-\epsilon^{0\al\be}\Big(\si T_{i\al\be}+\frac 1\mu R_{i\al\be}+\frac{1}{m^2}{\cal D}_\al f_{i\be} 
        +\eps_{ijk}e^j{}_\al \la^k{}_\be\Big) \, ,           \\
{\cal R}_i &=-\frac{1}{2m^2}\epsilon^{0\al\be}R_{i\al\be}\, ,            \\
\cT_i &=-\frac{1}{2}\epsilon^{0\al\be}T_{i\al\be}\,,                \\
D^\al&=\epsilon^{0\al\be}\left[ \om^i{}_0\left( \si e_{i\be}
        +\frac{1}{m^2}f_{i\be}\right)+e^i{}_0 \la_{i\be}\right]\,.
\end{align}
\end{subequations}

\paragraph{Secondary constraints.}
Going over to the total Hamiltonian,
\begin{subequations}
 \label{eq:appb6}
\eq{
\cH_T=\cH_c +u^i{}_0\phi_i{}^0+v^i{}_0\Phi_i{}^0
            +w^i{}_0 p_i{^0}+z^i{_0}P_i{^0}\, ,
}{eq:gCS84}
where $(u^i{_0},v^i{_0},w^i{_0},z^i{_0})$ are canonical
multipliers, we find that the consistency conditions of the primary
constraints $\pi_i{}^0$, $\Pi_i{}^0$, $p_i{}^0$ and $P_i{^0}$ yield the
secondary constraints:
\begin{align}
{\hat{\cal H}}_i &:=\cH_i+\frac{1}{m^2}e\left(e_i{^0}{\cal V}_K-\frac 12 f_{ik}(f^{k0}-fe^{k0})\right)\approx 0\, ,     \\
{\cal K}_i &\approx 0\, ,                                     \\
{\hat{\cal R}}_i &:={\cal R}_i+\frac{1}{2m^2}e(f_i{^0}-f e_i{^0})\approx 0\,,\\
\cT_i &\approx 0\, .
\end{align}
\end{subequations}
Let us note that  the canonical Hamiltonian can be rewritten in the form
\eq{
\cH_c= e^i{}_0{\hat{\cal H}}_i+\om^i{}_0{\cal K}_i+f^i{_0}{\hat{\cal R}}_i
      +\la^i{_0}\cT_i+\partial_\al D^\al\, .
}{A.5}

\paragraph{Ternary constraints.}
The consistency conditions of the secondary constraints lead to ternary ones:
\begin{subequations}
\begin{align}
&\theta_{0\be}:=f_{0\be}-f_{\be0}\approx 0\,, &\theta_{\al\be}:=f_{\al\be}-f_{\be\al}\approx 0\,, & \\
&\psi_{0\be}:=\la_{0\be}-\la_{\be0}\approx 0\,,&\psi_{\al\be}:=\la_{\al\be}-\la_{\be\al}\approx 0\,. &
\end{align}
\end{subequations}
In order to interpret the consistency conditions for $\theta_{0\be}$ and $\psi_{0\be}$ we introduce a change of variables in the total Hamiltonian  $\cH_T$:
\begin{align}\label{eq:gCS85}
\pi_i{^0}{}' &:=\pi_i{^0}+f_i{^k}P_k{^0}+\la_i{^k}p_k{^0}\, , \nonumber \\  
z^i{_0}{}' &:=z^i{_0}-f^i{_k}u^k{_0}\, ,\nonumber \\
w^i{_0}{}' &=w^i{_0}-\la^i{_k}u^k{_0}\,,
\end{align}
whereupon the $(\pi_i{^0},P_i{^0},p_i{^0})$ piece of the total Hamiltonian $\cH_T$ takes the form
\begin{eqnarray}
u^i{_0}\pi_i{^0}+z^i{_0}P_i{^0}+w^i{_0}p_i{^0}&=&u^i{_0}\pi_i{^0}{}'
  +z^i{_0}{}'P_i{^0}+w^i{_0}{}'p_i{^0}\, .                                  \nonumber
\end{eqnarray}
Consistency conditions of $\theta_{0\be}$ and $\psi_{0\be}$ lead to the determination
of the multipliers $z{}'_{\be0}$ and $w{}'_{\be0}$,
\begin{eqnarray}
\bar z{}'_{\be0}=e^i{_0}\bar z{}'_{i\be}\,,\qquad \bar w{}'_{\be0}=e^i{_0}\bar w{}'_{i\be}\,,
\end{eqnarray}
where  ${\bar z}'_{i\al}=\dot f_{i\al}-f_{ik}\dot e^k{_\al}$, ${\bar w}'_{i\al}=\dot \la_{i\al}-\la_{ik}\dot e^k{_\al}$ and $\dot\phi=\{\phi,H_T\}^*_1$.

\paragraph{Quaternary constraints.}
The consistency conditions of $\theta_{\al\be}$ and $\psi_{\al\be}$ lead to quaternary constraints:
\begin{subequations}
\begin{eqnarray}
&&\chi:=\la-\frac{f}{\mu}\approx 0\, ,\\
&&\varphi:=f+3{\mit\La}_0+\frac{1}{2m^2}{\cal V}_K\approx 0\,.
\end{eqnarray}
 \end{subequations}

\paragraph{End of the consistency procedure.}
The consistency condition of the quaternary constraints are given by:
\begin{subequations}
\begin{align}
\{\chi,H_T\}_1^* &= w'^\mu{_\mu}-\frac 1\mu z{}'^\mu{_\mu} \approx 0\,, \nonumber\\
\{\varphi,H_T\}_1^* &=\ECP^{\mu\nu}z'_{\mu\nu} \approx 0\, ,  \nonumber \\
\ECP^{\mu\nu} &:=\si g^{\mu\nu} +\frac{1}{4m^2}\left(f^{\mu\nu}-f g^{\mu\nu}\right)\, .
\end{align}
\end{subequations}
These relations determine the multipliers $z'_{00}$, $w'_{00}$ provided the coefficient $\ECP^{00}$ does not vanish.
We focus first on the case $\ECP^{00}\neq 0$.

\subsection{Classification of constraints}

Among the primary constraints those that appear in $\cH_T$ with arbitrary multipliers,
$\pi_i{^0}{}'$ and $\Pi_i{^0}$, are first class, while the remaining ones, $p_i{^0}$ and $P_i{^0}$, are second class.
Going to the secondary constraints, we use the same theorem as in the maintext below \eqref{eq:gCS19}, which implies that the secondary constraints ${\bar {\cal H}}_i:=-\{\pi_i{^0}{}',H_T\}^*_1$ and ${\bar {\cal K}}_i=-\{\Pi_i{^0},H_T\}^*_1$ are first class.
All other constraints turn out to be second class if $\ECP^{00}\neq 0$.
The complete classification of the constraints in the reduced space $R_1$ is given as in table \ref{tab:app}.

\begin{table}[h]
\begin{center}
\doublerulesep 1.8pt
\begin{tabular}{||l|l|l||}
                                                      \hline\hline
\rule{0pt}{12pt}
&~First class \phantom{x}&~Second class \phantom{x} \\
                                                      \hline
\rule[-1pt]{0pt}{15pt}
\phantom{x}Primary &~$\pi_i{^0}{}',\Pi_i{^0}$
            &~$p_i{^0},P_i{^0}$   \\
                                                      \hline
\rule[-1pt]{0pt}{15pt}
\phantom{x}Secondary\phantom{x} &~${\bar {\cal H}}_i,{\bar {\cal K}}_i$
           &~$\cT_i,{\hat{\cal R}}'_i$       \\
                                                      \hline
\rule[-1pt]{0pt}{15pt}
\phantom{x}Ternary\phantom{x}
                  & &~$\theta_{0\be},\theta_{\al\be},\psi_{0\be},\psi_{\al\be}$ \\
                                                      \hline
\rule[-1pt]{0pt}{15pt}
\phantom{x}Quaternary\phantom{x}
                 & &~$\chi,\varphi$  \\
                                                      \hline\hline
\end{tabular}
\end{center}
\caption{Classification of constraints in $R_1$ for $\ECP_{00}\neq 0$}
\label{tab:app}
\end{table}
\noindent Here, ${\hat{\cal R}}'_i$ is a suitable modification of ${\hat{\cal R}}_i$, defined so that
it does not contain $f_{i0}$:
$$
{\hat{\cal R}}'_i={\cal R}_i+\frac{e}{2m^2}\left[(g^{00}e_i{^\al}
  -g^{0\al}e_i{^0})f_{0\al}+g^{0\al}f_{i\al}-e_i{^0}f^\al{_\al}\right]\,.
$$

In order to prove the content of table \ref{tab:app}, we need to verify the second-class nature of the constraints in the last column.
This can be done similar to \cite{Blagojevic:2010ir} by calculating the determinant of their Dirac brackets.
According to the results in table \ref{tab:app}, we have a 48-dimensional phase space with 12 first class and 20 second class constraints.
Consequently, the theory for $\ECP_{00}\neq 0$ exhibits two local Lagrangian degrees of freedom, namely the massive bulk gravitons.

\subsection{Extra gauge symmetry}

For maximally symmetric solutions the  condition $\ECP_{\mu\nu}=0$ is equivalent to the choice of parameters ${\La_0}/m^2=-1$ \cite{Blagojevic:2010ir} that corresponds to the partially massless case.
At this point the constraint structure of GMG linearized around a maximally symmetric solution changes.
Namely, the constraint $\tilde P^{00}+\tilde p^{00}/\mu$ becomes first class similar to \cite{Blagojevic:2011qc}, which leads to an additional gauge symmetry.
In the second order formulation it corresponds to linearized Weyl rescaling of the metric [see also \eqref{eq:gCS80}]:
\begin{subequations}
\begin{eqnarray}
\de_W \tilde e^i{_\mu}&=&\Om\bar e^i{_\mu}\, ,                    \\
\de_W\tilde\om^i{_\mu}&=&
  \eps^{ijk}\bar e_{j\mu}\bar e_k{^\nu}\partial_\nu\Om\,,  \\
\de_W\tilde f^i{_\mu}&=&-2\bar{\cal D}_\mu(\bar e^{i\nu}\partial_\nu\Om)
                 -2m^2\si\Om\bar e^i{_\mu}\,,              \\
\de_W\tilde\la^i{_\mu}&=&\frac 1{\mu}\de_W\tilde f^i{_\mu}\, .
\end{eqnarray}
 \end{subequations}
Background variables are decorated by bars and small variations by tildes.
However, it is important to realize that this extra gauge symmetry does {\em not} persist in the full non-linear theory, except in a limit where CSG is recovered.
To verify this statement it is sufficient to find a specific background where the extra gauge symmetry is absent at the linearized level.
Such a background is provided e.g.~by warped AdS \cite{Tonni:2010gb}, which is not maximally symmetric and therefore has $\ECP\neq 0$.
It is possible to partially gauge-fix the Poincar{\'e} symmetries such that the component $\ECP_{00}\neq 0$.
Then, according to the general analysis above, no extra gauge symmetry exists besides residual Poincar{\'e} transformations.
Therefore, GMG does not exhibit an extra gauge symmetry beyond the linearisation around AdS, whereas CSG does have such an extra gauge symmetry \eqref{eq:Weyl}.

\section{Dirac bracket algebra of constraints in reduced phase space}\label{app:A}

The phase space $R_1$ is defined by imposing the second class constraints $\chi_I:=(\phi_i{^\al},\Phi_i{^\al},p_i{^\al})$ on the original phase space [see below \eqref{eq:gCS12}].
To construct the corresponding Dirac brackets, we consider the $18\times 18$ matrix $\De_1$ with matrix elements
\begin{subequations}
\begin{equation}\label{eq:app1}
\De_1(x,x')_{IJ}=\{\chi_I,\,\chi_J\}= \left( \begin{array}{ccc}
    \{\phi_i{^\al}(x),\,\phi_j{^\be}(x')\} & \{\phi_i{^\al}(x),\,\Phi_j{^\be}(x')\}
                               &\{\phi_i{^\al}(x),\,p_j{^\be}(x')\}   \\
    \{\Phi_i{^\al}(x),\,\phi_j{^\be}(x')\} & \{\Phi_i{^\al}(x),\,\Phi_j{^\be}(x')\}
                               &\{\Phi_i{^\al}(x),\,p_j{^\be}(x')\}   \\
    \{p_i{^\al}(x),\,\phi_j{^\be}(x')\} & \{p_i{^\al}(x),\,\Phi_j{^\be}(x')\}
                            & \{p_i{^\al}(x),\,p_j{^\be}(x')\}
           \end{array}
     \right)\,.
\end{equation}
The explicit form of $\De_1$ reads
\eq{
\De_1(x,x')=
     \left( \begin{array}{ccc}
                0 &  0  & -1 \\
                0 & -2  &  0 \\
               -1 &  0  &  0
               \end{array}
     \right)\otimes\epsilon^{0\al\be}\eta_{ij}\de^{(2)}(x-x')\,.
}{eq:app2}
The matrix $\De_1$ is regular, and its inverse has the form
\eq{
\De_1^{-1}(x,x')=
     \left( \begin{array}{ccc}
                       0 & 0             & 1 \\
                       0 &\frac{1}{2}& 0     \\
                       1 & 0             & 0
            \end{array}
     \right)\otimes\epsilon_{0\al\be}\eta^{ij}\de^{(2)}(x-x')\,.
}{eq:app3}
 \end{subequations}
The matrix $\De_1^{-1}$ defines the Dirac brackets in (partially) reduced phase space $R_1$:
\begin{equation}\label{eq:app4}
\{\phi(x),\,\psi(x')\}^*_1 =\{\phi(x),\,\psi(x')\} - \iint\limits_{y\,y'}\{\phi(x),\,\chi_I(y)\}(\De^{-1})^{IJ}(y,y')\{\chi_J(y'),\,\psi(x')\}\,.
\end{equation}
The explicit form of all non-trivial Dirac brackets is displayed in \eqref{3.3}.

The further reduced phase space $R_2$ is defined by the additional constraints $\zeta_I := (\ternary^y\,,\ternary^\varphi\,,p^{y 0}\,,p^{\varphi 0})$.
The $4\times 4$ matrix $\Delta_2:=\{\zeta_I,\,\zeta_J\}^*_1$ and its inverse respectively are given by
\begin{subequations}
\eq{
\De_2(x,x')=\left(\begin{array}{cccc}
0 & -2\la_{00} & 0 & 1 \\
2\la_{00} & 0 & -1 & 0 \\
0 & 1 & 0 & 0 \\
-1 & 0 & 0 & 0
\end{array}\right)\otimes \de^{(2)}(x-x')
}{eq:app5}
and
\eq{
\De_2^{-1}(x,x')=\left(\begin{array}{cccc}
0 & 0 & 0 & -1 \\
0 & 0 & 1 & 0 \\
0 & -1 & 0 & -2\la_{00} \\
1 & 0 & 2\la_{00} & 0
\end{array}\right)\otimes \de^{(2)}(x-x')\,.
}{eq:app6}
\end{subequations}
Consequently Dirac brackets in $R_2$ retain the same form as in $R_1$.
In $R_2$ the following relations are valid ($\de:=\de^{(2)}(x-x^\prime)$):
\begin{subequations}
\label{eq:gCS21}
\begin{align}
\{\pi^i{_0}(x),\la_{j0}(x')\}^*_2 &=-e_j{^\be}\la^i{}_{\be}\de \, ,  \\ 
\{e^i{_\al}(x),\la_{j0}(x')\}^*_2 &=-e^i{_0}e_j{^\be}\epsilon_{0\al\be}\de\,.
\end{align}
\end{subequations}

We display the most important Dirac brackets that facilitate the evaluation of the consistency requirements.
Dirac brackets between the secondary constraints read:
\begin{subequations}
\begin{align}
\{{\cal H}_i,\,{\cal H}_j\}^*_1&=\frac{1}{2}\epsilon^{0\al\be}\la_{i\al}\la_{j\be}\de \,,\\
\{{\cal H}_i,\,{\cal K}_j\}^*_1&=-\eps_{ijk}{\cal H}^k\de \,,\\
\{{\cal H}_i,\,{\cal T}_j\}^*_1 &= -\frac{1}{2}\eps_{ijk}{\cal K}^k-\frac12\epsilon^{0\al\be}
           (\eta_{ij}\la_{\al\be}+\la_{i\al}e_{j\be})\de \,,\\
\{{\cal K}_i,\,{\cal K}_j\}^*_1 &= -\eps_{ijk}{\cal K}^k\de \,,\\
    \{{\cal K}_i,\,{\cal T}_j\}^*_1 &= -\eps_{ijk}{\cal T}^k\de\,,\\
\{{\cal T}_i,\,{\cal T}_j\}^*_1 &= \frac{1}{2}\epsilon^{0\al\be}e_{i\al}e_{j\be}\de\,.
\end{align}
 \end{subequations}
The Dirac brackets between $\ternary^0$ and the secondary constraints read
\begin{subequations}
\begin{align}
&\{\ternary^0,\,{\cal H}_i\}^*_1={\cal H}_i\de \,,\\
&\{\ternary^0,\,{\cal K}_i\}^*_1=0 \,,\\
&\{\ternary^0,\,{\cal T}_i\}^*_1={\cal T}_i\de \,.
\end{align}
\end{subequations}

\section{Uniqueness of boundary conditions}\label{se:3.6}

We pose now the question to what extent the boundary conditions \eqref{eq:bc3} are unique.
For a start, we replace them by a looser set of conditions:
\eq{
\left(\begin{array}{lll}
h_{++}={\cal O}(1/y) & h_{+-} = {\cal O}(1/y) & h_{+y} = {\cal O}(1/y) \\
              & h_{--} = {\cal O}(1/y)        & h_{-y} = {\cal O}(1/y) \\
              &                               & h_{yy} = {\cal O}(1/y)
\end{array}\right)\,.
}{eq:bc4a}
To reduce clutter we consider only case~I explicitly.
The boundary conditions \eqref{eq:bc4a} are preserved by diffeomorphisms generated by a vector field $\xi$ satisfying
\begin{subequations}
 \label{eq:gCS32}
\begin{align}
 \xi^\pm &= \eps^\pm(x^\pm) + {\cal O}(y) \,, \\
 \xi^y &= \frac y2\,\big(\partial_+\eps^+ + \partial_-\eps^-\big) + {\cal O}(y^2) \,.
\end{align}
\end{subequations}
Additionally, we can allow for asymptotic Weyl rescalings \eqref{eq:Weyl} with
\eq{
\Om = {\cal O}(y) \,.
}{eq:gCS33}
If we relaxed the boundary conditions even further by replacing ${\cal O}(1/y)$ in \eqref{eq:bc4a} by ${\cal O}(1/y^\be)$, with $\be > 1$, then the Brown--York stress tensor becomes infinite.
We have not found any reasonable counterterms that would cancel these divergences, nor have we found any classical solutions that would require such boundary conditions.
Thus, it appears that the boundary conditions \eqref{eq:bc4a} cannot be made any looser.
We show now that the boundary conditions \eqref{eq:bc4a} actually are too loose, because they lead to incompatibilities with the following consistency requirements:
\begin{enumerate}
 \item The boundary condition preserving transformations \eqref{eq:gCS32}, \eqref{eq:gCS33} should leave the full action invariant, including boundary terms.
This is a stricter requirement than the one imposed by Porfyriadis and Wilczek in the context of Einstein gravity, who require finiteness of the first and second variation of the action \cite{Porfyriadis:2010vg}.
 \item For a flat boundary metric the Brown--York stress tensor should be conserved, as we expect no gravitational anomaly for this case.
 \item Since $c_L+c_R=0$ there is no trace anomaly and thus the Brown--York stress tensor must be traceless.
\end{enumerate}
We focus now on the first requirement.
The first variation of the action \eqref{eq:gCS67} under a diffeomorphism generated by a vector field $\xi$ is given by a boundary term:
\begin{equation}\label{1stbCSxi}
\de S_{\rm CSG}[\xi]=-\frac{k}{\pi}\,\int_{\partial M}\!\!\!\!\extd^2x\sqrt{-\gamma}\,\Big[n_{\mu} C^{\mu\nu}\xi_{\nu} + S^{\mu\nu}\nabla_\mu\xi_\nu + S^{\mu\nu}{}_\rho \big(\nabla_\nu\nabla_\mu\xi^\rho - R^\rho{}_{\mu\nu\si}\xi^\si\big)\Big] \stackrel{!}{=} 0 \,.
\end{equation}
Here $n^\mu$ is the unit normal vector, $S^{\mu\nu}$ is a symmetric traceless first order differential tensor operator and $S^{\mu\nu}{}_\rho$ is a non-tensorial quantity.
The first term vanishes on-shell and the remaining terms vanish if $\xi$ is a 
Killing vector.
However, we are also interested in situations where the metric is on-shell only asymptotically and $\xi$ only an asymptotic Killing vector.
Due to the asymptotic divergence of the background metric \eqref{eq:bc2} it is not guaranteed that \eqref{1stbCSxi} vanishes.
One can show that requiring the first variation of the action \eqref{1stbCSxi} to vanish for a metric \eqref{eq:bc1} with boundary conditions \eqref{eq:bc4a} and a diffeomorphism generated by \eqref{eq:gCS28} enforces the following restrictions on the asymptotic metric fluctuations:
\eq{
h_{\pm y} = {\cal O}(1)\,, \qquad\qquad h_{++} h_{--} = {\cal O}(1/y)\,.
}{eq:gCS34}
The first condition eliminates the possibility for $1/y$ behavior in $h_{\pm y}$.
The second condition has two different solutions of interest:
\begin{equation}\label{eq:gCS35}
h_{++} = {\cal O}(1)\quad h_{--} = {\cal O}(1/y) \qquad\textrm{or}\qquad h_{++}  = {\cal O}(1/y)\quad h_{--} = {\cal O}(1)\,.
\end{equation}
Thus, even though the conditions \eqref{eq:gCS34} are invariant under a chirality exchange $+\leftrightarrow -$, the solutions \eqref{eq:gCS35} break this invariance.
This is reminiscent of the behavior for TMG in the range $|\mu\ell|<1$ discovered by Henneaux, Martinez and Troncoso \cite{Henneaux:2010fy}.
It also resembles the situation encountered in partially massless gravity \cite{Grumiller:2010tj}.
Moreover, also the consistency of the boundary value problem requires to single out one chirality \cite{Guica:2010sw}.
We choose, with no loss of generality, the second solution $h_{++}={\cal O}(1/y)$ and $h_{--}={\cal O}(1)$.
This implies that not all partially massless excitations are allowed by the boundary conditions, see the discussion in section \ref{se:3.modes}.

So far we have considered solely restrictions from asymptotic diffeomorphisms.
Now we address asymptotic Weyl rescalings, see \eqref{eq:gCS38}.
Residual Weyl rescalings \eqref{eq:gCS33} leave the action invariant due to the fall-off behavior of $\Om$.
Even Weyl rescalings for cases II and III of the form \eqref{eq:gCS29} and \eqref{eq:3Weyl}, respectively, leave invariant the action, provided that $h_{\pm y}={\cal O}(1)$.
This can be shown as follows.
Consider first a Weyl rescaling \eqref{eq:gCS29} with constant $f$.
The variation of the action then vanishes asymptotically, provided we restrict $h_{\pm y}={\cal O}(1)$.
This is again a restriction encountered above already.
If the function $f$ in \eqref{eq:gCS29} is not constant then the variation of the action leads to a boundary term
\eq{
\de S_{\rm CSG}[\Om] = \frac{k}{2\pi}\,\int_{\partial M}\!\!\!\extd\; (f \extd \Omega)\,.
}{eq:gCS39}
However, if $\partial M$ is a smooth boundary its boundary vanishes and so does the variation of the action \eqref{eq:gCS39}.
Therefore, there is no further restriction from requiring invariance under appropriate Weyl rescalings.

To summarize our findings so far, we cannot consistently impose the loose boundary conditions \eqref{eq:bc4a} if we want the first variation of the action to vanish for all asymptotic symmetries.
Thus, we end up with the following set of boundary conditions for the metric excitations:
\eq{
\left(\begin{array}{lll}
h_{++}={\cal O}(1/y) & h_{+-} = {\cal O}(1/y) & h_{+y} = {\cal O}(1) \\
              & h_{--} = {\cal O}(1)          & h_{-y} = {\cal O}(1) \\
              &                               & h_{yy} = {\cal O}(1/y)
\end{array}\right)\,.
}{eq:bc5}
This is not quite the set of boundary conditions \eqref{eq:bc3} we proposed in the main text.

So far we have only imposed the first requirement.
Let us now address the second requirement.
If $h_{+-}$ or $h_{yy}$ have a ${\cal O}(1/y)$ contribution then $\ga^{(1)}_{+-}$ does not vanish in the Fefferman--Graham expansion \eqref{eq:gCS44}.
The non-vanishing components of the 1-point function are given by
\begin{align}
 T_{++} &= -\frac{k}{\pi}\,\big(\ga^{(2)}_{++}-\ga^{(1)}_{+-}\ga^{(1)}_{++}\big)\,,\\
 T_{--} &= \frac{k}{\pi}\,\ga^{(2)}_{--}\,,\\
 \PMR_{++} &= \frac{k}{2\pi}\,\ga_{++}^{(1)}\,.
\end{align}
Note in particular the bilinear term in $\ga^{(1)}$ in the $T_{++}$ component, in stark contrast to \eqref{eq:gCS48}.
By asymptotically solving the field equations it can be shown that there exist classical solutions where $\ga^{(1)}_{+-}$ is such that the Brown--York stress tensor is not conserved, $\partial_-T_{++}\neq 0$ (see appendix \ref{app:EOM}).
Thus, the second requirement implies that neither $h_{+-}$ nor $h_{yy}$ have a ${\cal O}(1/y)$ contribution.
The third requirement holds automatically.

In summary, demanding the first variation of the action to vanish for all transformations that preserve the boundary conditions and postulating conservation of the Brown--York stress tensor for a flat boundary metric reduces the loose set of boundary conditions \eqref{eq:bc4a} to the tighter set \eqref{eq:bc3}, up to an exchange of chiralities, $+\leftrightarrow -$.

We show now that the results above are easy to comprehend in terms of the normalizability condition
\eq{
\Big|\int\extd^3x e^{-3\phi}\sqrt{-g}\,h_{\mu\nu}h^{\mu\nu}\Big|<\infty\,.
}{eq:norm1}
The factor $e^{-3\phi}$ leads to a Weyl-invariant volume form and ensures that $\phi$ drops out completely of the condition \eqref{eq:norm1}.
Inserting for the background metric \eqref{eq:bc2} and keeping only the essential contributions we obtain
\begin{equation}\label{eq:norm2}
\Big|\int\limits_0^\eps\extd y \,y\,h_{++}h_{--}\Big|<\infty \, , \qquad
\Big|\int\limits_0^\eps\extd y \,y\,h_{+-}^2\Big|<\infty \, ,\qquad
\Big|\int\limits_0^\eps\extd y \,y\,h_{yy}^2\Big|<\infty \, .
\end{equation}
The first condition shows that it is not possible to have both $h_{++}$ and $h_{--}$ of order ${\cal O}(1/y)$.
Rather, we have to pick one chirality.
The second and third conditions show that neither $h_{+-}$ nor $h_{yy}$ can be of order ${\cal O}(1/y)$.
All these features are implemented in our boundary conditions \eqref{eq:bc3} and are consistent with the discussion above.
Thus, normalizability \eqref{eq:norm1} provides an independent rationale for the boundary conditions \eqref{eq:bc3}.

\section{Trivial gauge transformations} \label{app:gauge}

In this appendix we show that the transformations putting the two quantities $\zeta(x^+, x^-)$ and $b$ in \eqref{eq:bc2} and \eqref{eq:gCS30} to zero correspond to vanishing boundary charges. All other charges are also independent of $\zeta(x^+, x^-)$ showing that this quantity is pure gauge. 
However, we shall also see that although the variation of $b$ leads to no contributions, in case III
there is a charge that depends non-trivially on $b$, which is why we choose to keep $b$ arbitrary but fixed in the main text for this case. 

Including non-trivial $\zeta$ and $b$ gives the following expansion for the dreibein:
\eq{
e^i{}_\mu = e^{f} \,y^{b-1}\, \begin{pmatrix} e^{\zeta}&0&0\\0 & e^{\zeta}&0 \\ 0&0&1 \end{pmatrix}
 + e^{f-\zeta}\, y^{b}\, \begin{pmatrix} 0&0&0\\\bar e_{(1) +}^{\, -} &0 &0 \\ 0&0&0 \end{pmatrix} 
 + e^{f-\zeta}\, y^{b+1} \,\begin{pmatrix} \bar e_{(2)+}^+ & \bar e_{(2)-}^+& \bar e_{(2)y}^+\\ \bar e_{(2)+}^- & \bar e_{(2)-}^- & \bar e_{(2)y}^- \\ \bar e_{(2)+}^y & \bar e_{(2)-}^y & \bar e_{(2)y}^y \end{pmatrix} + \ldots
}{eq:Feffe_gauge}
Allowing for varying $b$ and $\zeta$ the transformations that preserve the boundary conditions differ appreciably from the ones in 3-dimensional Einstein gravity:
\begin{subequations}
 \label{eq:gaugeasy}
\begin{align}
\xi^\pm &= \eps^\pm(x^\pm)  -y^2e^{-2\zeta}\partial_\mp f_y(x^+,x^-) + {\cal O}(y^3)\\
\xi^y &= y\,f_y(x^+,x^-) + {\cal O}(y^2) \\
\weyl &= b_\weyl \,\ln y + f_\weyl(x^+,x^-) + {\cal O}(y)\, .
\end{align}
\end{subequations}
Here $\eps^\pm$, $f_y$ and $f_\weyl$ are arbitrary functions of the indicated variables and $b_\weyl$ is an arbitrary constant.
It is clear that the transformation parametrized by the constant $b_\weyl$ can be used to change the quantity $b$. A suitable combination
of $f_y$ and $f_\weyl$ can similarly be used to put $\zeta \equiv 0$. In fact, the transformation $\de[f_y]$
where only $f_y \neq 0$ (but $\eps^\pm = f_\weyl = b_\weyl = 0$) transforms the functions $f$ and $\zeta$ as
\eq{
\de[f_y] \, f = b f_y \qquad \de[f_y] \, \zeta = - f_y\,.}{eq:fy}
It is therefore clear that choosing $f_y = \zeta$ and $f_\weyl = (1-b)\zeta$ puts $\zeta \equiv 0$. We shall show below that 
the corresponding asymptotic charges vanish.

Let us first consider the transformation $b_\weyl$. Using the expansion \eqref{eq:Feffe_gauge} it is straightforward to obtain from \eqref{eq:weylcharge}
\eq{
Q_W[b_\weyl] = -\frac{k}{\pi} \int \limits_0^{2\pi} \extd \varphi \, e^{i\mu} \, \partial_\mu( b_\weyl\, \ln y) \, \de e_{i\varphi} = 
-\frac{k}{\pi} \int_0^{2\pi} \extd \varphi \, e^{iy}\,  \frac{b_\weyl}{y}\, \de e_{i\varphi} = 0\,.
}{eq:zero_charge}
Thus the charge associated with varying $b$ vanishes. In addition the Weyl charges are independent of $b$:
\eq{
\begin{split}
\de Q_W[f_\weyl] &= -\frac{k}{\pi}\int \limits_0^{2\pi} \extd \varphi y^{-b} \, e^{-f-\zeta}\, \partial_\varphi \Omega 
\, \de (y^{b} \, e^{f + \zeta}) =\\
&= -\frac{k}{\pi}\int_0^{2\pi} \extd \varphi \, \left( \de b\ln y + \de(f + \zeta) \right) \partial_\varphi \Omega 
 = -\frac{k}{\pi}\int_0^{2\pi} \extd \varphi \,  \de(f + \zeta) \, \partial_\varphi \Omega \, .
 \end{split}
}{eq:genweyl}
We shall however see below that there are diffeomorphism charges that depend on $b$.

Let us now return to the transformation $\de[f_y]$, i.e., a diffeomorphism along the vector field $\Xi$ with 
\begin{subequations}
 \label{eq:gauge1}
\begin{align}
\Xi^\pm &=  -y^2e^{-2\zeta}\partial_\mp f_y(x^+,x^-) \\
\Xi^y &= y\,f_y(x^+,x^-)\,. 
\end{align}
\end{subequations}
The metric \eqref{eq:bc1} transforms as
\eq{
\de[f_y] \, g_{\mu\nu} = {\cal L}_\Xi g_{\mu\nu} = 2 b f_y g_{\mu\nu} + \frac{e^{2(f+\zeta)}}{2y^2} 
\begin{pmatrix} 0 & -2f_y  & 0\\ -2f_y & 0 & 0 \\ 0 & 0 & 0 \end{pmatrix} + \ldots
}{eq:gauge2}
from which \eqref{eq:fy} can be read off. From the transformation of the subleading terms we record that
\eq{
\de[f_y] \, \ga_{++} = {\cal L}_\Xi \ga_{++} =y^{2b} \, e^{2 f}(2 b f_y \bar{\gamma}_{++}  + 2\partial_+ \zeta \partial_+ f_y - \partial_+^2 f_y)\, .
}{eq:gauge3}
We see that the first term on the right-hand side corresponds to the transformation \eqref{eq:fy} of $f$. 
Therefore it follows that 
\eq{
\de[f_y] \, \left(\bar{\ga}^{(2)}_{++} + (\partial_+ \zeta)^2 - \partial^2_+ \zeta \right) = 0\, .
}{eq:gauge4}
To compute the charges corresponding to these vector fields we need the corresponding Lorentz parameters $\theta^i[\Xi]$.
These are obtained by requiring that 
\eq{
\epsilon^i{}_{jk}e^j{}_\mu \theta^k[\Xi] + \partial_\mu \Xi^\nu e^i{}_\nu +  \Xi^\nu \partial_\nu e^i{}_\mu =
 (\de[f_y] f) e^i{}_\mu +  (\de[f_y] \zeta) \frac{e^{f + \zeta}}{y} \begin{pmatrix}1 & 0&0\\ 0&1&0\\0&0&0\end{pmatrix}^i_{\mu} + {\rm subleading},
}{eq:gauge5}
i.e., that \eqref{eq:fy} holds also in the first order formalism. This yields
\eq{
\theta^\pm [\Xi] = \pm 2\, y\, e^{-\zeta} \, \partial_\mp f_y \qquad \theta^{\hat y}[\Xi] = \cO(y).
}{eq:gauge6}
Computing the relevant quantities in order to use \eqref{eq:diffch3} yields
\begin{align}
 \de C_\pm &= \cO(1) & \de C_y &= (1-b)\partial_\varphi(\de f)\, , & \label{eq:gauge7} \\
 \de \omega_{\pm \varphi} &= \pm \frac{e^\zeta}{2y}\left[ (b-1)\de \zeta + \de b\right] & \de \omega_{\hat y\varphi} &= \partial_t(\de f + \de \zeta)\, . &
\label{eq:gauge8}
\end{align}
Putting everything together we get 
\eq{
\de Q_P[\Xi] = -\frac{k}{\pi}\int\limits_0^{2\pi} \extd \varphi \, \Big[ \Xi^+ \, \de C_+  + \Xi^-\, \de C_-  + \Xi^y\, \de C_y + \theta^{\hat y}[\Xi]\,\de\om_{\hat y\varphi}\Big] = -\frac{k}{\pi}\int\limits_0^{2\pi} \extd \varphi \, (b-1) (\de f + \de \zeta)(\partial_\varphi f_y) \, .
}{eq:fycharge}
The $\de b$ contribution vanishes because it is a total $\varphi$-derivative. Note that $\de Q_P[\Xi]$ depends on the choice of $b$.

Comparing the result \eqref{eq:fycharge} result with Eq.~\eqref{eq:genweyl} reveals that a combination of a diffeomorphism along $\Xi^\mu$ and a Weyl rescaling with
$f_\weyl = (1-b)f_y$ has zero charge:
\eq{
Q_P[\Xi] + Q_W[(1-b)f_y] = 0.
}{eq:gauge9}
This combined transformation is therefore potentially a gauge transformation, and it sends
\eq{
f \to f + f_y \qquad \zeta \to \zeta - f_y.
}{eq:gauge10}
A convenient gauge fixing condition is $\zeta = 0$, which is what is used in the main text. An alternative would be to set $f = 0$, which for the case $b=0$
corresponds to AdS boundary conditions, but with a curved boundary metric. Because of the Weyl invariance of the bulk theory, the former choice is however technically simpler. Note that the combination $\tilde{f} \equiv f + \zeta$ is gauge
invariant, explaining why this is the combination occurring in \eqref{eq:genweyl} and \eqref{eq:fycharge}. 
To show that all charges corresponding to the transformations \eqref{eq:gaugeasy} are gauge invariant, we only have left to check the transformations
corresponding to $\eps^\pm$. Consider therefore a diffeomorphism parametrized by $\eps^+(x^+)\neq 0$ and $f_y = f_\weyl 
= b_\weyl = \eps^- = 0$.

Computing again the relevant quantities produces
\eq{
\de C_\pm =  \de\left( \bar{\gamma}^{(2)}_{\pm\pm} + (\partial_\pm \zeta)^2 - \partial_\pm^2 \zeta   \right) + \partial_\pm\partial_\varphi \de \tilde{f} - (\partial_\pm\tilde{f})(\partial_\varphi \de \tilde{f} ) 
\qquad \de C_y = (1-b)\partial_\varphi(\de f).
}{eq:gCS240}
Note that $\de C_\pm$ is gauge invariant on its own by \eqref{eq:gauge4}. 
The variation of the connection is still given by \eqref{eq:gauge8}, but the Lorentz parameter is now given by
\eq{
\theta^\pm = \cO(y^2) \qquad \theta^{\hat y} = \frac{1}{2} \partial_+ \eps^+  + \cO(y).
}{eq:gCS241}
Since the variation $\de \omega_{\hat y\varphi}$ is gauge invariant, the whole charge
\eq{
\de Q_P[\eps^+] = -\frac{k}{\pi}\int\limits_0^{2\pi} \extd \varphi \, \Big[\eps^+ \de C_+ + \theta^{\hat y} \de \omega_{\hat y\varphi} \Big]
}{eq:gCS242}
is also gauge invariant. The discussion for the transformation parametrized by $\eps^-$ is completely parallel. Thus we have shown that all charges are independent of how we fix the gauge for $\zeta$.

Finally let us show the following. In case II treated in the main text all asymptotic charges are independent of $b$. Therefore,
it is only in the most general case III where we need to keep $b$ arbitrary. Indeed, 
we have the following asymptotic symmetries in case II
\eq{
\xi^{\pm} = \eps^\pm - y^2 e^{-2\zeta} \partial_\mp A(x^+, x^-) \qquad \xi^y = yA(x^+, x^-) \qquad \Omega = -\eps\cdot\partial f - b A(x^+, x^-)\, ,
}{asyII}
where $\Omega$ is a compensating Weyl rescaling to ensure that $f$ remains unchanged. The function $A$ is given by 
\eq{
A = \eps \cdot \partial \zeta + \frac{1}{2}\partial \cdot \eps\, .
}{eq:gCS243}
The asymptotic symmetry group in \eqref{asyII} is a combination of a diffeomorphism by $\eps^\pm$, a diffeomorphism of the form \eqref{eq:gauge1} with parameter $f_y = A$ and
a Weyl rescaling. The total charge is the sum of these pieces:
\eq{
\de Q_{\rm tot}^{II} = \de Q_P[\eps^+] + \de Q_P[\eps^-] + \de Q_P[f_y = A] + \de Q_W[f_\weyl = -\eps\cdot\partial f - b A(x^+, x^-)]\, .
}{eq:gCS244}
We know already that the two first contributions are gauge invariant, and independent from $b$. Computing the other two yields
\eq{
\de Q_P[f_y = A] + \de Q_W[f_\weyl = -\eps\cdot\partial f - b A(x^+, x^-)] = 
-\frac{k}{\pi}\int\limits_0^{2\pi} \extd \varphi \,
\partial_\varphi\left( -\eps \cdot \partial \tilde{f} - \frac{1}{2}\partial \cdot \eps \right)\de \tilde{f}\, , 
}{eq:gCS245}
the $b$-dependence canceling and the $f$ and $\zeta$ contributions arranging themselves nicely into $\tilde{f}$.
Therefore, in case II it is permissible to fix $b = 0$, which is done in the main text.

Let us also note in passing that the full result for $Q_{\rm tot}^{II}$ simplifies to
\eq{
\de Q_{\rm tot}^{II} = -\frac{k}{\pi}\int\limits_0^{2\pi} \extd \varphi \,
\eps^+ \de\left( \bar{\gamma}^{(2)}_{++} + (\partial_+ \zeta)^2 - \partial_+^2 \zeta   \right)
+ \eps^- \de\left( \bar{\gamma}^{(2)}_{--} + (\partial_- \zeta)^2 - \partial_-^2 \zeta   \right)\, .
}{eq:gCS246}
These charges are conserved by the asymptotic equations of motion \eqref{eq:assolgen}
\begin{equation}
 \partial_{\mp}\bar \ga_{\pm\pm}^{(2)} = -2(\partial_{\pm}\zeta)(\partial_+\partial_-\zeta)+\partial_{\mp}\partial^2_{\pm}\zeta \, .
\end{equation}

\section{Classical and asymptotic analysis}\label{app:EOM}

The set of all classical solutions of CSG consists of all conformally flat metrics in three dimensions.
It is thus convenient to split the metric into an (arbitrary) conformal factor and some conformal class.
Restricting to stationary and axi-symmetric configurations all solutions for the conformal class can be found in closed form.
A Kaluza-Klein reduction (e.g.~on an $S^1$) to two dimensions reduces CSG to a specific non-linear Maxwell-Einstein theory \cite{Guralnik:2003we}.
This theory in turn can be mapped to a specific Dilaton-Maxwell-Einstein theory, whose classical solutions can be found globally \cite{Grumiller:2003ad}.
Apart from AdS$_3$ there are two types of solutions.
One is the BTZ black hole (``generic solutions'' in the classification of \cite{Grumiller:2003ad}), which we present for unit AdS length $\ell=1$:
\eq{
\extd s^2 = -\frac{(r^2-r_+^2)(r^2-r_-^2)}{r^2}\,\extd t^2 + \frac{r^2}{(r^2-r_+^2)(r^2-r_-^2)}\,\extd r^2 + 
r^2\big(\extd\varphi - \frac{r_+r_-}{r^2}\extd t\big)^2\, .
}{eq:BTZ}
The other type of solution is warped AdS, which in the present case simplifies to AdS$_2\times S^1$:
\eq{
\extd s^2 =  \mp\cosh^2\!\rho\,\extd\tau^2 + \extd\rho^2 \pm \extd\varphi^2\, .
}{eq:eom2}
Note that the spacetime \eqref{eq:eom2} admits a covariantly constant vector field $j=\partial_\varphi$.
The Ricci tensor is covariantly constant as well.
Up to the choice of conformal factor the solutions \eqref{eq:BTZ}, \eqref{eq:eom2} exhaust all stationary axi-symmetric solutions (locally and globally).
The conformal factor can further modify the global structure \cite{Oliva:2009hz}.
This can lead to new solutions, for instance hairy black holes
($r_+ > r_- > 0$) \cite{Oliva:2009ip}
\eq{
\extd s^2 = -(r-r_+)(r-r_-)\,\extd t^2 + \frac{\extd r^2}{(r-r_+)(r-r_-)} + r^2\,\extd\varphi^2
}{eq:hairyBH}
or solitons ($\alpha > -1$) \cite{Oliva:2009ip}
\eq{
\extd s^2 = -(\alpha + \cosh\rho)^2\,\extd t^2 + \extd\rho^2 + \sinh^2\!\rho\,\extd\varphi^2\,. }{eq:soliton}
The latter may be interpreted as partially massless graviton condensate on an AdS background; note, however, that this condensate violates the boundary conditions \eqref{eq:bc3}.

There is also a class of non-stationary supersymmetry-preserving solutions.
They can be found by taking the limit $\mu\rightarrow 0$ of the pp-wave solutions in
Refs.~\cite{Dereli:2001kq,AyonBeato:2004fq,Olmez:2005by,AyonBeato:2005qq,Carlip:2008eq,Garbarz:2008qn},
\eq{
\extd s^2= \ell^2\,\frac{\extd r^2}{r^2}+r^2\,\extd x^+\extd x^- + \ell\,r F(x^{\pm})\,(\extd x^{\pm})^2\,,
}{eq:pp-wave}
where $F(x^{\pm})$ is an arbitrary function, $\ell$ is some length scale and $x^\pm$ are dimensionless light cone coordinates.
The solutions \eqref{eq:pp-wave} are supersymmetric due to a null Killing vector $\partial_\mp$ \cite{Gibbons:2008vi}.
Note that for the upper sign in \eqref{eq:pp-wave} these solutions are compatible with the boundary conditions \eqref{eq:bc1}-\eqref{eq:bc3} for case~I.

We solve now the field equations \eqref{2.3a} asymptotically in conformal Gaussian coordinates \eqref{eq:bc1}-\eqref{eq:bc3} with trivial Weyl factors $\phi=0=\zeta$.
For simplicity we set $h_{y\mu}=0$.
We allow additionally for a trace-part to order $1/y$, denoted by $\ga_{+-}^{(1)}$, so that we can address the essential aspects of the looser set of boundary conditions \eqref{eq:bc5}.
We indicate expressions containing this term by a slash.
\begin{subequations}
\begin{align}
 h_{++} &= \frac{1}{y}\,\ga_{++}^{(1)} + \ga_{++}^{(2)} + y\,\ga_{++}^{(3)} + {\cal O} (y^2) \\
 h_{+-} &= \cancel{\frac{1}{y}\,\ga_{+-}^{(1)}} + \ga_{+-}^{(2)} + y\, \ga_{+-}^{(3)} + {\cal O} (y^2) \\
 h_{--} &= \phantom{\frac{1}{y}\,\ga_{++}^{(1)} + }\;\, \ga_{--}^{(2)} + y\, \ga_{--}^{(3)} + {\cal O} (y^2)
\end{align}
\end{subequations}
The Cotton tensor vanishes up to terms of ${\cal O}(y^2)$ if and only if the following relations hold:
\begin{subequations}
\label{eq:assol}
\begin{align}
 \partial_-^2\ga_{++}^{(1)} &= \ga_{++}^{(1)}\ga_{--}^{(2)}\, , \\
 \ga_{++}^{(2)} &= \tilde\ga_{++}^{(2)}(x^+) +  \cancel{\frac32\,\ga_{+-}^{(1)} \ga_{++}^{(1)}}\, , \\
 \ga_{--}^{(2)} &= \tilde \ga_{--}^{(2)}(x^-)\, ,\\
 \ga_{++}^{(3)} &= \frac53\,\ga_{++}^{(1)}\ga_{+-}^{(2)}-\frac13\,\partial_+\partial_-\ga_{++}^{(1)} + \cancel{\frac13\,\partial_+^2\ga_{+-}^{(1)} - \frac12\,(\ga_{+-}^{(1)})^2\ga_{++}^{(1)} + \ga_{+-}^{(1)} \tilde\ga_{++}^{(2)}} \, , \\
 \ga_{--}^{(3)} &= 0 + \cancel{\frac13\,\partial_-^2\ga_{+-}^{(1)} + \ga_{+-}^{(1)}\tilde\ga_{--}^{(2)}}\, .
\end{align}
\end{subequations}
All slashed terms vanish for the boundary conditions \eqref{eq:bc3} since the latter require $\ga_{+-}^{(1)}=0$.
The first condition implies a useful relation between the response functions \eqref{eq:response}.
The second and third conditions imply conservation of the stress-energy tensor \eqref{eq:gCS46} for the boundary conditions \eqref{eq:bc3}.
Note that $\ga^{(2)}_{+-}$ can be set to zero by a residual Weyl rescaling with an asymptotically trivial Weyl factor \eqref{eq:gCS31} that has a non-vanishing $y^2$-term,
accompanied by a residual diffeomorphism generated by an asymptotically trivial vector field $\xi={\cal O}(y^3)$ so that the total transformation preserves Gaussian quasi-normal coordinates to leading order, and similarly for $\ga^{(3)}_{+-}$.

If the boundary Weyl factor is non-vanishing, $\zeta\neq 0$, then the asymptotic solutions \eqref{eq:assol} with boundary conditions \eqref{eq:bc3} generalize to
\begin{subequations}
\label{eq:assolgen}
\begin{align}
 \partial_-^2\ga_{++}^{(1)} &= \ga_{++}^{(1)}\ga_{--}^{(2)} + 2 (\partial_-\zeta)\partial_-\ga_{++}^{(1)}\, ,\\
 \ga_{++}^{(2)} &= \tilde\ga_{++}^{(2)}(x^+) + \partial_+^2\zeta-(\partial_+\zeta)^2 \, ,\\
 \ga_{--}^{(2)} &= \tilde \ga_{--}^{(2)}(x^-) + \partial_-^2\zeta-(\partial_-\zeta)^2  \, ,\\
 \ga_{++}^{(3)} &= e^{-2\zeta}\Big(\frac53\,\ga_{++}^{(1)}\ga_{+-}^{(2)}-\frac13\,\partial_+\partial_-\ga_{++}^{(1)} -\frac23\,\ga_{++}^{(1)}\partial_+\partial_-\zeta +\frac43\,\partial_-\ga_{++}^{(1)}\partial_+\zeta\Big) \, ,\\
 \ga_{--}^{(3)} &= 0 \, .
\end{align}
\end{subequations}
Finally, we collect asymptotic expansions for metric, extrinsic curvature and related quantities in Gaussian normal coordinates \eqref{eq:gCS43}.
\begin{subequations}
\begin{align}
\ga_{\al\be} &= e^{2\rho}\ga_{\al\be}^{(0)} + e^\rho \ga_{\al\be}^{(1)} + \ga_{\al\be}^{(2)} + {\cal O}(e^{-\rho}) \, ,\\
\ga^{\al\be} &= e^{-2\rho}\ga^{\al\be}_{(0)} - e^{-3\rho} \ga^{\al\be}_{(1)} - e^{-4\rho}\big(\ga^{\al\be}_{(2)}-\ga^\al_{(1)\,\ga}\ga^{\ga\be}_{(1)}\big) + {\cal O}(e^{-5\rho})\, , \\
 K_{\al\be} &= e^{2\rho}\ga_{\al\be}^{(0)} + \frac12\,e^\rho \ga_{\al\be}^{(1)} + {\cal O}(e^{-\rho}) \, ,\\
 K^{\al\be} &= e^{-2\rho}\ga^{\al\be}_{(0)} - \frac32\,e^{-3\rho} \ga^{\al\be}_{(1)} - 2e^{-4\rho}\big(\ga^{\al\be}_{(2)}-\ga^\al_{(1)\,\ga}\ga^{\ga\be}_{(1)}\big) + {\cal O}(e^{-5\rho})\, , \\
 K &= 2-\frac12\,e^{-\rho}\ga_{\al\be}^{(1)}\ga^{\al\be}_{(0)} -e^{-2\rho}\big(\ga_{\al\be}^{(2)}\ga^{\al\be}_{(0)}-\frac12\,\ga_{\al\be}^{(1)}\ga^{\al\be}_{(1)}\big) + {\cal O}(e^{-3\rho})\, , \\
 k_{\al\be}^{L/R} &= \frac12\big(\de_\al^\ga\pm\eps_\al{}^\ga\big)\big(-\frac12\,e^\rho(\ga_{\ga\be}^{(1)}-\frac12\,\ga_{\ga\be}^{(0)}\ga_{\de\si}^{(1)}\ga^{\de\si}_{(0)})\nonumber \\
& -\ga_{\ga\be}^{(2)}+\frac14\,\ga_{\ga\be}^{(1)}\ga_{\de\si}^{(1)}\ga^{\de\si}_{(0)} +\frac12\,\ga_{\ga\be}^{(0)}(\ga_{\de\si}^{(2)}\ga^{\de\si}_{(0)}-\frac12\,\ga_{\de\si}^{(1)}\ga^{\de\si}_{(1)})\big)+{\cal O}(e^{-\rho})\, , \\
 \partial_\rho K_\al{}^\be &= \frac12\,e^{-\rho}\ga_\al^{(1)\,\be}+2e^{-2\rho}\big(\ga_\al^{(2)\,\be}-\frac12\,\ga_{\al\ga}^{(1)}\ga^{\ga\be}_{(1)}\big) +{\cal O}(e^{-3\rho})\, ,\\
 K_{\al\ga}K^{\ga\be} &= \de_\al^\be-e^{-\rho}\ga_\al^{(1)\,\be}-2e^{-2\rho}\big(\ga_\al^{(2)\,\be}-\frac58\,\ga_{\al\ga}^{(1)}\ga^{\ga\be}_{(1)}\big) + {\cal O}(e^{-3\rho})\, ,\\
 k_{\al\be}^L k^{\al\be}_R &= \frac18\,e^{-2\rho}\big(\ga_{\al\be}^{(1)}\ga^{\al\be}_{(1)}-\frac12\,(\ga_{\al\be}^{(1)}\ga^{\al\be}_{(0)})^2\big) + {\cal O}(e^{-3\rho}) \, .
\end{align}
\end{subequations}
The absence of ${\cal O}(1)$ and ${\cal O}(e^{-\rho})$ terms in the last equation is particularly noteworthy.

\section{Weyl rescaling formulas}\label{app:Weyl}

Under a Weyl rescaling \eqref{eq:Weyl} various geometric bulk expressions transform as follows:
\begin{subequations}
\begin{align}
 & g_{\mu\nu} = e^{2\Om}\, \bar g_{\mu\nu} \, , \\
 & \Ga^\la{}_{\mu\nu} =  \bar\Ga^\la{}_{\mu\nu} + \de^\la_\mu\partial_\nu\weyl+\de^\la_\nu\partial_\mu\weyl-\bar g_{\mu\nu} \bar g^{\la\si}\partial_\si\weyl  \, ,\\
 & L_{\mu\nu} = \bar L_{\mu\nu} - \bar\nabla_\mu\partial_\nu\Om + (\partial_\mu\Om)(\partial_\nu\Om) - \frac12\,\bar g_{\mu\nu} \bar g^{\la\si} (\partial_\la\Om)(\partial_\si\Om) \, ,\\
 & C_{\mu\nu} = \bar C_{\mu\nu} \, ,\\
 & \CS(\Ga) = \overline{\CS}(\bar\Ga) + \partial_\mu \big(\epsilon^{\mu\nu\la}\bar g^{\rho\si} (\partial_\nu\bar g_{\rho\la})(\partial_\si\Om)\big)\, .
\end{align}
\end{subequations}
The quantities $g$, $\Ga$, $L$, $C$ and CS are, respectively, metric, Christoffel connection, Schouten tensor, Cotton tensor and Chern--Simons density.

Various geometric boundary expressions transform as follows:
\begin{subequations}
\begin{align}
 & \ga_{\mu\nu} = e^{2\Om}\,\bar\ga_{\mu\nu}\, ,\\
 & n^\mu = e^{-\Om}\,\bar n^\mu \, ,\\
 & K_{\mu\nu} =  e^\Om\,\big(\bar K_{\mu\nu} +\bar\ga_{\mu\nu} \bar n^\la\partial_\la\Om\big)\, ,\\
 & K_{\mu\nu} - \frac12\,K\,\ga_{\mu\nu} = e^\Om\,\big(\bar K_{\mu\nu} - \frac12\,\bar K\,\bar\ga_{\mu\nu}\big) , \\
 & \sqrt{-\ga}\,k_{\mu\nu}^L k^{\mu\nu}_R = \sqrt{-\bar\ga}\,\bar k_{\mu\nu}^L \bar k^{\mu\nu}_R\, .
\end{align}
\end{subequations}
The quantities $\ga$, $n$, $K$ and $k^{L/R}$ are, respectively, boundary metric, unit normal vector, extrinsic curvature tensor and its chiral projections \eqref{eq:gCS68}.
The last expression proves Weyl invariance of the boundary term appearing in \eqref{eq:gCS67}.

Various first order expressions transform as follows:
\begin{subequations}
\begin{align}
 & e^i{}_\mu = e^\Om\bar e^i{}_\mu \, ,\\
 & \om^i{}_\mu = \bar\om^i{}_\mu + \eps^{ijk}\,\bar e_{j\mu} \bar e_k{}^\nu \partial_\nu\Om \, ,\\
 & \la^i{}_\mu = e^{-\Om} \Big[\bar\la^i{}_\mu + 2\bar e^{i\nu}\big(\bar\nabla_\mu\partial_\nu\Om - (\partial_\mu\Om)(\partial_\nu\Om) + \frac12\,\bar g_{\mu\nu}\bar g^{\la\si}(\partial_\la\Om)(\partial_\si\Om)\big)\Big] \, ,\\
 & T^i_{\mu\nu} = e^\Om\,\bar T^i_{\mu\nu}\, .
\end{align}
\end{subequations}
The quantities $e$, $\om$, $\la$ and $T$ are, respectively, triad, dualized connection, Lagrange multiplier 1-form and torsion 2-form.

\end{appendix}


\bibliographystyle{apsrev}

\begin{thebibliography}{10}

\bibitem{Afshar:2011yh}
H.~Afshar, B.~Cvetkovic, S.~Ertl, D.~Grumiller, and N.~Johansson, ``{Holograms
  of Conformal Chern-Simons Gravity},'' {\em Phys.Rev.} {\bf D84} (2011)
  041502(R), \href{http://www.arXiv.org/abs/1106.6299}{{\tt 1106.6299}}.

\bibitem{Staruszkiewicz:1963zz}
A.~Staruszkiewicz, ``{Gravitation theory in three-dimensional space},'' {\em
  Acta Phys. Polon.} {\bf 24} (1963) 734.

\bibitem{Deser:1982vy}
S.~Deser, R.~Jackiw, and S.~Templeton, ``Three-dimensional massive gauge
  theories,'' {\em Phys. Rev. Lett.} {\bf 48} (1982)
975--978.

\bibitem{Deser:1982wh}
S.~Deser, R.~Jackiw, and S.~Templeton, ``Topologically massive gauge
  theories,'' {\em Ann. Phys.} {\bf 140} (1982)
372--411.

\bibitem{Deser:1982a}
S.~Deser, R.~Jackiw, and S.~Templeton, ``Topologically massive gauge
  theories,'' {\em Erratum-ibid.} {\bf 185} (1988) 406.

\bibitem{Witten:1988hc}
E.~Witten, ``(2+1)-dimensional gravity as an exactly soluble system,'' {\em
  Nucl. Phys.} {\bf B311} (1988)
46.

\bibitem{Banados:1992wn}
M.~Banados, C.~Teitelboim, and J.~Zanelli, ``The black hole in
  three-dimensional space-time,'' {\em Phys. Rev. Lett.} {\bf 69} (1992)
  1849--1851,
\href{http://www.arXiv.org/abs/hep-th/9204099}{{\tt hep-th/9204099}}.

\bibitem{Banados:1992gq}
M.~Banados, M.~Henneaux, C.~Teitelboim, and J.~Zanelli, ``Geometry of the (2+1)
  black hole,'' {\em Phys. Rev.} {\bf D48} (1993) 1506--1525,
\href{http://www.arXiv.org/abs/gr-qc/9302012}{{\tt gr-qc/9302012}}.

\bibitem{Strominger:1997eq}
A.~Strominger, ``Black hole entropy from near-horizon microstates,'' {\em JHEP}
  {\bf 02} (1998) 009,
\href{http://www.arXiv.org/abs/hep-th/9712251}{{\tt hep-th/9712251}}.

\bibitem{Carlip:1998uc}
S.~Carlip, {\em Quantum gravity in 2+1 dimensions}.
\newblock Cambridge University Press, 1998.

\bibitem{Deser:1984tn}
S.~Deser, R.~Jackiw, and G.~'t~Hooft, ``Three-dimensional einstein gravity:
  Dynamics of flat space,'' {\em Ann. Phys.} {\bf 152} (1984)
220.

\bibitem{Witten:2007kt}
E.~Witten, ``{Three-Dimensional Gravity Revisited},''
\href{http://www.arXiv.org/abs/0706.3359}{{\tt 0706.3359}}.

\bibitem{Li:2008dq}
W.~Li, W.~Song, and A.~Strominger, ``{Chiral Gravity in Three Dimensions},''
  {\em JHEP} {\bf 04} (2008) 082,
\href{http://www.arXiv.org/abs/0801.4566}{{\tt 0801.4566}}.

\bibitem{Grumiller:2008qz}
D.~Grumiller and N.~Johansson, ``{Instability in cosmological topologically
  massive gravity at the chiral point},'' {\em JHEP} {\bf 07} (2008) 134,
\href{http://www.arXiv.org/abs/0805.2610}{{\tt 0805.2610}}.

\bibitem{Ertl:2009ch}
S.~Ertl, D.~Grumiller, and N.~Johansson, ``{Erratum to `Instability in
  cosmological topologically massive gravity at the chiral point',
  arXiv:0805.2610},''
\href{http://www.arXiv.org/abs/0910.1706}{{\tt 0910.1706}}.

\bibitem{Grumiller:2010tj}
D.~Grumiller, N.~Johansson, and T.~Zojer, ``{Short-cut to new anomalies in
  gravity duals to logarithmic conformal field theories},'' {\em JHEP} {\bf
  1101} (2011) 090, \href{http://www.arXiv.org/abs/1010.4449}{{\tt 1010.4449}}.

\bibitem{Paulos:2010ke}
M.~F. Paulos, ``{New massive gravity extended with an arbitrary number of
  curvature corrections},'' {\em Phys.Rev.} {\bf D82} (2010) 084042,
  \href{http://www.arXiv.org/abs/1005.1646}{{\tt 1005.1646}}.

\bibitem{Deser:1982sv}
S.~Deser, ``{Cosmological Topological Supergravity},'' in {\em Quantum Theory
  Of Gravity}, S.~M. Christensen, ed., pp.~374--381.
\newblock Adam Hilger, Bristol, 1984.
\newblock \href{http://www.arXiv.org/abs/Print-82-0692 (Brandeis)}{{\tt
  Print-82-0692 (Brandeis)}}.

\bibitem{Bergshoeff:2009hq}
E.~A. Bergshoeff, O.~Hohm, and P.~K. Townsend, ``{Massive Gravity in Three
  Dimensions},'' {\em Phys. Rev. Lett.} {\bf 102} (2009) 201301,
\href{http://www.arXiv.org/abs/0901.1766}{{\tt 0901.1766}}.

\bibitem{Bergshoeff:2009aq}
E.~A. Bergshoeff, O.~Hohm, and P.~K. Townsend, ``{More on Massive 3D
  Gravity},'' {\em Phys. Rev.} {\bf D79} (2009) 124042,
\href{http://www.arXiv.org/abs/0905.1259}{{\tt 0905.1259}}.

\bibitem{Sinha:2010ai}
A.~Sinha, ``{On the new massive gravity and AdS/CFT},'' {\em JHEP} {\bf 06}
  (2010) 061,
\href{http://www.arXiv.org/abs/1003.0683}{{\tt 1003.0683}}.

\bibitem{Gullu:2010pc}
I.~Gullu, T.~Cagri~Sisman, and B.~Tekin, ``{Born-Infeld extension of new
  massive gravity},'' {\em Class.Quant.Grav.} {\bf 27} (2010) 162001,
  \href{http://www.arXiv.org/abs/1003.3935}{{\tt 1003.3935}}.

\bibitem{Deser:1998rj}
S.~Deser and G.~Gibbons, ``{Born-Infeld-Einstein actions?},'' {\em
  Class.Quant.Grav.} {\bf 15} (1998) L35--L39,
  \href{http://www.arXiv.org/abs/hep-th/9803049}{{\tt hep-th/9803049}}.

\bibitem{Einstein:1915by}
A.~Einstein, ``Zur {A}llgemeinen {R}elativit{\"a}tstheorie,'' {\em Sitzungsber.
  Preuss. Akad. Wiss. Berlin (Math. Phys.)} {\bf 1915} (1915)
778--786.

\bibitem{Hilbert:1915tx}
D.~Hilbert, ``{Die Grundlagen der Physik. 1.},'' {\em Gott.Nachr.} {\bf 27}
  (1915) 395--407.

\bibitem{Horne:1988jf}
J.~H. Horne and E.~Witten, ``Conformal gravity in three-dimensions as a gauge
  theory,'' {\em Phys. Rev. Lett.} {\bf 62} (1989)
501--504.

\bibitem{Liu:2009pha}
Y.~Liu and Y.-W. Sun, ``{On the Generalized Massive Gravity in $AdS_3$},'' {\em
  Phys. Rev.} {\bf D79} (2009) 126001,
\href{http://www.arXiv.org/abs/0904.0403}{{\tt 0904.0403}}.

\bibitem{Liu:2009kc}
Y.~Liu and Y.-W. Sun, ``{Consistent Boundary Conditions for New Massive Gravity
  in $AdS_3$},'' {\em JHEP} {\bf 05} (2009) 039,
\href{http://www.arXiv.org/abs/0903.2933}{{\tt 0903.2933}}.

\bibitem{Grumiller:2009sn}
D.~Grumiller and O.~Hohm, ``{AdS$_3$/LCFT$_2$ -- Correlators in New Massive
  Gravity},'' {\em Phys. Lett.} {\bf B686} (2010) 264--267,
\href{http://www.arXiv.org/abs/0911.4274}{{\tt 0911.4274}}.

\bibitem{Oliva:2009ip}
J.~Oliva, D.~Tempo, and R.~Troncoso, ``{Three-dimensional black holes,
  gravitational solitons, kinks and wormholes for BHT masive gravity},'' {\em
  JHEP} {\bf 07} (2009) 011,
\href{http://www.arXiv.org/abs/0905.1545}{{\tt 0905.1545}}.

\bibitem{Deser:2009hb}
S.~Deser, ``{Ghost-free, finite, fourth order D=3 (alas) gravity},'' {\em Phys.
  Rev. Lett.} {\bf 103} (2009) 101302,
\href{http://www.arXiv.org/abs/0904.4473}{{\tt 0904.4473}}.

\bibitem{Maloney:2009ck}
A.~Maloney, W.~Song, and A.~Strominger, ``{Chiral Gravity, Log Gravity and
  Extremal CFT},'' {\em Phys. Rev.} {\bf D81} (2010) 064007,
\href{http://www.arXiv.org/abs/0903.4573}{{\tt 0903.4573}}.

\bibitem{Sachs:2008yi}
I.~Sachs, ``{Quasi-Normal Modes for Logarithmic Conformal Field Theory},'' {\em
  JHEP} {\bf 09} (2008) 073,
\href{http://www.arXiv.org/abs/0807.1844}{{\tt 0807.1844}}.

\bibitem{Grumiller:2008es}
D.~Grumiller and N.~Johansson, ``{Consistent boundary conditions for
  cosmological topologically massive gravity at the chiral point},'' {\em Int.
  J. Mod. Phys.} {\bf D17} (2009) 2367--2372,
\href{http://www.arXiv.org/abs/0808.2575}{{\tt 0808.2575}}.

\bibitem{Henneaux:2009pw}
M.~Henneaux, C.~Martinez, and R.~Troncoso, ``{Asymptotically anti-de Sitter
  spacetimes in topologically massive gravity},'' {\em Phys. Rev.} {\bf D79}
  (2009) 081502R,
\href{http://www.arXiv.org/abs/0901.2874}{{\tt 0901.2874}}.

\bibitem{Skenderis:2009nt}
K.~Skenderis, M.~Taylor, and B.~C. van Rees, ``{Topologically Massive Gravity
  and the AdS/CFT Correspondence},'' {\em JHEP} {\bf 09} (2009) 045,
\href{http://www.arXiv.org/abs/0906.4926}{{\tt 0906.4926}}.

\bibitem{Skenderis:2009kd}
K.~Skenderis, M.~Taylor, and B.~C. van Rees, ``{AdS boundary conditions and the
  Topologically Massive Gravity/CFT correspondence},''
\href{http://www.arXiv.org/abs/0909.5617}{{\tt 0909.5617}}.

\bibitem{Grumiller:2009mw}
D.~Grumiller and I.~Sachs, ``{AdS$_3$/LCFT$_2$ -- Correlators in Cosmological
  Topologically Massive Gravity},'' {\em JHEP} {\bf 03} (2010) 012,
\href{http://www.arXiv.org/abs/0910.5241}{{\tt 0910.5241}}.

\bibitem{Grumiller:2010rm}
D.~Grumiller and N.~Johansson, ``{Gravity duals for logarithmic conformal field
  theories},'' {\em J. Phys. Conf. Ser.} {\bf 222} (2010) 012047,
\href{http://www.arXiv.org/abs/1001.0002}{{\tt 1001.0002}}.

\bibitem{Giribet:2010ed}
G.~Giribet and M.~Leston, ``{Boundary stress tensor and counterterms for
  weakened AdS$_3$ asymptotic in New Massive Gravity},'' {\em JHEP} {\bf 09}
  (2010) 070,
\href{http://www.arXiv.org/abs/1006.3349}{{\tt 1006.3349}}.

\bibitem{Gaberdiel:2010xv}
M.~R. Gaberdiel, D.~Grumiller, and D.~Vassilevich, ``{Graviton 1-loop partition
  function for 3-dimensional massive gravity},'' {\em JHEP} {\bf 1011} (2010)
  094, \href{http://www.arXiv.org/abs/1007.5189}{{\tt 1007.5189}}.

\bibitem{Hosseiny:2011ct}
A.~Hosseiny and A.~Naseh, ``{On Holographic Realization of Logarithmic GCA},''
  \href{http://www.arXiv.org/abs/1101.2126}{{\tt 1101.2126}}.

\bibitem{Percacci:2010yk}
R.~Percacci and E.~Sezgin, ``{One Loop Beta Functions in Topologically Massive
  Gravity},'' {\em Class.Quant.Grav.} {\bf 27} (2010) 155009,
  \href{http://www.arXiv.org/abs/1002.2640}{{\tt 1002.2640}}.

\bibitem{Deser:1990bj}
S.~Deser and Z.~Yang, ``{IS TOPOLOGICALLY MASSIVE GRAVITY RENORMALIZABLE?},''
  {\em Class.Quant.Grav.} {\bf 7} (1990)
1603--1612.

\bibitem{Maloney:2007ud}
A.~Maloney and E.~Witten, ``{Quantum Gravity Partition Functions in Three
  Dimensions},'' {\em JHEP} {\bf 1002} (2010) 029,
  \href{http://www.arXiv.org/abs/0712.0155}{{\tt 0712.0155}}.

\bibitem{Deser:1983mm}
S.~Deser and R.~I. Nepomechie, ``Gauge invariance versus masslessness in de
  sitter space,'' {\em Ann. Phys.} {\bf 154} (1984)
396.

\bibitem{Deser:2001pe}
S.~Deser and A.~Waldron, ``{Gauge invariances and phases of massive higher
  spins in (A)dS},'' {\em Phys. Rev. Lett.} {\bf 87} (2001) 031601,
\href{http://www.arXiv.org/abs/hep-th/0102166}{{\tt hep-th/0102166}}.

\bibitem{Deser:2001us}
S.~Deser and A.~Waldron, ``{Partial masslessness of higher spins in (A)dS},''
  {\em Nucl. Phys.} {\bf B607} (2001) 577--604,
\href{http://www.arXiv.org/abs/hep-th/0103198}{{\tt hep-th/0103198}}.

\bibitem{Blagojevic:2010ir}
M.~Blagojevic and B.~Cvetkovic, ``{Hamiltonian analysis of BHT massive
  gravity},'' {\em JHEP} {\bf 1101} (2011) 082,
  \href{http://www.arXiv.org/abs/1010.2596}{{\tt 1010.2596}}.

\bibitem{Maldacena:1997re}
J.~M. Maldacena, ``{The large N limit of superconformal field theories and
  supergravity},'' {\em Adv. Theor. Math. Phys.} {\bf 2} (1998) 231--252,
\href{http://www.arXiv.org/abs/hep-th/9711200}{{\tt hep-th/9711200}}.

\bibitem{Aharony:1999ti}
O.~Aharony, S.~S. Gubser, J.~M. Maldacena, H.~Ooguri, and Y.~Oz, ``{Large N
  field theories, string theory and gravity},'' {\em Phys. Rept.} {\bf 323}
  (2000) 183--386,
\href{http://www.arXiv.org/abs/hep-th/9905111}{{\tt hep-th/9905111}}.

\bibitem{Carlip:2008qh}
S.~Carlip, ``{The Constraint Algebra of Topologically Massive AdS Gravity},''
  {\em JHEP} {\bf 10} (2008) 078,
\href{http://www.arXiv.org/abs/0807.4152}{{\tt 0807.4152}}.

\bibitem{Deser:1991qk}
S.~Deser and X.~Xiang, ``{Canonical formulations of full nonlinear
  topologically massive gravity},'' {\em Phys.Lett.} {\bf B263} (1991)
39--43.

\bibitem{Kraus:2005zm}
P.~Kraus and F.~Larsen, ``{Holographic gravitational anomalies},'' {\em JHEP}
  {\bf 01} (2006) 022,
\href{http://www.arXiv.org/abs/hep-th/0508218}{{\tt hep-th/0508218}}.

\bibitem{Garcia:2003bw}
A.~Garcia, F.~W. Hehl, C.~Heinicke, and A.~Macias, ``{The Cotton tensor in
  Riemannian spacetimes},'' {\em Class. Quant. Grav.} {\bf 21} (2004)
  1099--1118,
\href{http://www.arXiv.org/abs/gr-qc/0309008}{{\tt gr-qc/0309008}}.

\bibitem{Faddeev:1988qp}
L.~D. Faddeev and R.~Jackiw, ``Hamiltonian reduction of unconstrained and
  constrained systems,'' {\em Phys. Rev. Lett.} {\bf 60} (1988)
1692.

\bibitem{Grumiller:2008pr}
D.~Grumiller, R.~Jackiw, and N.~Johansson, ``{Canonical analysis of
  cosmological topologically massive gravity at the chiral point},'' in {\em
  {Fundamental Interactions - A Memorial Volume for Wolfgang Kummer}}.
\newblock World Scientific, 2009.
\newblock
\href{http://www.arXiv.org/abs/0806.4185}{{\tt 0806.4185}}.
\newblock

\bibitem{Castellani:1981us}
L.~Castellani, ``Symmetries in constrained hamiltonian systems,'' {\em Annals
  Phys.} {\bf 143} (1982) 357.

\bibitem{Brown:1986nw}
J.~D. Brown and M.~Henneaux, ``{Central Charges in the Canonical Realization of
  Asymptotic Symmetries: An Example from Three-Dimensional Gravity},'' {\em
  Commun. Math. Phys.} {\bf 104} (1986)
207--226.

\bibitem{Guica:2010sw}
M.~Guica, K.~Skenderis, M.~Taylor, and B.~C. van Rees, ``{Holography for
  Schrodinger backgrounds},'' {\em JHEP} {\bf 1102} (2011) 056,
  \href{http://www.arXiv.org/abs/1008.1991}{{\tt 1008.1991}}.

\bibitem{Grumiller:2002nm}
D.~Grumiller, W.~Kummer, and D.~V. Vassilevich, ``Dilaton gravity in two
  dimensions,'' {\em Phys. Rept.} {\bf 369} (2002) 327--429,
\href{http://arXiv.org/abs/hep-th/0204253}{{\tt hep-th/0204253}}.

\bibitem{Sachs:2008gt}
I.~Sachs and S.~N. Solodukhin, ``{Quasi-Normal Modes in Topologically Massive
  Gravity},'' {\em JHEP} {\bf 08} (2008) 003,
\href{http://www.arXiv.org/abs/0806.1788}{{\tt 0806.1788}}.

\bibitem{Afshar:2010ii}
H.~R. Afshar, M.~Alishahiha, and A.~E. Mosaffa, ``{Quasi-Normal Modes of
  Extremal BTZ Black Holes in TMG},'' {\em JHEP} {\bf 1008} (2010) 081,
  \href{http://www.arXiv.org/abs/1006.4468}{{\tt 1006.4468}}.

\bibitem{Solodukhin:2005ah}
S.~N. Solodukhin, ``{Holography with Gravitational Chern-Simons Term},'' {\em
  Phys. Rev.} {\bf D74} (2006) 024015,
\href{http://www.arXiv.org/abs/hep-th/0509148}{{\tt hep-th/0509148}}.

\bibitem{Kiritsis:2007}
E.~Kiritsis, {\em String theory in a nutshell}.
\newblock Princeton University Press, 2007.

\bibitem{Witten:2010cx}
E.~Witten, ``{Analytic Continuation Of Chern-Simons Theory},''
  \href{http://www.arXiv.org/abs/1001.2933}{{\tt 1001.2933}}.

\bibitem{Bertin:2011jk}
M.~Bertin, D.~Grumiller, D.~Vassilevich, and T.~Zojer, ``{Generalised massive
  gravity one-loop partition function and AdS/(L)CFT},'' {\em JHEP} {\bf 1106}
  (2011) 111, \href{http://www.arXiv.org/abs/1103.5468}{{\tt 1103.5468}}.

\bibitem{Castro:2011ui}
A.~Castro, T.~Hartman, and A.~Maloney, ``{The Gravitational Exclusion Principle
  and Null States in Anti-de Sitter Space},'' {\em Class.Quant.Grav.} {\bf 28}
  (2011) 195012, \href{http://www.arXiv.org/abs/1107.5098}{{\tt 1107.5098}}.

\bibitem{diFrancesco}
P.~Di~Francesco, P.~Mathieu, and D.~Senechal, {\em Conformal Field Theory}.
\newblock Springer, 1997.

\bibitem{Perez:2011qp}
A.~Perez, D.~Tempo, and R.~Troncoso, ``{Gravitational solitons, hairy black
  holes and phase transitions in BHT massive gravity},'' {\em JHEP} {\bf 1107}
  (2011) 093, \href{http://www.arXiv.org/abs/1106.4849}{{\tt 1106.4849}}.

\bibitem{Ammon:2011nk}
M.~Ammon, M.~Gutperle, P.~Kraus, and E.~Perlmutter, ``{Spacetime Geometry in
  Higher Spin Gravity},'' \href{http://www.arXiv.org/abs/1106.4788}{{\tt
  1106.4788}}.

\bibitem{Castro:2011fm}
A.~Castro, E.~Hijano, A.~Lepage-Jutier, and A.~Maloney, ``{Black Holes and
  Singularity Resolution in Higher Spin Gravity},''
  \href{http://www.arXiv.org/abs/1110.4117}{{\tt 1110.4117}}.

\bibitem{Chen:2011vp}
B.~Chen, J.~Long, and J.-b. Wu, ``{Spin-3 Topological Massive Gravity},''
  \href{http://www.arXiv.org/abs/1106.5141}{{\tt 1106.5141}}.

\bibitem{Bagchi:2011vr}
A.~Bagchi, S.~Lal, A.~Saha, and B.~Sahoo, ``{Topologically Massive Higher Spin
  Gravity},'' \href{http://www.arXiv.org/abs/1107.0915}{{\tt 1107.0915}}.

\bibitem{Bagchi:2011td}
A.~Bagchi, S.~Lal, A.~Saha, and B.~Sahoo, ``{One loop partition function for
  Topologically Massive Higher Spin Gravity},''
  \href{http://www.arXiv.org/abs/1107.2063}{{\tt 1107.2063}}.

\bibitem{Chen:2011xx}
B.~Chen and J.~Long, ``{High Spin Topologically Massive Gravity},''
  \href{http://www.arXiv.org/abs/1110.5113}{{\tt 1110.5113}}.

\bibitem{Chu:2009gi}
X.~Chu and B.~E. Nilsson, ``{Three-dimensional topologically gauged N=6 ABJM
  type theories},'' {\em JHEP} {\bf 1006} (2010) 057,
  \href{http://www.arXiv.org/abs/0906.1655}{{\tt 0906.1655}}.

\bibitem{Chu:2010fk}
X.~Chu, H.~Nastase, B.~E. Nilsson, and C.~Papageorgakis, ``{Higgsing M2 to D2
  with gravity: N=6 chiral supergravity from topologically gauged ABJM
  theory},'' {\em JHEP} {\bf 1104} (2011) 040,
  \href{http://www.arXiv.org/abs/1012.5969}{{\tt 1012.5969}}.

\bibitem{Aharony:2008ug}
O.~Aharony, O.~Bergman, D.~L. Jafferis, and J.~Maldacena, ``{N=6 superconformal
  Chern-Simons-matter theories, M2-branes and their gravity duals},'' {\em
  JHEP} {\bf 0810} (2008) 091, \href{http://www.arXiv.org/abs/0806.1218}{{\tt
  0806.1218}}.

\bibitem{Deser:1983sw}
S.~Deser and J.~H. Kay, ``Topologically massive supergravity,'' {\em Phys.
  Lett.} {\bf B120} (1983)
97--100.

\bibitem{vanNieuwenhuizen:1985cx}
P.~van Nieuwenhuizen, ``D = 3 conformal supergravity and {C}hern-{S}imons
  terms,'' {\em Phys. Rev.} {\bf D32} (1985)
872.

\bibitem{Mannheim:1988dj}
P.~D. Mannheim and D.~Kazanas, ``Exact vacuum solution to conformal weyl
  gravity and galactic rotation curves,'' {\em Astrophys. J.} {\bf 342} (1989)
635--638.

\bibitem{Lu:2011zk}
H.~Lu and C.~Pope, ``{Critical Gravity in Four Dimensions},'' {\em
  Phys.Rev.Lett.} {\bf 106} (2011) 181302,
  \href{http://www.arXiv.org/abs/1101.1971}{{\tt 1101.1971}}.

\bibitem{'tHooft:2011we}
G.~'t~Hooft, ``{A class of elementary particle models without any adjustable
  real parameters},'' \href{http://www.arXiv.org/abs/1104.4543}{{\tt
  1104.4543}}.

\bibitem{Maldacena:2011mk}
J.~Maldacena, ``{Einstein Gravity from Conformal Gravity},''
  \href{http://www.arXiv.org/abs/1105.5632}{{\tt 1105.5632}}.

\bibitem{Blagojevic:2011qc}
M.~Blagojevic and B.~Cvetkovic, ``{Extra gauge symmetries in BHT gravity},''
  {\em JHEP} {\bf 1103} (2011) 139,
  \href{http://www.arXiv.org/abs/1103.2388}{{\tt 1103.2388}}.

\bibitem{Tonni:2010gb}
E.~Tonni, ``{Warped black holes in 3D general massive gravity},'' {\em JHEP}
  {\bf 1008} (2010) 070, \href{http://www.arXiv.org/abs/1006.3489}{{\tt
  1006.3489}}.

\bibitem{Porfyriadis:2010vg}
A.~P. Porfyriadis and F.~Wilczek, ``{Effective Action, Boundary Conditions, and
  Virasoro Algebra for AdS$_3$},''
\href{http://www.arXiv.org/abs/1007.1031}{{\tt 1007.1031}}.

\bibitem{Henneaux:2010fy}
M.~Henneaux, C.~Martinez, and R.~Troncoso, ``{More on Asymptotically Anti-de
  Sitter Spaces in Topologically Massive Gravity},'' {\em Phys.Rev.} {\bf D82}
  (2010) 064038, \href{http://www.arXiv.org/abs/1006.0273}{{\tt 1006.0273}}.

\bibitem{Guralnik:2003we}
G.~Guralnik, A.~Iorio, R.~Jackiw, and S.~Y. Pi, ``{Dimensionally reduced
  gravitational Chern-Simons term and its kink},'' {\em Ann. Phys.} {\bf 308}
  (2003) 222--236,
\href{http://www.arXiv.org/abs/hep-th/0305117}{{\tt hep-th/0305117}}.

\bibitem{Grumiller:2003ad}
D.~Grumiller and W.~Kummer, ``{The classical solutions of the dimensionally
  reduced gravitational Chern-Simons theory},'' {\em Ann. Phys.} {\bf 308}
  (2003) 211--221,
\href{http://www.arXiv.org/abs/hep-th/0306036}{{\tt hep-th/0306036}}.

\bibitem{Oliva:2009hz}
J.~Oliva, D.~Tempo, and R.~Troncoso, ``{Static spherically symmetric solutions
  for conformal gravity in three dimensions},'' {\em Int.J.Mod.Phys.} {\bf A24}
  (2009) 1588--1592, \href{http://www.arXiv.org/abs/0905.1510}{{\tt
  0905.1510}}.

\bibitem{Dereli:2001kq}
T.~Dereli and O.~Sarioglu, ``Supersymmetric solutions to topologically massive
  gravity and black holes in three dimensions,'' {\em Phys. Rev.} {\bf D64}
  (2001)
027501.

\bibitem{AyonBeato:2004fq}
E.~Ayon-Beato and M.~Hassaine, ``{pp waves of conformal gravity with
  self-interacting source},'' {\em Annals Phys.} {\bf 317} (2005) 175--181,
  \href{http://www.arXiv.org/abs/hep-th/0409150}{{\tt hep-th/0409150}}.

\bibitem{Olmez:2005by}
S.~Olmez, O.~Sarioglu, and B.~Tekin, ``{Mass and angular momentum of
  asymptotically ads or flat solutions in the topologically massive gravity},''
  {\em Class.Quant.Grav.} {\bf 22} (2005) 4355--4362,
  \href{http://www.arXiv.org/abs/gr-qc/0507003}{{\tt gr-qc/0507003}}.

\bibitem{AyonBeato:2005qq}
E.~Ayon-Beato and M.~Hassaine, ``{Exploring AdS waves via nonminimal
  coupling},'' {\em Phys.Rev.} {\bf D73} (2006) 104001,
  \href{http://www.arXiv.org/abs/hep-th/0512074}{{\tt hep-th/0512074}}.
  Dedicated to the memory of Professor Jerzy Plebanski.

\bibitem{Carlip:2008eq}
S.~Carlip, S.~Deser, A.~Waldron, and D.~K. Wise, ``{Topologically Massive AdS
  Gravity},'' {\em Phys. Lett.} {\bf B666} (2008) 272--276,
\href{http://www.arXiv.org/abs/0807.0486}{{\tt 0807.0486}}.

\bibitem{Garbarz:2008qn}
A.~Garbarz, G.~Giribet, and Y.~Vasquez, ``{Asymptotically AdS$_3$ Solutions to
  Topologically Massive Gravity at Special Values of the Coupling Constants},''
  {\em Phys. Rev.} {\bf D79} (2009) 044036,
\href{http://www.arXiv.org/abs/0811.4464}{{\tt 0811.4464}}.

\bibitem{Gibbons:2008vi}
G.~W. Gibbons, C.~N. Pope, and E.~Sezgin, ``{The General Supersymmetric
  Solution of Topologically Massive Supergravity},'' {\em Class. Quant. Grav.}
  {\bf 25} (2008) 205005,
\href{http://www.arXiv.org/abs/0807.2613}{{\tt 0807.2613}}.

\end{thebibliography}

\providecommand{\href}[2]{#2}\begingroup\raggedright\endgroup

\end{document}